\documentclass[journal, onecolumn, draftclsnofoot, letterpaper, twoside, 11pt]{IEEEtran}

\usepackage{hyperref}
\usepackage{graphicx}
\usepackage{subfig}
\usepackage[utf8]{inputenc}
\usepackage{graphics}
\usepackage{amsmath}
\usepackage{amssymb}
\usepackage{mathtools}
\usepackage{color}
\usepackage[a4paper,total={6in,10in}]{geometry}
\usepackage{babel}[english]
\usepackage{makecell}
\usepackage{amsthm}
\usepackage{verbatim}
\usepackage{enumitem}
\usepackage{algorithm}
\usepackage{algpseudocode}
\usepackage{multirow}
\usepackage{marginnote}

\newtheorem{definition}{Definition}[section]
\newtheorem{proposition}{Proposition}[section]
\newtheorem{theorem}{Theorem}[section]
\newtheorem{corollary}{Corollary}[theorem]
\newtheorem{lemma}{Lemma}[subsection]

\usepackage{titlesec}
\titleformat{\paragraph}{\normalfont\normalsize\bfseries}
{\theparagraph}{1em}{}
\titlespacing*{\paragraph}{0pt}{3.25ex plus 1ex minus .2ex}{1.5ex plus .2ex}

\newcommand{\n}{n^*}
\newcommand{\thre}{c^*}
\newcommand{\cA}{\mathcal{A}}
\newcommand{\cC}{\mathcal{C}}
\newcommand{\cE}{\mathcal{E}}
\newcommand{\cF}{\mathcal{F}}
\newcommand{\cL}{\mathcal{L}}
\newcommand{\Pro}{\mathsf{P}}
\newcommand{\Exp}{\mathsf{E}}
\newcommand{\bN}{\mathbb{N}}
\newcommand{\bR}{\mathbb{R}}
\newcommand{\ST}{T^\prime}

\newcommand{\GMT}{\hat T}

\newcommand{\alpham}{\alpha_m}
\newcommand{\betam}{\beta_m}
\newcommand{\bcE}{\boldsymbol{\cE}}
\newcommand{\bcL}{\boldsymbol{\cL}}
\newcommand{\sorted}{\mathsf{N}}
\newcommand{\sortedc}{\mathsf{C}}
\newcommand{\sortedl}{\mathsf{L}}
\newcommand{\myremark}{\noindent\underline{\textbf{Remark}}: }

\begin{document}

\title{Signal Recovery With Multistage Tests \\ And Without  Sparsity Constraints}

\author{Yiming~Xing~
and~Georgios~Fellouris
\thanks{Yiming Xing and Georgios Fellouris are with the Department of Statistics, University of Illinois at Urbana-Champaign, Urbana-Champaign, IL, USA (email: yimingx4@illinois.edu, fellouri@illinois.edu).}
\thanks{This research was supported in part by the US National Science
Foundation under grant ATD-1737962 through the University of
Illinois at Urbana-Champaign.}
\thanks{A portion of this work was presented in 2022 IEEE International Symposium on Information Theory in Espoo, Finland \cite{our_conf}.}}


\maketitle
\begin{abstract}
A  signal recovery problem is considered, where the same  binary testing problem is posed over multiple, independent data streams. The goal is  to identify all signals, i.e., streams where the  alternative hypothesis is correct, and noises, i.e., streams where the null  hypothesis is correct,  subject to  prescribed bounds on the  classical or generalized familywise error probabilities.   It is not   required that the exact number of signals be a priori known, only upper bounds on the number of signals and noises are assumed instead.   A decentralized formulation is adopted, according to which  the sample size and the decision for each testing problem  must be based only on observations  from the corresponding data stream. A novel  multistage testing procedure  is proposed for this problem and is  shown  to  enjoy a high-dimensional asymptotic optimality property. Specifically, it achieves the optimal, average over all streams, expected sample size, uniformly in the true number of signals, as the  maximum possible numbers of signals and noises go to infinity at arbitrary rates, in the class of all sequential tests with the same global error control.    In contrast, existing multistage tests in the literature are shown to  achieve this  high-dimensional asymptotic optimality property only  under additional sparsity or symmetry  conditions. These results are based on an asymptotic analysis for the fundamental binary testing problem  as the two error probabilities go to zero. For this problem, unlike existing multistage tests in the literature,  the proposed  test  achieves the optimal expected sample size  under both hypotheses,  in the class of all sequential tests with the same error control, as the two error probabilities go to zero at arbitrary rates.  These results  are further supported by simulation studies and  extended to problems with non-iid data and composite hypotheses.
\end{abstract}



\IEEEpeerreviewmaketitle

\section{Introduction}
Multistage tests are testing procedures in which the sampling process can  be terminated   only at a small number of time instances. As they can be more efficient than fixed-sample-size tests, which do not allow for early stopping, and  more practical than fully sequential tests,  which allow for stopping at every time instance, they are commonly applied  in  areas such as inspection control \cite{Dodge1929} and clinical trials \cite{Pocock_1977, Lan_DeMets_1983, Jennison_1987, Lai_Shih_2004, Bartroff_Lai_2011}. For general textbook references on multistage tests, we refer to \cite{SamplingInspectionBook, Jennison_Turnbull_Book, Bartroff_Book_Clinicaltrials}.

For the fundamental problem of testing two simple hypotheses, multistage tests with at most 3 or 4 stages  and deterministic stage sizes have been shown, in the case of iid observations \cite{Lorden_1983} and more generally \cite{PaperI}, to achieve the optimal expected sample size under both hypotheses, in the class of all sequential tests with the same control of the type-I and type-II error probabilities, as the two error probabilities go to zero \textit{as long as they do not do so very asymmetrically}. The first goal of the present work is to introduce a multistage test, to which we refer as the \textit{General Multistage Test}, that is asymptotically optimal under both hypotheses as the two error probabilities go to zero \emph{at arbitrary rates}. The proposed test adds, if necessary,  to the 3-Stage Test in \cite{Lorden_1983, PaperI} a  number of opportunities either only  to accept or only to  reject the null hypothesis. This  number is specified explicitly and depends on the relative magnitude of the two user-specified error probabilities.

The second goal of the present work is to apply the proposed multistage test to  a high-dimensional signal recovery problem, where  a large number of pairs of hypotheses are tested  simultaneously and the problem is to correctly identify the  data streams in which the alternative (resp.  null) hypothesis holds,  to which we refer as \textit{signals} (resp.  \textit{noises}).  This problem arises  in various scientific and engineering applications, e.g., genetics \cite{ST9, Zehetmayer_2005, zehetmayer2008optimized, ST8}, spectrum sensing in cognitive radio \cite{tajer2012adaptive, geng2016quickest}, searching for regions of interest (ROI) in an image or other mediums \cite{ST23, ST24}.

In the present work, we  consider a formulation for the  signal recovery problem that generalizes the one adopted in  \cite{Malloy_Nowak_2014}.  In the latter,  (i) there are multiple, independent data streams, (ii) each of them  generates iid data, (iii)  the same binary testing problem is posed for each of them,  (iv) the same testing procedure must be  applied to each testing problem, using data only from the corresponding data stream, (v)   the  misclassification probability, i.e., the probability of at least one error of either kind,  is controlled, and (vi) the \textit{exact} numbers of  signals and noises  are assumed to be known a priori. In \cite{Malloy_Nowak_2014}, a  multistage test, termed \textit{Sequential Thresholding}, is introduced  and  shown  to achieve the optimal, average over all streams,  expected sample size  as the target misclassification probability remains fixed and the number of data streams goes to infinity,  \textit{under a  sparsity condition on the a priori known number of signals}.

The signal recovery problem that we consider  relaxes features (v) and (vi) in \cite{Malloy_Nowak_2014}.  Specifically, we require control below distinct, arbitrary, user-specified levels  of the  probabilities of at least one type-I error and at least one type-II error (classical familywise error probabilities) or, more generally, of  the probabilities  of at least $\kappa$ type-I errors and at least $\iota$ type-II errors (generalized familywise error probabilities \cite{lehmann2012generalizations}), where $\kappa$ and $\iota$ are user-specified positive integers that can be larger than one. Moreover, we do not assume that the number of signals is known a priori, which is a rather unrealistic assumption, especially when the number of streams is large. Instead, we only require that the \textit{maximum} possible numbers of signals and noises be specified.    

In this more general setup, we formulate a novel asymptotic optimality criterion, according to which the   optimal, average over all streams,  expected sample size, in the class of all sequential tests with the same global  error control, is achieved as the number of streams goes to  infinity \textit{uniformly in the true number of  signals}. We show that   the proposed multistage test, i.e.,  the General Multistage Test, enjoys this property as long as the maximum possible numbers of signals and noises go to infinity. On the other hand, Sequential Thresholding (as well as a modification of this test that we introduce in this work)  requires an additional sparsity condition on the maximum possible number of signals, whereas the 3-stage test in \cite{Lorden_1983} requires an additional symmetry condition on the maximum possible numbers of signals and noises.  

The theoretical results in this work are supported by two numerical studies, one for the binary testing problem and one for the signal recovery problem. Finally, they  are   extended to setups with non-iid data or composite hypotheses. Indeed,  all results in this work (apart from some that refer to the Sequential Thresholding), as well as their proofs, are shown to remain valid for many testing problems with neither  independent nor identically distributed observations. Moreover,  we show that for  the one-sided testing problem for a one-parameter exponential family,  the asymptotic optimality  theory developed  in this work  applies to the   optimal \textit{worst-case} expected sample size under each hypothesis. 

The remainder of this work is organized as follows: in Sections \ref{sec: problem formulation of one-dim testing}-\ref{section, ST}  we focus on the  binary testing problem, where we formulate the problem and present some preliminary results in Section \ref{sec: problem formulation of one-dim testing}, revisit the 3-Stage Test of \cite{Lorden_1983} in Section \ref{section, 3ST},  introduce and analyze the proposed multistage test   in Section \ref{section, GMT}, and revisit  the Sequential Thresholding in \cite{Malloy_Nowak_2014} and propose a modified version for it in Section \ref{section, ST}. In Section \ref{sec: problem formulation about high-dim} we consider the high-dimensional signal recovery problem. In Section \ref{sec: numerical study} we present the numerical studies, in Section \ref{sec: generalizations} we discuss  generalizations  of the present work, and in Section \ref{sec: conclusion} we conclude and pose some open problems. The proofs of most results are presented in  Appendices \ref{proofs about FSST}-\ref{proofs about hign-dim}.

We end this introductory section with some  notations that we use throughout the paper. We denote by $\bN$ the set of positive integers and by $\bR$ the set of real numbers. For any $n \in \{0\}\cup\bN$ we set:
$$[n]\equiv 
\begin{cases}
\begin{aligned}
    & \emptyset, \quad && \text{if}  \; n=0, \\
    & \{1,\ldots,n\}, \quad && \text{if}  \; n \in \bN.
\end{aligned}
\end{cases}
$$
For any $x\in\bR$, we denote by $\lceil x \rceil$ the smallest integer that is greater than or equal to $x$ and by $\lfloor x \rfloor$ the greatest integer that is less than or equal to $x$. For any $x,y\in\bR$, we set $x\vee y \equiv  \max\{x,y\}$ and $x\wedge y \equiv \min\{x,y\}$. For a sequence of real numbers $\{x_n, n\in\bN\}$ and integers $l,u$, we make the convention that
$$\sum_{n=l}^u x_n  \equiv 0\ \quad \text{if} \quad  l>u.$$
For two sequences of positive real numbers $\{x_n,n\in\bN\}$ and $\{y_n, n\in\bN\}$, $x_n\sim y_n$ stands for   $\lim (x_n/y_n)=1$, $x_n\gtrsim y_n$ stands for  $\liminf (x_n/y_n) \geq 1$, $x_n\lesssim y_n$    stands for   $\limsup (x_n/y_n) \leq 1$, $x_n\ll y_n$   stands for   $\lim (x_n/y_n)=0$, $x_n\gg y_n$  stands for  $\lim (x_n/y_n)=\infty$, $x_n=O(y_n)$  means that there exists a  $C>0$ such that  $x_n\leq C\, y_n$  for  all $n\in\bN$, and $x_n=\Theta(y_n)$  means that there exists a  $C>0$ such that   $y_n/C\leq x_n \leq C\,y_n$ for  all $n\in\bN$. Finally, for a function $g:\bR\to \bR$ we denote by $g(x+)$ its right limit at $x$ and by  $g(x-)$ its left limit at $x\in\bR$,  when they are  well-defined.

\section{Binary testing} \label{sec: problem formulation of one-dim testing}
\subsection{Problem formulation}
We let  $X\equiv \{X_n,\,n\in\bN\}$ be a sequence of independent random elements with common density, $f$, with respect to a $\sigma$-finite measure, $\nu$, and we  consider the problem of testing two simple hypotheses about $f$:
\begin{equation} \label{one-dim testing problem}
    H_0: f=f_0 \quad \text{ versus } \quad H_1: f=f_1.
\end{equation}
The only assumption regarding $f_0$ and $f_1$ is that their Kullback-Leibler  divergences are positive and finite, i.e., 
\begin{align} \label{KL information numbers}
\begin{aligned}
    I_0 &\equiv \int \log \left( \frac{f_0}{f_1} \right) f_0 \, d \nu\in (0,\infty),\\
    I_1 &\equiv \int \log \left( \frac{f_1}{f_0} \right) f_1 \, d \nu \in (0,\infty).
\end{aligned}
\end{align} 
For each  $i \in \{0,1\}$,  we denote by $\Pro_i$ the distribution of $X$, and by $\Exp_i$ the corresponding expectation, when $f=f_i$. For each $n \in \bN$, we denote  by $\cF_n$  the $\sigma$-algebra generated by the first $n$ observations, i.e., $\cF_n\equiv\sigma(X_1,\ldots,X_n)$, by $\Lambda_n$ the corresponding log-likelihood ratio, i.e., 
\begin{equation} \label{def: LLR}
    \Lambda_n \equiv \sum_{i=1}^n \log  \left( \frac{f_1(X_i)}{f_0(X_i)} \right),
\end{equation}
and by $\bar\Lambda_n$ the corresponding  average log-likelihood ratio, i.e.,  $\bar\Lambda_n \equiv  \Lambda_n/n$.

A \textit{sequential test}, or simply, a \textit{test}, for this testing problem consists of a random time, $T$, that represents the number of  observations until a decision is made, and  a  Bernoulli random variable, $D$, that represents the decision, i.e.,   $H_0$  is rejected if and only if  $D=1$. The determination at each time instance whether to  stop sampling and, if so, which hypothesis to select must depend only on the observations that have been collected up to this time.  Therefore, we say that  a  pair $(T,D)$ is a test,  if
\begin{itemize}
\item $T$ is an $\{\cF_n, \, n\in \bN\}$-stopping time, i.e., 
$\{T=n\} \in \cF_n$ for every $n \in \bN$, 
\item $D$ is  an $\cF_T$-measurable Bernoulli random variable, i.e.,  $\{T=n, \, D=i\} \in \cF_n$
for every  $n \in \bN$ and $i \in \{0,1\}$.
\end{itemize}

We denote by $\cE$ the family of all  tests and, for  $\alpha,\beta\in (0,1)$, by $\cE(\alpha,\beta) $ the subfamily of tests whose  type-I and type-II  error probabilities  do not exceed $\alpha$ and $\beta$ respectively, i.e.,
\begin{equation} \label{class(alpha,beta)}
    \cE(\alpha,\beta)\equiv \big\{ (T,D)\in \cE: \; \Pro_0(D=1)\leq\alpha \; \text{ and }\; \Pro_1(D=0)\leq \beta \big\}, 
\end{equation}
and by $\cL_i(\alpha,\beta)$  the optimal expected sample size under $\Pro_i$  in $\cE(\alpha,\beta)$, i.e.,  
\begin{equation} \label{one-dim, def: optimal performance}
     \cL_i(\alpha,\beta) \equiv \inf_{(T,D)\in \cE(\alpha,\beta)} \Exp_i[T], \quad \text{where} \;\; \;  i \in \{0,1\}.
\end{equation}

We refer to a test  $(T,D) \in \cE$  as  \textit{fully sequential} if its stopping time, $T$, can take any value in $\bN$, and as \textit{multistage} if $T$ can only take a small number of values.  The first goal of the present work is to introduce a \textit{multistage} test,  the first in the literature to the best of our knowledge,  that  achieves both  infima in  \eqref{one-dim, def: optimal performance}  to  a first-order asymptotic approximation as $\alpha, \beta \to 0$ \emph{without any assumption on the decay rates of $\alpha$ and $\beta$}.   To be precise, we state  the following definition of asymptotic optimality.

\begin{definition} \label{definition of asy opt in one-dim}
A family of  tests,
\begin{align} \label{family}
    \chi^*  &\equiv\left\{ \big(T^*(\alpha, \beta),D^*(\alpha,\beta)\big) \in \cE(\alpha,\beta): \;    \alpha,\beta \in (0,1) \right \}, 
 \end{align}
is \emph{asymptotically optimal}    under  the null hypothesis if, as  $\alpha, \beta \to 0$, 
\begin{equation} \label{def of AO null}
    \Exp_{0}[T^*(\alpha,\beta)]  \sim \cL_0(\alpha, \beta),
\end{equation}
and under the alternative hypothesis if, as  $\alpha, \beta \to 0$,
\begin{equation} \label{def of AO alternative}
    \Exp_{1}[T^*(\alpha,\beta)]   \sim \cL_1(\alpha, \beta).
\end{equation}
\end{definition}   

\subsection{The Sequential Probability Ratio Test}
It is well known (see, e.g., \cite[Chapter 3.2]{Tartakovsky_Book}) that both infima in  \eqref{one-dim, def: optimal performance} are achieved by a fully sequential test, the  Sequential Probability Ratio Test (SPRT), i.e., 
\begin{align} \label{def: SPRT}
\begin{split}
    T &\equiv \inf\{n\in \bN: \Lambda_n\notin (-B,A)\},\\
    D &\equiv 1\{ \Lambda_{T}\geq A \},
\end{split}
\end{align}
when  $A,B$ are selected as functions of $\alpha$ and $\beta$ so that the error constraints be satisfied with equality. In what follows,  for any  $\alpha, \beta \in (0,1)$ we denote by  $$\big(\widetilde{T}(\alpha, \beta), \widetilde{D}(\alpha, \beta)\big)$$  the test in \eqref{def: SPRT} with
$$ A=|\log\alpha| \quad \text{and} \quad B=|\log\beta|. $$
For this  selection of thresholds, it is well known (see, e.g., \cite[Chapter 3.1]{Tartakovsky_Book})  that,  for any  $\alpha, \beta \in (0,1)$,  
$$\big(\widetilde{T}(\alpha, \beta), \widetilde{D}(\alpha, \beta)\big) \in \cE(\alpha, \beta),$$
and that, as $\alpha,\beta\to 0$,
\begin{align} \label{optimal performance}
\begin{split}
    \Exp_{0}[\widetilde{T}(\alpha, \beta)]   &\sim \cL_0(\alpha,\beta)\sim \frac{|\log\beta|}{I_0}, \\
    \Exp_{1}[\widetilde{T}(\alpha, \beta)] & \sim \cL_1(\alpha,\beta)\sim \frac{|\log\alpha|}{I_1}.
\end{split}
\end{align}
As a result, according to Definition \ref{definition of asy opt in one-dim}, the family of SPRTs, 
\begin{align} \label{family SPRTs}
\widetilde{\chi} &\equiv\left\{ \big(\widetilde{T}(\alpha, \beta), \widetilde{D}(\alpha,\beta)\big): \;    \alpha,\beta \in (0,1) \right \},
\end{align}
is asymptotically optimal under both hypotheses.

\subsection{Asymptotic optimality with respect to a mixture}
Generalizing the notation for $\Pro_i$ and $\cL_{i}$ with $i \in \{0,1\}$,   for any $\pi \in [0,1]$  we  introduce  the  mixture distribution 
\begin{align} \label{mixture}
    \Pro_\pi \equiv  (1-\pi) \, \Pro_0 + \pi \, \Pro_1,
\end{align} 
and denote by $\cL_\pi(\alpha,\beta)$ the optimal expected sample size in $\cE(\alpha, \beta)$ under  $\Pro_\pi$, i.e., 
\begin{equation}\label{optimal_mixture}
    \cL_{\pi}(\alpha,\beta) \equiv  \inf_{(T,D) \in   \cE(\alpha, \beta)}\Exp_{\pi}[T],
\end{equation}
where $\Exp_\pi$ denotes the expectation under $\Pro_\pi$. This  notation and the following proposition  will be useful for the formulation and analysis of the  signal recovery problem in  Section \ref{sec: problem formulation about high-dim}. 

\begin{proposition} \label{prop: sprt_mixture_optimality}
If the family of tests, $\chi^*$, defined in \eqref{family}, is asymptotically optimal under both hypotheses, then,  as $\alpha, \beta \to 0$,
$$ \Exp_\pi[T^*(\alpha,\beta)] \sim \cL_\pi(\alpha,\beta) \sim (1-\pi)\frac{|\log\beta|}{I_0} + \pi\frac{|\log\alpha|}{I_1} \text{ uniformly in } \pi\in[0,1]. $$
\end{proposition}
\begin{IEEEproof}
See Appendix \ref{proofs about FSST}. \\
\end{IEEEproof}

\subsection{The fixed-sample-size test} \label{sec: FSST}
The  building block for all multistage tests we consider in this work  is the fixed-sample-size test that rejects the null hypothesis if and only if  the \emph{average} log-likelihood ratio at a predetermined time instance exceeds a predetermined threshold.  Specifically, for any  $\alpha,\beta\in (0,1)$, we denote by $\n(\alpha,\beta)$ the smallest  sample size such a test  can have in order to belong to  $\cE(\alpha,\beta)$, i.e.,
\begin{align} \label{def: n*(alpha,beta)}  
    \n(\alpha,\beta) &\equiv \min \, \left\{n \in \bN: \; \exists \; c \in \bR \; \text{ so that } \; \Pro_0( \bar\Lambda_n > c)\leq \alpha \; \text{ and } \; \Pro_1( \bar\Lambda_n \leq c) \leq \beta \right\},
\end{align}
by $\thre(\alpha,\beta)$ the smallest such threshold, i.e.,
\begin{equation} \label{def: c*(alpha,beta)}
    \thre(\alpha,\beta)\equiv \min\left\{ c\in\bR: \Pro_0(\bar\Lambda_{\n(\alpha,\beta)}>c)\leq \alpha \; \text{ and } \; \Pro_1(\bar\Lambda_{\n(\alpha,\beta)}\leq c)\leq \beta \right\}.
\end{equation}
In Proposition \ref{well-definedness of n* and c*} in Appendix \ref{proofs about FSST} we show that both these quantities are well-defined and, in what follows, we set  
\begin{equation} \label{sf FSS}
    {{\sf{FSST}}}(\alpha,\beta) \equiv \big( \n(\alpha,\beta), \, \thre(\alpha,\beta) \big).
\end{equation}


Next, we establish a non-asymptotic upper bound on $\n(\alpha,\beta)$, which we use extensively in the analysis of the multistage tests we consider in this work. By the Chernoff bound it follows  that, for any $c\in (-I_0,I_1)$ and $n\in\bN$,
\begin{align} \label{LD upper bounds}
\begin{aligned}
    \Pro_0(\bar\Lambda_n>c) &\leq \exp\{-n\psi_0(c) \},\quad \forall \;  c \geq -I_0, \\
    \Pro_1(\bar\Lambda_n\leq c) &\leq  \exp\{-n\psi_1(c) \},\quad \forall \;  c \leq I_1,
\end{aligned}
\end{align}
where
\begin{align} \label{def: psi}
\begin{split}
    \psi_0(c) &\equiv \sup_{\theta\geq 0} 
    \left\{\theta c - \log \left(\int f_1^\theta f_{0}^{1-\theta} \; d \nu \right) \right\}, \quad c\geq -I_0, \\
    \psi_1(c) &\equiv \sup_{\theta\leq 0} 
    \left\{ \theta c- \log \left( \int f_0^\theta f_{1}^{1-\theta} \; d \nu \right) \right\}, \quad c\leq I_1.
\end{split}
\end{align}
A well-known (see, e.g, \cite[Corollary 3.4.6]{Dembo_Zeitouni_LDPBook}) upper bound on $\n(\alpha,\beta)$ (see  \eqref{non-asy 2} below) can be obtained in terms of the \textit{Chernoff information}:
\begin{equation*}
    \cC\equiv \sup_{\theta\geq 0} 
    \left\{- \log \left(\int  f^{\theta}_1  f_{0}^{1-\theta} \; d \nu \right) \right\} = \psi_0(0)=\psi_1(0).
\end{equation*}
For our purposes  in this work, we will need a sharper upper bound. To state it, we first need to introduce the following function:
\begin{equation} \label{def of g}
    g(c) \equiv  \frac{\psi_0(c)}{\psi_1(c)}, \qquad c\in (-I_0,I_1).
\end{equation}
Since (see, e.g., \cite[Chapter 2.2 $\&$ Chapter 3.4]{Dembo_Zeitouni_LDPBook}) 
\begin{itemize}
    \item $\psi_0$ (resp. $\psi_1$) is convex and continuous in $[-I_0,\infty)$ (resp. $(-\infty,I_1]$),
    \item $\psi_0$ is strictly increasing in $[-I_0,\infty)$ with $\psi_0(-I_0)=0,\,\psi_0(I_1)=I_1$,
    \item $\psi_1$ is strictly decreasing in $(-\infty,I_1]$ with $\psi_1(-I_0)=I_0,\,\psi_1(I_1)=0$,
\end{itemize}
the function $g$ is continuous and strictly increasing in $(-I_0,I_1)$, with 
$$g(-I_0+)=0 \quad \text{and}  \quad  g(I_1-) = \infty, $$
and thus, its inverse   $g^{-1}:(0,\infty)\to (-I_0,I_1)$ is well defined. The upper bound we establish  on $\n(\alpha,\beta)$ is in terms  of the following quantities:
\begin{equation} \label{definition of hi}
    h_i(\alpha, \beta)  \equiv \psi_i \left( g^{-1} \left(  \frac{|\log\alpha|}{|\log\beta|} \right)  \right), \quad i \in \{0,1\}.
\end{equation}

\begin{theorem} \label{Theorem, two non-asy upper bounds on n*(alpha,beta)}
For any $\alpha,\beta\in(0,1)$,
\begin{align}
    \n(\alpha,\beta) &\leq\frac{|\log\beta|}{h_1(\alpha,\beta)} + 1 = \frac{|\log\alpha|}{h_0(\alpha,\beta)} + 1, \label{non-asy 1}
\end{align}
and, consequently,
\begin{align}
    \n(\alpha,\beta) &\leq \frac{|\log(\alpha\wedge \beta)|}{\cC} + 1. \label{non-asy 2}
\end{align}
\end{theorem}

\begin{IEEEproof}
    See Appendix \ref{proofs about FSST}. \\
\end{IEEEproof}

Using \eqref{non-asy 1}, we next  obtain a generalization of Stein's lemma (see, e.g., \cite[Lemma 3.4.7]{Dembo_Zeitouni_LDPBook}), according to which, as $\alpha, \beta \to 0$ such that  $|\log\alpha|/|\log\beta|$ goes to $0$ (resp. infinity), the fixed-sample-size test is asymptotically optimal under the null (resp. alternative) hypothesis at the expense  of severe performance loss  (relative to the optimal) under the other hypothesis.

\begin{corollary} \label{Corollary, asy upper bounds on n*(alpha,beta)}
    \begin{enumerate}
        \item [(i)] If $\alpha,\beta\to 0$ so that $|\log\alpha|\ll |\log\beta|$, then $h_1(\alpha,\beta) \to I_0 $ and 
        \begin{equation} \label{r=0}
            \cL_1(\alpha,\beta) \ll  \n(\alpha,\beta)\sim \cL_0(\alpha,\beta).
        \end{equation}
        \item [(ii)] If $\alpha,\beta\to 0$ so that $|\log\alpha|\gg|\log\beta|$, then 
        $ h_0(\alpha,\beta) \to  I_1$ and 
        \begin{equation} \label{r=infty}
            \cL_0(\alpha,\beta) \ll  \n(\alpha,\beta)\sim  \cL_1(\alpha,\beta).
        \end{equation}    
    \end{enumerate}
\end{corollary}
\begin{IEEEproof}
See Appendix \ref{proofs about FSST}. \\
\end{IEEEproof}

\subsection{A Gaussian example} \label{subsec: a gaussian example}
We illustrate the above quantities in the special case of testing the mean $\mu$ of a  Gaussian distribution with unit variance, $N(\mu, 1)$, i.e.,  
\begin{equation} \label{Normal testing problem}
    H_0: \mu=-\eta \qquad \text{ versus } \qquad H_1: \mu=\eta
\end{equation}
for some $\eta>0$. In this case,  the Kullback-Leibler divergences and  the Chernoff information take the following form:
\begin{align*} 
   I_0=I_1=2\eta^2\equiv I, \qquad \cC=\eta^2/2=I/4,
\end{align*}  
and we have an explicit form for the  fixed-sample-size test: 
\begin{equation} \label{exact formulas in testing Gaussian mean}
\begin{split}
    \n(\alpha,\beta) &= \left\lceil\frac{1}{4\eta^2} \left( z_\alpha+z_\beta \right)^2\right\rceil,  \\
    \thre(\alpha,\beta) &= \frac{z_\alpha-z_\beta}{2\sqrt{\n(\alpha,\beta)}},
\end{split}
\end{equation}
where $z_\alpha$ is the upper $\alpha$-quantile of the standard Gaussian distribution.
Moreover, the functions in \eqref{def: psi}-\eqref{def of g}  take the following form: 
\begin{align*}
    \psi_0(c)&=\frac{1}{4I}(I+c)^2,  \quad c\geq -I, \\
    \psi_1(c)&= \frac{1}{4I} (I-c)^2,  \quad c\leq I, \\
    g(c) &=\left( \frac{I+c}{I-c} \right)^2, \quad c\in (-I,I),
\end{align*}
consequently, 
\begin{align*}
    \psi_0\left(g^{-1}(u)\right) &= \frac{u}{(1+\sqrt u)^2} \, I,   \quad u\in \bR,\\
    \psi_1\left(g^{-1}(u)\right) &= (1+\sqrt u)^{-2} \, I, \quad u\in \bR,
\end{align*}
and,  for any $\alpha,\beta\in (0,1)$, the functions in  \eqref{definition of hi}   take the following form: 
\begin{equation} \label{hi in Gaussian case}
\begin{aligned}
    h_0(\alpha,\beta) & = I \cdot\left( 1+\sqrt{\frac{|\log\beta|}{|\log\alpha|}} \right)^{-2} ,\\
    h_1(\alpha,\beta) & = I \cdot \left( 1+\sqrt{\frac{|\log\alpha|}{|\log\beta|}} \right)^{-2}.
\end{aligned}
\end{equation}

\section{Lorden's 3-Stage Test} \label{section, 3ST}
The  multistage test that we propose in the present paper (Section \ref{section, GMT})  generalizes the test in  \cite[Section 2]{Lorden_1983}. In this section, we review the latter test and its asymptotic optimality property. 

\subsection{Description}
The test in  \cite[Section 2]{Lorden_1983} provides two opportunities to accept and two opportunities  to reject the null hypothesis.  To be specific, we denote by  $N_{i,0}$  the  number of observations  until the first opportunity to  select $H_i$, where $i \in \{0,1\}$, and  by $N$  the  maximum possible number of observations, where  $N_{0,0}, N_{1,0}, N$ are deterministic positive integers such that $N_{0,0}\vee N_{1,0}\leq N$. Then, assuming it has not done so earlier, this  test terminates
\begin{itemize}
    \item after $N_{0,0}$ observations  if   $\bar\Lambda_{N_{0,0}}\leq C_{0,0}$, in which case it accepts $H_0$,
    \item after  $N_{1,0}$ observations if $\bar\Lambda_{N_{1,0}} > C_{1,0}$, in which case it rejects $H_0$,
    \item after $N$ observations,  rejecting $H_0$ if and only if $\bar\Lambda_N>C$,
\end{itemize}
where $C_{0,0}, C_{1,0}, C$ are real-valued thresholds, to be specified together with $N_{0,0},N_{1,0},N$. To avoid overlap between an acceptance region and a rejection region, we require that 
$$C_{0,0}\leq C_{1,0} \quad \text{if} \quad N_{0,0}=N_{1,0},$$ 
and we make the  convention that if  $N_{i,0}=N$ for some $i\in\{0,1\}$, then  $C_{i,0}$ is ignored and $C$ is the only effective threshold.  Since at most 3 stages are needed for the  implementation of this test,  in what follows  we refer to it as the \textit{3-Stage Test}.

\subsection{Analysis}
By an application of the union bound it follows that, for any $\alpha, \beta \in (0,1)$,  the 3-Stage Test belongs to $\cE(\alpha,\beta)$ when  its parameters in the first opportunities to accept and reject the null hypothesis are selected so that 
\begin{align*}
    \Pro_0(\bar\Lambda_{N_{1,0}}> C_{1,0}) \leq \alpha/2 \quad \text{ and } \quad 
    \Pro_1(\bar\Lambda_{N_{0,0}}\leq C_{0,0}) \leq\beta/2 , 
\end{align*}
and its parameters in the last possible stage are selected so that
\begin{align*}
   \Pro_0(\bar\Lambda_N>C) \leq \alpha/2 \quad \text{ and } \quad \Pro_1(\bar\Lambda_N\leq C) \leq \beta/2.
\end{align*}
The latter constraint is satisfied, by definition, when 
\begin{equation} \label{Lorden's, N, C}
    (N,C)={\sf{FSST}}\left(\alpha /2 , \, \beta/2\right).
\end{equation}
Moreover, by an application of Markov's inequality it follows that 
\begin{equation} \label{by Markov's}
    \Pro_0(\bar\Lambda_{N_{1,0}}> C_{1,0}) \leq e^{-N_{1,0} C_{1,0}} \quad \text{ and } \quad   \Pro_1(\bar\Lambda_{N_{0,0}}\leq C_{0,0}) \leq e^{N_{0,0} C_{0,0}}, 
\end{equation}
thus, the former  constraint is satisfied when
\begin{equation} \label{Lornde's way of controlling error probs at n0, n1}
    C_{1,0} \, N_{1,0}= |\log(\alpha/2)|  \quad \text{ and } \quad  C_{0,0} \, N_{0,0} =-|\log(\beta/2)|.
\end{equation}  
Equations \eqref{Lorden's, N, C}-\eqref{Lornde's way of controlling error probs at n0, n1} leave two free parameters, which  can be  selected  to guarantee  the  asymptotic optimality  of the  3-Stage Test under both hypotheses  as $\alpha$ and $\beta $  go to 0, \textit{as long as they do not do so very asymmetrically}. 

\begin{theorem} \label{thm: asy opt of 3ST} 
For any $\alpha, \beta \in (0,1)$,  let  $\left(\check{T}(\alpha, \beta),\check{D}(\alpha, \beta)\right)$ denote the sample size and decision of the 3-Stage Test when its  parameters are selected   according to  \eqref{Lorden's, N, C}, \eqref{Lornde's way of controlling error probs at n0, n1},  and 
$$N_{0,0}=\check N_{0,0}\wedge N, \quad \quad N_{1,0}=\check N_{1,0}\wedge N, $$ 
where 
\begin{equation*} \label{Lorden's selection of N00, N10}
    \check N_{0,0}=\left\lceil\frac{|\log(\beta/2)|}{(1-\epsilon_0)\,I_0}\right\rceil, \quad \quad \check N_{1,0} =\left\lceil\frac{|\log(\alpha/2)|}{(1-\epsilon_1)\,I_1}\right\rceil,
\end{equation*}
and $\epsilon_0,\epsilon_1\in(0,1)$ are arbitrary functions of $\alpha$ and $\beta$ such that, as  $\alpha,\beta\to 0$, 
\begin{equation} \label{Lorden's constraint on epsilon0, epsilon1}
\begin{aligned}
    & \epsilon_0\to 0 \quad \text{and} \quad  \Pro_0\left(\bar\Lambda_{\check N_{0,0}}>-(1-\epsilon_0)\,I_0\right)\to 0, \\
    & \epsilon_1\to 0 \quad \text{and} \quad  \Pro_1\left(\bar\Lambda_{\check N_{1,0}}\leq (1-\epsilon_1)\,I_1\right)\to 0.
\end{aligned}
\end{equation}
Then,  the family $$\check{\chi} \equiv 
\left\{ \big(\check{T}(\alpha, \beta), \check{D}(\alpha, \beta)\big): \alpha, \beta \in (0,1) \right\}$$
is asymptotically optimal under both hypotheses as long as   $\alpha,\beta\to 0$ so that  
\begin{equation} \label{constraint on alpha and beta for A.O. of 3ST}
    |\log\alpha|= \Theta(|\log\beta|). 
\end{equation}
\end{theorem}
\begin{IEEEproof}This theorem was established in  \cite[Section 2]{Lorden_1983}, and its  proof is presented in Appendix  \ref{proofs about 3ST} for completeness. \\
\end{IEEEproof}

While it is always possible to find $\epsilon_0$ and $\epsilon_1$  that satisfy  \eqref{Lorden's constraint on epsilon0, epsilon1} under the assumption of \eqref{KL information numbers}, the above theorem does not provide a concrete selection for $N_{0,0},C_{0,0}$ and $N_{1,0},C_{1,0}$. Such a selection is proposed in \cite{PaperI}, where  it is also shown that constraint  \eqref{constraint on alpha and beta for A.O. of 3ST} on the decay rates  of $\alpha$ and $\beta$ can be somewhat relaxed.  In the same work,  a test with at most four stages is   proposed and its asymptotic optimality under both hypotheses is established under an even weaker constraint on the decay rates of $\alpha$ and $\beta$. In the next section  we introduce a multistage test with a  deterministic  maximum number of stages, which is a function of the  relative magnitude of $\alpha$ and $\beta$,  that is  asymptotically  optimal under both hypotheses \textit{without any constraint on the decay rates of $\alpha$ and $\beta$}.

\section{The General Multistage Test} \label{section, GMT}
In this section we  introduce and analyze the proposed multistage test  in this work.

\subsection{Description}
The proposed test generalizes the 3-Stage Test of the previous section in that  it provides $K_i$ additional opportunities to select $H_i$, where  $K_i $ is a  deterministic, \textit{non-negative}  integer and $i\in\{0,1\}$.  To be specific, in addition to $K_0, K_1$ and   the parameters that are  present in the 3-Stage Test,   i.e., 
\begin{align*} 
N, \, C \quad & \text{ and } \quad  N_{i,0}, \, C_{i,0} \quad  i \in \{0,1\}, 
\end{align*}
for each $i\in\{0,1\}$  we need to determine an increasing sequence of positive integers, $$\{N_{i,j}, \, j\in[K_i]\},$$ \textit{between $N_{i,0}$ and $N$}, i.e.,
\begin{align} \label{cond}
N_{i,0}\leq N_{i,1} \leq \cdots \leq N_{i,K_i} \leq N,
\end{align}
as well as  a sequence of real numbers, 
$$\{C_{i,j}, \, j\in [K_i]\},$$
so that, for each $j\in[K_i]$, $N_{i,j}$ represents the total number of observations collected at the time of the $(j+1)^{\text{th}}$ opportunity to select $H_i$ and $C_{i,j}$ the corresponding threshold. Then,  assuming it has not  done so earlier,  the  proposed test terminates
\begin{itemize}
    \item  after $N_{0,j}$ observations if   $\bar\Lambda_{N_{0,j}}\leq C_{0,j}$,  for some $j\in \{0\}\cup[K_0]$, in which case it accepts $H_0$,
    \item after  $N_{1,j}$ observations if $\bar\Lambda_{N_{1,j}} > C_{1,j}$, for some $j\in \{0\}\cup[K_1]$,  in which case it rejects $H_0$,
    \item  after $N$ observations,  rejecting $H_0$ if and only if $\bar\Lambda_N>C$.
\end{itemize}

By the description of the test it follows that 
when $N_{i,j}=N_{i,k}$ for some $j, k \in \{0\}\cup[K_i]$, $i \in \{0,1\}$, then only the maximum (resp. minimum) of the corresponding thresholds,  $C_{i,j}$ and $C_{i,k}$,  is effective when $i=0$ (resp. $i=1$).  However, in order to avoid overlap of an acceptance region and a rejection region, we need to require that 
\begin{equation} \label{cond on c's}
    C_{0,j} \leq C_{1,k} \; \text{ if } \; N_{0,j} = N_{1,k}, \; \forall \; j \in \{0\}\cup[K_0] \; \text{ and } \; k \in \{0\}\cup[K_1], 
\end{equation}
and also  make the convention that if  $N_{i,j}=N$ for some $ j \in \{0\}\cup[K_i], \;  i\in\{0,1\},$  then $C_{i,j}$ is ignored and the only  effective threshold is $C$.

This testing procedure can be implemented using at most $3+K_0+K_1$ stages and it reduces to the 3-Stage Test when $K_0=K_1=0$. Due to its general structure when compared to the 3-Stage Test or the 4-Stage Test in \cite{PaperI}, we refer to it as the \textit{General Multistage Test (GMT)}. \\

\myremark In Algorithm   \ref{Algorithm, GMT} we provide an  algorithmic description  of the GMT. To this end, we denote by  $\{\sorted_j: \, j\in [K_0+K_1+2]\}$ the increasingly ordered version of
$$\big\{N_{i,j}: \;    j\in \{0\}\cup[K_i], \;  i \in \{0,1\}\big\},$$
and by  $\{\sortedc_j: \, j\in [K_0+K_1+2]\}$ the corresponding thresholds. Moreover, to each $j\in [K_0+K_1+2]$ we assign a  label, $\sortedl_j$,  that is equal to ``$+$" (resp.  ``$-$")  if $(\sorted_j,\sortedc_j)$ corresponds to an opportunity to  reject (resp. accept) the null  hypothesis. We stress, however, that this notation is used only in Algorithm   \ref{Algorithm, GMT}.

\begin{algorithm}
\caption{General  Multistage Test (GMT)} \label{Algorithm, GMT}
\begin{algorithmic}
    \State Input: $K_0$, $K_1$; $(\sorted_j,\sortedc_j,\sortedl_j), \, j\in [K_0+K_1+2]; \; (N,C)$.
    \State Initialize: $\sorted_0=0$, $j=1$. 
    \While{$j\leq K_0+K_1+2$}
        \State take $\sorted_j-\sorted_{j-1}$ samples
        \If{$\sortedl_j=$``$-$" and $\bar\Lambda_{\sorted_j}\leq                    \sortedc_j$}
            \State stop and accept the null
            \ElsIf{$\sortedl_j=$``$+$" and $\bar\Lambda_{\sorted_j}>\sortedc_j$
            }
                \State stop and reject the null
                \Else
                    \State $j=j+1$
        \EndIf
    \EndWhile
    \If{$j=K_0+K_1+3$}
        \State take $N-\sorted_{j-1}$ samples
            \If{$\bar\Lambda_N\leq C$}
                \State stop and accept the null
            \Else
                \State stop and reject the null
            \EndIf
    \EndIf
\end{algorithmic}
\end{algorithm}

\subsection{Error control}
Given  $K_0$ and $K_1$, the GMT has $2 \cdot (3+K_0+K_1)$ parameters that need to be determined.  We start by specifying  a  design which guarantees that, for any given $\alpha, \beta \in (0,1)$,  the test satisfies \eqref{cond}-\eqref{cond on c's} and belongs to $\cE(\alpha, \beta)$.   

The first feature of this  design is that at the last possible stage  a fixed-sample-size test is performed whose parameters  do not depend on $K_0$ or $K_1$. 
Specifically, we set
\begin{equation} \label{N}
    (N,C) = {\sf{FSST}} \left(\alpha/4, \, \beta/4\right).
\end{equation}

The second feature is that the type-II (resp. type-I) error probabilities in the $K_0$ (resp. $K_1$) intermediate opportunities to accept (resp. reject) the null hypothesis decay exponentially fast. Specifically,  we set
\begin{align}  
    (N_{0,j}, C_{0,j}) = {\sf{FSST}}  \left(\gamma_{0,j},\; \left(\beta/4\right)^j\right),  \quad j \in [K_0], \label{GMT, general design, 0} \\
    (N_{1,j}, C_{1,j}) = {\sf{FSST}}  \left(  \left(\alpha/4\right)^j, \gamma_{1,j}\right),  \quad j \in [K_1], \label{GMT, general design, 1}
\end{align}
where  $\{\gamma_{0,j}, j \in \bN\}$ and $\{\gamma_{1,j}, j \in \bN\}$  are two \textit{infinite}  sequences in $(0,1)$ to  be determined.   
  
The third feature  is that all  remaining  type-II (resp. type-I)  error probability  is  assigned  to the first opportunity to accept (resp. reject) the null hypothesis. Specifically,  we set
\begin{align} 
    (N_{0,0}, C_{0,0}) = {\sf{FSST}} \left(\gamma_{0,0},\; \; \frac{3\beta}{4}-\sum_{j=1}^{K_0} \left(\frac{\beta}{4}\right)^j\right), \label{GMT, general design, 00}\\
    (N_{1,0}, C_{1,0}) ={\sf{FSST}} \left( \frac{ 3\alpha}{4}-\sum_{j=1}^{K_1} \left(\frac{\alpha}{4}\right)^j,\; \; \gamma_{1,0}\right),  \label{GMT, general design, 10}
\end{align}
where  $\gamma_{0,0}$ and $\gamma_{1,0}$ are two additional free parameters in $(0,1)$.

In order to guarantee that the above design satisfies conditions \eqref{cond} and \eqref{cond on c's}, we need to impose some constraints on its free parameters,
\begin{align} \label{GMT_free_parameters}
\begin{split}
    & \gamma_{0,0},\quad \{\gamma_{0,j}, \, j \in \bN \}, \quad K_0, \\
    & \gamma_{1,0}, \quad \{\gamma_{1,j}, \, j \in \bN \}, \quad K_1.
\end{split}
\end{align}
Once these conditions are satisfied, the error control follows directly by an application of the union bound. 
\begin{proposition} \label{GMT, make sense and error control}
For any $\alpha,\beta\in(0,1)$, if the GMT is designed according to \eqref{N}-\eqref{GMT, general design, 10} and the free parameters in \eqref{GMT_free_parameters} satisfy 
\begin{align} 
    & \gamma_{0,0}\geq 3\alpha/4 \quad \text{ and } \quad \gamma_{1,0}\geq 3\beta/4, \label{gamma0 >= 3alpha/4 and gamma1 >= 3beta/4} \\
    & \gamma_{i,0} > \gamma_{i,j} > \gamma_{i,j+1}, \quad \forall\;  j \in \bN, \; i \in \{0,1\}, \label{all gammai's strictly decrease} \\
    & K_0\leq \widehat{K_0} \quad \text{ and } \quad K_1\leq \widehat{K_1}, \label{M0 <= M0hat and M1 <= M1hat}
    \end{align}
where 
\begin{equation} \label{def of M0hat and M1hat}
\begin{split}
    \widehat{K_0} &\equiv \max\left\{ j\in\bN: \n\left(\gamma_{0,j},\left(\beta/4\right)^j\right)\leq \n\left(\alpha/4, \beta/4\right) \quad \text{and} \quad \gamma_{0,j} \geq 3\alpha/4 \right\}, \\
    \widehat{K_1} &\equiv \max\left\{ j\in\bN: \n\left(\left(\alpha/4\right)^j, \gamma_{1,j} \right)\leq \n\left(\alpha/4, \beta/4\right) \quad \text{and} \quad \gamma_{1,j} \geq 3\beta/4 \right\},
\end{split}
\end{equation}
then conditions  \eqref{cond}-\eqref{cond on c's} hold and the GMT belongs to $\cE(\alpha,\beta)$.
\end{proposition}
\begin{IEEEproof} 
Appendix  \ref{proofs about GMT}. \\
\end{IEEEproof}

\myremark (i) We stress that only finitely many terms of the sequences 
\begin{equation} \label{sequences}
    \{\gamma_{0,j},\, j \in \bN\} \quad \text{ and } \quad \{\gamma_{1,j},\, j \in \bN\} 
\end{equation}  
appear in the implementation of the testing procedure. It is  convenient, however,  to introduce these two \textit{infinite} sequences  as free parameters, because in this way once these two sequences have been specified, we  automatically obtain the upper bounds for $K_0$ and $K_1$  in \eqref{def of M0hat and M1hat}.

(ii) By \eqref{gamma0 >= 3alpha/4 and gamma1 >= 3beta/4} and \eqref{all gammai's strictly decrease} it follows that 
\begin{equation} \label{range for gamma0 and gamma1}
    \gamma_{0,0}\in\big( (3\alpha/4)\vee\gamma_{0,1},1 \big)  \quad \text{ and } \quad \gamma_{1,0}\in\big( (3\beta/4)\vee\gamma_{1,1},1 \big).
\end{equation}

(iii) We refer to $\gamma_{0,j},\,j\in \{0\}\cup [K_0]$ \big(resp. $\gamma_{1,j},\, j\in\{0\}\cup [K_1]$\big) as the \textit{inactive} type-I (resp. type-II) error probabilities, and to the other arguments of $\sf{FSST}$ in \eqref{GMT, general design, 0}-\eqref{GMT, general design, 10} as the \textit{active} type-II (resp. type-I) error probabilities.

\subsection{A robustness property}
With its maximum possible sample size  selected according to  \eqref{N},  the GMT enjoys  an interesting robustness property when compared to the  SPRT, independently of how  its other  parameters are chosen. Indeed, by \eqref{N} and  \eqref{non-asy 2} it follows that, for any $\alpha, \beta \in (0,1)$,  the maximum sample size of the GMT cannot exceed 
\begin{equation} \label{robustness of GMT}
    \frac{|\log(\alpha\wedge\beta)|+\log 4}{\cC}+1.
\end{equation}
On the other hand,  the SPRT  has  an inflated expected sample size when the true distribution is ``between'' $\Pro_0$ and $\Pro_1$ and  $\alpha$ and $\beta$ are small enough  (see, e.g.,  \cite[Chapter 3.1.1.2]{Tartakovsky_Book}). Indeed,   if $\Pro$ is a distribution under which $\{\Lambda_n,n\in\bN\}$ is a random walk whose increments have \emph{zero} mean and finite variance $\sigma^2$, then the expected sample size of $\widetilde{T}(\alpha, \beta)$  is, ignoring the overshoot over the boundary,  equal to $$ |\log\alpha|\cdot|\log\beta|\,/\,\sigma^2. $$
Comparing   with   \eqref{robustness of GMT}, we can see that  the GMT will perform much better than the SPRT under such a $\Pro$ when $\alpha$ and $\beta$ are small enough. This phenomenon is also illustrated in Figure \ref{Figure, one-dim, four tests}.

\subsection{Specification of free parameters}
We continue with a   concrete specification of the free parameters in \eqref{GMT_free_parameters}. For this, we need an upper  bound on the expected sample size  of GMT under $\Pro_0$ (resp. $\Pro_1$), which  is obtained by ignoring all opportunities  to   reject (resp. accept) the null hypothesis and depends only on $\gamma_{0,0}, \{\gamma_{0,j}, j \in \bN\},  K_0$ \big(resp. $\gamma_{1,0}, \{\gamma_{1,j}, j \in \bN\}, K_1$\big).  

\begin{proposition} \label{prop: ESS upper bound}
Fix $\alpha,\beta\in(0,1)$. If the parameters of GMT are selected as in \eqref{N}-\eqref{GMT, general design, 10} such that \eqref{cond}-\eqref{cond on c's} hold, then its expected sample size under $\Pro_0$ is upper bounded  by 
\begin{equation} \label{GMT, ESS under P0}
    \n\left( \gamma_{0,0}, \frac{3\beta}{4}-\sum_{j=1}^{K_0} \left(\frac{\beta}{4}\right)^j \right)+\sum_{j=1}^{K_0} \n\left( \gamma_{0,j}, \left(\frac{\beta}{4}\right)^j \right) \cdot \gamma_{0,j-1}+\n\left(\frac{\alpha}{4}, \frac{\beta}{4}\right) \cdot\gamma_{0,K_0},
\end{equation}
and  its expected sample size under $\Pro_1$ is upper bounded by
\begin{equation} \label{GMT, ESS under P1}
   \n\left( \frac{ 3\alpha}{4}-\sum_{j=1}^{K_1} \left(\frac{\alpha}{4}\right)^j, \gamma_{1,0} \right) + \sum_{j=1}^{K_1} \n\left( \left(\frac{\alpha}{4}\right)^j, \gamma_{1,j} \right) \cdot \gamma_{1,j-1} + \n\left(\frac{\alpha}{4}, \frac{\beta}{4}\right) \cdot \gamma_{1,K_1}.
\end{equation}
\end{proposition}
\begin{IEEEproof}
By the definition of GMT it follows that, by ignoring all stages where it is possible only to reject the null hypothesis, its  expected sample size under $\Pro_0$ is upper bounded by
\begin{equation*} 
\begin{aligned}
    N_{0,0} + \sum_{j=1}^{K_0} N_{0,j} \cdot \Pro_0\left(\bar\Lambda_{N_{0,j-1}}>C_{0,j-1}\right) + N \cdot \, \Pro_0\left(\bar\Lambda_{N_{0,K_0}}>C_{0,K_0}\right).
\end{aligned}
\end{equation*}
The upper bound in  \eqref{GMT, ESS under P0} then follows  by   \eqref{N}-\eqref{GMT, general design, 10} and the definition of the FSST. The upper bound in \eqref{GMT, ESS under P1} can be proved similarly. \\
\end{IEEEproof}

Let us start the description of the proposed specification of the free parameters in \eqref{GMT_free_parameters}  by assuming  that  the two infinite sequences in \eqref{sequences} have  already been specified. As we mentioned earlier, this specification  determines $\widehat{K_0}$ and $\widehat{K_1}$ in  \eqref{def of M0hat and M1hat}, i.e.,  the maximum  possible numbers of additional, relative to the 3-Stage Test, opportunities to accept and reject the null hypothesis respectively. Then, recalling \eqref{range for gamma0 and gamma1}, we can select
$$K_0\in\{0,\ldots,\widehat{K_0}\} \quad \text{ and } \quad \gamma_{0,0}\in\big( (3\alpha/4)\vee\gamma_{0,1},1 \big)$$
to jointly minimize \eqref{GMT, ESS under P0}, and
$$K_1\in\{0,\ldots,\widehat{K_1}\} \qquad \text{and}
\qquad \gamma_{1,0}\in\big( (3\beta/4)\vee\gamma_{1,1},1 \big)$$ to jointly minimize \eqref{GMT, ESS under P1}.
Thus, $K_0, \gamma_{0,0}$ and $K_1,\gamma_{1,0}$ are completely determined by  the  sequences in \eqref{sequences}, and it remains to show how to select the latter. 

For each $i \in \{0,1\}$ and $j \in \bN$, a very small value of $\gamma_{i,j}$ may unnecessarily increase the size of the corresponding stage, whereas  a very large value may too frequently allow continuation to the next stage.  To solve this trade-off,  we select each inactive error probability in \eqref{GMT, general design, 0}-\eqref{GMT, general design, 1} to match the corresponding active  error probability, i.e.,  we set
\begin{align} \label{GMT, opt condition} 
\begin{split}
    \gamma_{0,j} &=\left(\beta / 4 \right)^j, \quad j\in \bN, \\
    \gamma_{1,j} &= \left(\alpha/ 4 \right)^j, \quad j\in \bN.
\end{split}
\end{align}   
Then, $ \widehat{K_0}$ and  $ \widehat{K_1}$ in \eqref{def of M0hat and M1hat} take the following form:
\begin{align}  \label{M} 
\begin{split}
    \widehat{K_0} &= \max\left\{ j\geq 0: \n\left(\left(\beta/4\right)^j,\left(\beta/4\right)^j\right)\leq \n\left(\alpha/4, \beta/4\right) \quad \text{and} \quad  \left(\beta/4\right)^j \geq 3\alpha/4  \right\}, \\
    \widehat{K_1} &= \max\left\{ j\geq 0: \n\left(\left(\alpha/4\right)^j, \left(\alpha/4\right)^j \right)\leq \n\left(\alpha/4, \beta/4\right)  \quad \text{and} \quad  \left(\alpha/4\right)^j \geq 3\beta/4\right\},
\end{split}
\end{align}
and we have the following specification of the  free parameters in \eqref{GMT_free_parameters}: select
\begin{itemize}
    \item $\{\gamma_{i,j}:\,  j \in \bN\}$, $i\in\{0,1\}$  according to  \eqref{GMT, opt condition}, 
    \item $K_0 \in \{0,\ldots, \widehat{K_0}\}$  and   $\gamma_{0,0} \in \big((3\alpha\vee\beta)/4,1\big)$ that jointly minimize \eqref{GMT, ESS under P0},
    \item $K_1 \in \{0,\ldots, \widehat{K_1}\}$  and   $\gamma_{1,0}\in\big((\alpha\vee 3\beta)/4,1\big)$ that jointly minimize   \eqref{GMT, ESS under P1},
\end{itemize}
where $\widehat{K_0}$ and $\widehat{K_1}$  are given by \eqref{M}.  
 
This specification requires, for each $i \in \{0,1\}$, $\widehat{K_i}+1$ optimizations for  the selection of $\gamma_{i,0}$.  Even though this  task is not  prohibitive from a computational point of view, for both  practical and theoretical  purposes it  suffices, for each $i \in \{0,1\}$, to  select $K_i$ equal to its largest possible value, $\widehat{K_i}$,  and  perform only a single  optimization for the specification of  $\gamma_{i,0}$. In fact, as we show next,  this choice for $K_0$ and $K_1$ leads to a relatively small number of stages  and  guarantees the asymptotic optimality of GMT under both hypotheses  as $\alpha,\beta\to 0$ at arbitrary rates, even with a suboptimal selection of $\gamma_{0,0}$ and $\gamma_{1,0}$.

\subsection{The number of stages}
By their definition in \eqref{M}, we  obtain  the following upper bounds for $\widehat{K_0}$ and $\widehat{K_1}$:
\begin{equation*}
\begin{aligned}
    & \widehat{K_0}\leq \max\left\{ j\geq 0: (\beta/4)^j\geq 3\alpha/4 \right\} = \left\lfloor \frac{|\log(3\alpha/4)|}{|\log(\beta/4)|} \right\rfloor \\
    & \widehat{K_1}\leq \max\left\{ j\geq 0: (\alpha/4)^j\geq 3\beta/4 \right\} = \left\lfloor \frac{|\log(3\beta/4)|}{|\log(\alpha/4)|} \right\rfloor,
\end{aligned}
\end{equation*}
which do not depend on the hypotheses being tested. 
These bounds imply that  
\begin{itemize}
    \item $\widehat{K_0}=\widehat{K_1}=0$ when $\beta/3<\alpha<3\beta$, 
    \item $\widehat{K_1}=0$ (resp. $\widehat{K_0}=0$)   when $\alpha\leq \beta/3$ (resp. $\alpha\geq 3\beta$).
\end{itemize} 
That is,  when  $K_0$ and $K_1$ are selected as  $\widehat{K_0}$ and $\widehat{K_1}$ in  \eqref{M}, the GMT reduces to the 3-Stage Test when $\alpha$ and $\beta$ are not very different and, otherwise,  it adds to it  opportunities either only to accept or only to reject the null hypothesis. The number of these opportunities depends on the level of asymmetry between $\alpha$ and $\beta$, and it can be much smaller than the above upper bound.  For example,  in the Gaussian mean testing problem of Subsection \ref{subsec: a gaussian example} with $\eta=0.5$, 
$$ \widehat{K_0} \;\; (\text{resp. $\widehat{K_1}$}) \leq\begin{cases}
1,\quad  \text{when} \; r\leq 3, \\
2,\quad  \text{when} \; r\leq 6, \\ 
3,\quad  \text{when} \; r\leq 9,
\end{cases} $$
where $$ r=\frac{|\log(3\alpha/4)|}{|\log(\beta/4)|} \quad \left(\text{resp. } \frac{|\log(3\beta/4)|}{|\log(\alpha/4)|}\right). $$

\myremark  When $\widehat{K_0}=\widehat{K_1}=0$, the GMT reduces to the 3-Stage Test, but the selection of the parameters $N_{i,0}, C_{i,0}$, $i \in \{0,1\}$   according to  \eqref{GMT, general design, 00} and \eqref{GMT, general design, 10} is  different from the one  in  Section \ref{section, 3ST}, which relies on an application of Markov's inequality in \eqref{by Markov's}. Specifically, the  design in this section  is less conservative in terms of  error control, and when  $\gamma_{0,0}$ and $\gamma_{1,0}$ are selected to minimize \eqref{GMT, ESS under P0} and \eqref{GMT, ESS under P1} respectively, it also leads to  smaller expected sample size under both hypotheses.

\subsection{Asymptotic optimality}
We now state the  main theoretical result of this work in the context of the  binary testing problem, which is the  asymptotic optimality of the GMT  under both hypotheses  as $\alpha,\beta\to 0$ \emph{at arbitrary rates}.

\begin{theorem} \label{Theorem, optimality of GMT}  
For any $\alpha,\beta\in (0,1)$,     let 
$\big(\widehat{T}(\alpha, \beta),\widehat{D}(\alpha, \beta)\big)$ denote the sample size and the decision of GMT when its  parameters are selected   according to \eqref{N}-\eqref{GMT, general design, 10}, with 
\begin{itemize}
    \item  $\{\gamma_{i,j}:\, j \in \bN\}$, $i\in\{0,1\}$ given by  \eqref{GMT, opt condition}, 
    \item $K_i=\widehat{K}_i$,  $i\in\{0,1\}$, given by \eqref{M},
    \item  $\gamma_{0,0}\in\big((3\alpha\vee\beta)/4,1\big)$ and $\gamma_{1,0}\in\big((\alpha\vee 3\beta)/4,1\big)$ that minimize \eqref{GMT, ESS under P0}  and  \eqref{GMT, ESS under P1} respectively or, more generally, as functions of $\alpha$ and $\beta$  such  that, as $\alpha,\beta\to 0$, 
    \begin{align}  \label{condition on gamma} 
    \begin{split}
        & \gamma_{0,0} \to 0 \quad \text{ and } \quad |\log\gamma_{0,0}| \ll |\log\beta|, \\
        & \gamma_{1,0} \to 0 \quad \text{ and } \quad  |\log\gamma_{1,0}| \ll |\log\alpha|.  
    \end{split}
    \end{align}
\end{itemize}
Then, the family  
$$\widehat{\chi} \equiv 
\left\{ \big(\widehat{T}(\alpha, \beta), \widehat{D}(\alpha, \beta)\big): \alpha, \beta \in (0,1) \right\}$$
is asymptotically optimal under both hypotheses as $\alpha,\beta\to 0$ at arbitrary rates.
\end{theorem}
\begin{IEEEproof}
See Appendix \ref{proofs about GMT}. \\
\end{IEEEproof}  
  
In the special case of  the Gaussian mean testing problem of Subsection \ref{subsec: a gaussian example}, we can  derive  a second-order asymptotic upper bound on the expected sample size of GMT under each of the two hypotheses. These upper bounds coincide with the ones derived for the 3-Stage Test in \cite[Section 2]{Lorden_1983}, for the same testing problem,  under  condition  \eqref{constraint on alpha and beta for A.O. of 3ST} on the decay rates of $\alpha$ and $\beta$. 

\begin{proposition} \label{proposition, GMT}
Consider the Gaussian mean testing problem of Subsection \ref{subsec: a gaussian example} and, for any $\alpha,\beta\in (0,1)$, let  $\big(\widehat{T}(\alpha, \beta),\widehat{D}(\alpha, \beta)\big)$   be defined as in  Theorem \ref{Theorem, optimality of GMT},  with the only difference that  \eqref{condition on gamma}  is replaced by 
\begin{align} \label{Higher-order, selection of gamma}
\begin{split}
    \gamma_{0,0} &= \Theta\left(1/\sqrt{|\log\beta|}\right), \\
    \gamma_{1,0}&=\Theta\left(/\sqrt{|\log\alpha|}\right).
\end{split}
\end{align}
Then, as $\alpha,\beta\to 0$, 
\begin{equation}  
\begin{aligned}
    \Exp_0[\GMT(\alpha,\beta)] & \leq \frac{|\log\beta|}{I}  \left(1+O\left( \sqrt{\frac{|\log|\log\beta||}{|\log\beta|}} \right)\right), \\
    \Exp_1[\GMT(\alpha,\beta)] & \leq \frac{|\log\alpha|}{I}  \left(1+O\left( \sqrt{\frac{|\log|\log\alpha||}{|\log\alpha|}} \right)\right).
    \end{aligned}
\end{equation}
\end{proposition}
\begin{IEEEproof}
See Appendix \ref{proofs about GMT}. \\
\end{IEEEproof}

\section{Sequential Thresholding} \label{section, ST}
In this section we revisit the test that was proposed in  \cite{Malloy_Nowak_2014}, termed  \emph{Sequential Thresholding (ST)}. This is also a  multistage test with  deterministic  stage sizes  and maximum number of stages. Its two main characteristics are that (i)  the null hypothesis can be accepted at every stage but rejected only at the last possible stage, and (ii) all previous data  are discarded at the beginning of every stage. From the results in \cite{Malloy_Nowak_2014} it  follows that, with an appropriate selection of  the maximum  number of stages, this test  is asymptotically optimal under the null hypothesis as the error probabilities go to zero at arbitrary rates. However, as we show in this section,  this  comes at the price of severe performance loss under the alternative hypothesis.  Motivated by this phenomenon, we  introduce a  modification of ST, to which we refer as  \emph{Modified Sequential Thresholding (mod-ST)},  which does not discard data from previous stages and, as a result, turns out to have substantially better  performance under the alternative hypothesis.

\subsection{Description}
We describe the two tests, ST and mod-ST, in parallel.  For both  of them,    we denote by $K$ the maximum number of stages and, for each $j\in [K]$,  we denote by  $m_j$ the sample size, by $b_j$ the threshold, and by  $\Lambda'_{j}$ the test statistic that is utilized at  the $j^{th}$ stage, which is 
\begin{align} \label{Lambda_prime}
\Lambda'_{j} &\equiv
\begin{cases}
\begin{aligned}
    & \frac{1}{m_j} (\Lambda_{M_j}- \Lambda_{M_{j-1}}) && \text{in ST} ,\\
    & \bar\Lambda_{M_{j}} && \text{in mod-ST} ,
\end{aligned}
\end{cases}
\end{align}
where 
$M_j \equiv  m_1+\cdots+m_{j},\, M_0=0$.  That is,  $\Lambda'_j$ is the average log-likelihood ratio of the observations collected \emph{only during the $j^{th}$ stage} for ST, whereas   it  is the average log-likelihood ratio of all observations that have been collected \emph{up to and including the $j^{th}$ stage}  for mod-ST.  An algorithmic description for both tests is provided in Algorithm \ref{Algorithm of ST}.  Clearly, they both   reduce to the fixed-sample-size test when $K=1$.

\begin{algorithm} \caption{Sequential Thresholding (ST) and Modified Sequential Thresholding (mod-ST)} \label{Algorithm of ST}
\begin{algorithmic} 
    \State Input: $K$; $(m_j,b_j),\, j\in [K]$
    \State Initialize: $j=1$
    \While{$j\leq K-1$}
        \State take $m_j$ samples 
        \If{$\Lambda'_{j}\leq b_j$}
            \State stop and accept the null
        \Else
            \State $j=j+1$
        \EndIf
    \EndWhile
    \If{$j=K$}
        \State take $m_K$ samples 
        \If{$\Lambda'_K\leq b_K$}
            \State stop and accept the null
        \Else
            \State stop and reject the null
        \EndIf
    \EndIf
\end{algorithmic}
\end{algorithm}

\subsection{Selection of parameters}
Given the maximum number of stages, $K$, for each of the two tests, ST and mod-ST,  there are $2K$ parameters,   $(m_j,b_j),\,j\in[K]$, that need to be determined.  For each test, the events of rejecting and accepting the null hypothesis can be written respectively as
\begin{equation*} 
    \bigcap_{j=1}^K \left\{\Lambda'_j>b_j\right\} \qquad \text{and}  \qquad \bigcup_{j=1}^K \left\{\Lambda'_j\leq b_j\right\}.
\end{equation*}
Therefore, for each of them to belong to $\cE(\alpha, \beta)$ it suffices that its parameters are selected so that
\begin{align}
    \Pro_0\left(   \bigcap_{j=1}^{K}  \left\{  \Lambda'_j>b_j  \right\}   \right)  & \leq \alpha \label{ST and MST, type-I error} 
\end{align}
and 
\begin{align}
   \sum_{j=1}^K  \Pro_1\left(\Lambda'_j \leq b_j  \right) & \leq \beta. \label{ST and MST, type-II error} 
\end{align}

Similarly to  the GMT, we  require that the active type-II error probabilities after the first stage  decay exponentially with the number of stages, i.e., 
\begin{align}
    \Pro_1\left( \Lambda'_j\leq b_j \right) &\leq (\beta/2)^j, \quad j \in \{2, \ldots, K \},\label{ST and MST, type-II error in latter stages} 
\end{align}
and that all remaining type-II error probability be assigned to the first stage, i.e.,
\begin{align}
    \Pro_1\left(\Lambda'_1 \leq b_1  \right) &\leq \beta-   \sum_{j=2}^K  (\beta/2)^j.   \label{ST and MST, type-II error in the first stage} 
\end{align}
Moreover, we  require that  the type-I error probability (inactive in the first $K-1$ stages and active in the last stage) be distributed evenly among the $K$  stages,  so that  \eqref{ST and MST, type-I error} is strengthened   to
\begin{align} \label{ST and MST, joint}
    \Pro_0\left(  \bigcap_{i=1}^{j} \left\{ \Lambda'_i>b_i \right\}  \right) & \leq \alpha^{j/K}, \quad j \in [K]. 
\end{align}

Inequalities \eqref{ST and MST, type-II error in latter stages}-\eqref{ST and MST, joint} provide $2K$ constraints for the specification of the  $2K$ parameters of each test.  Specifically, from  \eqref{ST and MST, type-II error in the first stage}  and \eqref{ST and MST, joint} with $j=1$ we obtain the following  specification for the test parameters in the \textit{first} stage:
\begin{align} \label{m1, b1}
    (m_1, b_1) & ={\sf FSST} \left( \alpha^{1/K}, \, \beta-\sum_{j=2}^K\left(\beta/2\right)^j \right),
\end{align}
which is common for  ST and mod-ST. However, the parameters  of the two tests in the remaining stages  differ.  Indeed,  for ST, the test statistics $\Lambda'_j,\,j\in[K]$ are independent, thus,  \eqref{ST and MST, joint}  is equivalent to
\begin{equation*} \label{ST, type-I error in each stage}
    \Pro_0(\Lambda'_j>b_j) \leq \alpha^{1/K}, \quad j\in [K],
\end{equation*}
which, combined with \eqref{ST and MST, type-II error in latter stages}, implies that  
\begin{align} \label{mj, bj}
    (m_j, b_j) & ={\sf FSST} \left( \alpha^{1/K}, \, \left(\beta/2\right)^j \right), \quad  j \in \{2, \ldots, K\}. 
\end{align}
On the other hand,  the remaining parameters of mod-ST have to be specified recursively.  Indeed, suppose that $(m_1,b_1), \ldots, (m_{j-1}, b_{j-1})$  have been specified for some $j \in \{2, \ldots, K\}$ and set
$$M_i = m_1+\ldots+m_i \quad \text{ for  } \;  i \in [j-1].$$
Then, $m_j$ is the minimum non-negative integer such that  
\eqref{ST and MST, type-II error in latter stages} and \eqref{ST and MST, joint} hold simultaneously, i.e.,
\begin{equation} \label{mj defined by joint probs}
\begin{aligned}
    m_j = \min\bigg\{ n\in\bN: & \; \exists \, b\in\bR \text{ such that } \\
    & \Pro_0\bigg( \bar\Lambda_{M_{j-1}+n}> b,\; \bigcap_{i=1}^{j-1} 
    \left\{\bar\Lambda_{M_i}> b_i  \right\}\bigg)\leq \alpha^{j/K} \quad \text{and} \\ 
    & \Pro_1\left( \bar\Lambda_{M_{j-1}+n}\leq b \right)\leq (\beta/2)^j  \bigg\},
\end{aligned}
\end{equation}
and $b_j$ is the minimum of such thresholds, i.e., 
\begin{align} \label{bj defined by joint probs}
    b_j=\min \left\{ b\in\bR: \Pro_0\bigg(  \bigcap_{i=1}^{j} 
    \left\{\bar\Lambda_{M_i}> b_i  \right\}\bigg) \leq \alpha^{j/K}  \quad \text{and} \quad 
    \Pro_1\left( \bar\Lambda_{M_{j}}\leq b_j \right)\leq (\beta/2)^j \right\},
\end{align}
where $M_j =M_{j-1}+m_j.$ 

\subsection{The  number of stages}
To complete the specification of  ST and mod-ST, we need to select the maximum number of stages, $K$. In contrast to the  case of  the  3-Stage Test or the  GMT, this choice  has to strike a balance between the relative cost of sampling under the two hypotheses. Indeed,  since ST and mod-ST  allow for rejecting the null hypothesis only at the final stage, one should clearly select $K=1$, i.e., apply a fixed-sample-size test, when the absolute priority is to have  small expected sample size  under the alternative hypothesis.  A larger value of $K$ can lead to  smaller expected sample size under the null  hypothesis \textit{at the expense of performance loss relative to the FSST under the alternative  hypothesis}.  To resolve this trade-off, one  may select the largest value of $K$ for which  the increase  in the expected sample size  under the alternative, relative to the FSST, can be tolerated.  Alternatively,  $K$ can be selected to minimize the expected sample size under the mixture  distribution $\Pro_\pi$, defined in \eqref{mixture}, for some given  $\pi \in [0,1]$. A natural choice for this $\pi$  arises in the signal recovery problem of Section  \ref{sec: problem formulation about high-dim}, as we discuss in Subsection \ref{subsec: second study}. We stress, however,  that no such external criterion is needed for the  design of the 3-Stage Test or the GMT.

\subsection{Asymptotic optimality}
We next show  that when $K$ is selected appropriately as a function of $\alpha$ and $\beta$, ST is  asymptotically optimal under the null hypothesis as $\alpha,\beta\to 0$ at arbitrary rates. This  was shown in \cite{Malloy_Nowak_2014} under a second-moment assumption on the log-likelihood ratio statistic, whereas here we only require finiteness of the first moment, i.e., \eqref{KL information numbers},  which is our standing assumption throughout this paper. However, at the same time we show that, with this selection of $K$, the  expected sample size of ST  under the alternative hypothesis  is  asymptotically larger than the optimal by a factor that is much  larger than $(K+1)/2$. That is, the asymptotic optimality of ST under the null hypothesis comes at the price of severe performance loss  under the alternative hypothesis.  Finally, we show that this performance loss is substantially mitigated when using mod-ST instead of ST. Specifically,  we show that mod-ST enjoys the same asymptotic optimality property  as ST under the null hypothesis, while its expected sample size under the alternative hypothesis is  smaller than that of ST by a factor that is not smaller than $(K+1)/2$. 

\begin{theorem} \label{Theorem, ST}
For any $\alpha,\beta\in (0,1)$, let $\big(T'(\alpha,\beta),D'(\alpha,\beta)\big)$ and $\big(T''(\alpha,\beta),D''(\alpha,\beta)\big)$ denote the sample size and the decision of ST and mod-ST  respectively when their parameters are selected according to \eqref{ST and MST, type-II error in latter stages}-\eqref{ST and MST, joint}, with $K$ being a function of $\alpha$ and $\beta$ such that, as $\alpha,\beta\to 0$,
\begin{equation} \label{ST, condition on K, transformation}
    \alpha^{1/K} \to 0   \quad \text{ and } \quad |\log\alpha^{1/K}| \ll |\log\beta|
\end{equation}    
or, equivalently, 
\begin{equation} \label{ST, condition on K}
    \frac{|\log\alpha|}{|\log\beta|} \ll  K \ll  |\log\alpha|.
\end{equation}
Then, the families 
 \begin{align*}
    \chi'  &\equiv 
    \big\{ \big(T'(\alpha, \beta), D'(\alpha, \beta)\big): \alpha, \beta \in (0,1) \big\} \\
    \chi'' & \equiv 
    \big\{ \big(T''(\alpha, \beta), D''(\alpha, \beta)\big): \alpha, \beta \in (0,1) \big\}
\end{align*}
are both asymptotically optimal under the null hypothesis. Moreover,    as $\alpha,\beta\to 0$,
\begin{align} 
    \Exp_1[T'(\alpha,\beta)] &\sim \frac{K(K+1)}{2}\,  \frac{|\log\beta|}{I_0} \gg \frac{K+1}{2}  \cL_1(\alpha,\beta)   \label{ST, asy upper bound on E1[T']}\\
    \Exp_1[T''(\alpha,\beta)]  &\lesssim  K \frac{|\log\beta|}{I_0} \sim  \frac{2}{K+1}  \, \Exp_1[T'(\alpha,\beta)]. \label{ST, asy upper bound on E1[T'']}
\end{align}  
\end{theorem}
\begin{IEEEproof}
See Appendix \ref{proofs about ST}. \\
\end{IEEEproof}

With a more specific selection of $K$, we obtain   the same second-order asymptotic upper bound for the expected sample sizes  of ST and mod-ST under the null hypothesis  as for GMT in Proposition \ref{proposition, GMT} in the Gaussian mean testing problem.

\begin{proposition} \label{Proposition: higher-order, ST and mod-ST}
Consider the Gaussian mean testing problem of Subsection \ref{subsec: a gaussian example} and  for any $\alpha,\beta\in (0,1)$ let $\big(T'(\alpha,\beta),D'(\alpha,\beta)\big)$ and $\big(T''(\alpha,\beta),D''(\alpha,\beta)\big)$ be defined as in Theorem \ref{Theorem, ST} with  \eqref{ST, condition on K, transformation}  replaced by 
\begin{equation} \label{ST, selection of K}
    \alpha^{1/K} =  \Theta\left(1/\sqrt{|\log\beta|}\right),
\end{equation} 
or equivalently
$$K = \frac{2 |\log\alpha|}{|\log|\log\beta||+\Theta(1)}.$$
Then, as $\alpha, \beta  \to 0$, 
\begin{equation} \label{higher-order upper bound on ST and MST}
  \Exp_0[T'(\alpha, \beta)], \; \Exp_0[T''(\alpha,\beta)] \leq \frac{|\log\beta|}{I} \left(1+O\left( \sqrt{\frac{|\log|\log\beta||}{|\log\beta|}} \right)\right).
\end{equation}
\end{proposition}
\begin{IEEEproof}
See Appendix \ref{proofs about ST}. \\
\end{IEEEproof}

\myremark Comparing \eqref{m1, b1} and \eqref{GMT, general design, 00}, we can see that $\alpha^{1/K}$ is the inactive type-I error probability at the first opportunity to accept the null hypothesis for ST and mod-ST and, in that sense, it plays the same role as $\gamma_{0,0}$ for the GMT. In view of this,  conditions   \eqref{ST, condition on K, transformation} and \eqref{ST, selection of K}  are completely analogous to  \eqref{condition on gamma} and  \eqref{Higher-order, selection of gamma}.

\section{High-dimensional testing} \label{sec: problem formulation about high-dim}
In this section we consider the problem of simultaneously testing $m\in\bN$ copies of the binary testing problem  in Section \ref{sec: problem formulation of one-dim testing}. Thus, we let   $ X^1, \ldots, X^m$ be independent streams of  iid random elements, each  with  density either $f_0$ or $f_1$, where, as before, our standing and only  assumption regarding these densities is  \eqref{KL information numbers}.   We refer to a stream as   \emph{``noise''}  if its  density is $f_0$ and as  \emph{``signal''}   if its  density is $f_1$, and we  denote by  $\Pro_\cA$ and  $\Exp_\cA$  the probability and the   expectation  respectively when the subset  of signals is  $\cA \subseteq [m]$.   

We restrict ourselves to multiple testing procedures  that apply  the  same  binary test, i.e., test in $\cE$, to each stream. For such procedures, the  expected average sample size  over all $m$ streams  depends only on the number of signals, but not on the actual subset of signals. Indeed, for any  $(T,D) \in \cE$, let   $(T^j, D^j)$ denote its version that is applied to the $j^{th}$ data stream, $X^j$, where  $j\in [m]$. Then,    for any  $\cA\subseteq[m]$,   
\begin{align} \label{EASS}
\begin{split}
    \frac{1}{m}\,\sum_{j=1}^m \Exp_\cA\left[  T^j \right] &= \frac{1}{m}\,\left(\sum_{j\notin \cA} \Exp_0\left[  T \right]  + \sum_{j\in \cA} \Exp_1\left[  T \right]\right) \\
    &= \left(1-\frac{|\cA|}{m}\right) \, \Exp_0[T] + \frac{|\cA|}{m} \, \Exp_1[T] \\
    &= \Exp_{|\cA|/m} [T],
    \end{split}
\end{align}
recalling that  $\Exp_\pi$ denotes the expectation under the mixture distribution $\Pro_\pi$, defined in \eqref{mixture}. Our main goal in this section  is to find multiple testing procedures composed by multistage tests that minimize the \textit{expected average sample size over all $m$ streams} as $m\to\infty$, in the class of all multiple testing procedures that satisfy certain global error constraints. Specifically, we consider \textit{classical}  familywise error control  in Subsection \ref{subsec: familywise}  and  \textit{generalized} familywise error control in  Subsection \ref{subsec: generalized_familywise}. In both setups, we assume that  there is  a user-specified lower bound, $l_m$, and a user-specified upper bound, $u_m$,  on the number of signals, where  
\begin{align} \label{conditions for l_m, u_m}
    0\leq l_m\leq u_m\leq m,  \quad u_m> 0, \quad l_m< m.
\end{align}

\myremark The  problem formulation  in  Subsection \ref{subsec: familywise} generalizes the one in \cite{Malloy_Nowak_2014} in two ways. First, by not requiring the number of signals to be a priori known and, second, by allowing for distinct control of the probabilities of at least one type-I error and at least one type-II error. The  formulation  in  Subsection \ref{subsec: generalized_familywise} is  a further generalization, which provides additional flexibility in the design of multiple testing procedures.  

\subsection{Controlling classical familywise error probabilities} \label{subsec: familywise}
For any $(T,D)\in\cE$ and $\cA\subseteq[m]$, we denote by $\text{FWE-I}_{\cA}(T,D)$ and $\text{FWE-II}_{\cA}(T,D)$ the familywise type-I and type-II error probabilities when the  subset of signals is $\cA$, i.e.,
\begin{align*}
    \text{FWE-I}_{\cA} (T,D) &\equiv \Pro_\cA\left( \exists \,  j \notin \cA :  D^j=1 \right) ,\\
    \text{FWE-II}_{\cA}(T,D) &\equiv 
    \Pro_\cA\left(  \exists \,  j \in \cA :  \,  D^j=0 \right). 
\end{align*}
For any $\alpha, \beta \in (0,1)$ and $m\in \bN$, we denote by $\cE_m(\alpha,\beta)$ the family of  tests  whose familywise type-I and type-II error probabilities do not exceed  $\alpha$ and $\beta$ respectively, i.e.,
\begin{align*}
    \cE_m(\alpha,\beta) \equiv \big\{ 
    (T,D)\in \cE: \;  \text{FWE-I}_{\cA}(T,D) \leq \alpha\; & \text{ and } \; 
    \text{FWE-II}_{\cA}(T,D) \leq \beta,\;\\
    & \forall \, \cA\subseteq [m],\, l_m \leq |\cA|\leq u_m
    \big\}.
\end{align*}
For any $\alpha, \beta \in (0,1)$ and $m\in\bN$, this family of tests can be expressed in terms of the one in  \eqref{class(alpha,beta)} in the following way (see Lemma \ref{lemma for equivalence of classes} in Appendix \ref{proofs about hign-dim}):  
\begin{align} \label{alpham, betam}
\begin{split}
    \cE_m(\alpha,\beta) &=\cE\big(\alpham, \betam\big), \\
    \text{where} \quad \alpham &\equiv 1- (1-\alpha)^{1/(m-l_m)},  \\
    \betam &\equiv 1-(1-\beta)^{1/u_m}.
\end{split}
\end{align}
As a result, by \eqref{EASS} and \eqref{alpham, betam} it follows that the optimal expected average sample size  in $\cE_m(\alpha, \beta)$ when the number of signals is  $s$, where $s \in\{l_m,\ldots,u_m\}$,  is
\begin{align*} \label{optimal_ESS_multiple}
    \inf_{(T,D) \in \cE_m(\alpha, \beta)}\Exp_{s/m}[T]&=   \inf_{(T,D) \in \cE(\alpha_m, \beta_m)}\Exp_{s/m}[T] = \cL_{s/m}(\alpha_m, \beta_m),
\end{align*}
where in the second equality we apply the definition of $\cL_\pi$  in \eqref{optimal_mixture}. Our goal in this subsection is to find families  of tests that  \textit{achieve this infimum,  for every $\alpha, \beta \in (0,1)$,   uniformly in the  possible number of signals, $s\in\{l_m,\ldots,u_m\}$, as the number of streams, $m$,  goes to infinity.} This is expressed by the  following notion of asymptotic optimality.

\begin{definition} \label{def: AO in high-dim sense}
A family of tests, $\chi^*$,  defined as  in \eqref{family},  is  asymptotically optimal in the high-dimensional sense if, for any $\alpha,\beta\in(0,1)$, we have, as $m \to \infty$,
\begin{equation} \label{def of AO in high-dim, uniform}
    \Exp_{s/m}[T^*(\alpham,\betam)]\sim \cL_{s/m}(\alpham,\betam) \text{ uniformly in } s\in\{l_m,\ldots,u_m\},
\end{equation}
i.e., 
\begin{equation} \label{def of AO in high-dim, max}
    \max_{s\in \{l_m, \ldots, u_m\}} 
    \frac{\Exp_{s/m}[T^*(\alpham,\betam)]}{\cL_{s/m}(\alpham,\betam) } \longrightarrow 1.
\end{equation}
\end{definition}

We start by showing that in order to establish such a high-dimensional asymptotic optimality property for a family of tests other than the SPRT, the maximum possible number of signals, $u_m$, and the maximum possible number of noises, $m-l_m$, should both go to infinity as $m \to \infty$.

\begin{theorem} \label{thm: sharpness}
Consider a family of tests, $\chi^*$,  defined in \eqref{family}, 
that is asymptotically optimal in the high-dimensional sense. 
Recall the  family of SPRTs, $\widetilde{\chi}$, defined in \eqref{family SPRTs}.
    \begin{itemize}
        \item If $u_m\not\to \infty$ as $m\to \infty$, then there exists a strictly increasing sequence of positive integers, $(m_k)_{k\in\bN}$, such that as $k\to\infty$,
        $m_k \to \infty$   and 
        \begin{equation} \label{only possible for the SPRT}
            \Exp_0[T^*(\alpha_{m_k},\beta)]\lesssim \Exp_0[\widetilde{T}(\alpha_{m_k},\beta)], \quad \forall\;\alpha,\beta\in(0,1).
        \end{equation}
        \item If  $m-l_m \not\to \infty$ as $m\to \infty$, then there exists a strictly increasing sequence of positive integers, $(m_k)_{k\in\bN}$,  such that as $k\to\infty$,$m_k \to \infty$ and 
        \begin{equation} \label{only possible for the SPRT,2}
            \Exp_1[T^*(\alpha,\beta_{m_k})]\lesssim \Exp_1[\widetilde{T}(\alpha,\beta_{m_k})], \quad \forall\;\alpha,\beta\in(0,1).
        \end{equation}
\end{itemize}
\end{theorem}
\begin{IEEEproof}
    See Appendix \ref{proofs about hign-dim}. \\
\end{IEEEproof}

\myremark For any $\alpha,\beta\in(0,1)$, $\Exp_0[\widetilde{T}(\alpha_{m_k},\beta)]$ is bounded as $\alpha_{m_k}\to 0$ and $\Exp_1[\widetilde{T}(\alpha_{m_k},\beta)]$ is bounded  as $\beta_{m_k}\to 0$. Therefore, \eqref{only possible for the SPRT} (resp. \eqref{only possible for the SPRT,2}) implies that  in order to achieve asymptotic optimality in the high-dimensional sense when $u_m \not \to \infty$ (resp. $m-l_m \not \to \infty$), $\chi^*$ should perform as well as $\widetilde{\chi}$ in \textit{a non-asymptotic sense} under the null (resp. alternative) hypothesis.  Moreover, the  proof in Appendix \ref{proofs about hign-dim} applies with $\tilde{\chi}$  replaced by  the family of SPRTs with the smallest possible thresholds that satisfy the corresponding error constraints, which  is  exactly optimal under both hypotheses when these constraints are satisfied with equality. This suggests that letting  both $u_m$ and $m-l_m$ go to infinity as $m\to\infty$  is a necessary condition  for any family of tests, other than the SPRT, to achieve asymptotic optimality in the high-dimensional sense, and it will be assumed in the rest of this subsection. \\

We next characterize the optimal asymptotic performance and provide criteria for a family to achieve it. 

\begin{theorem} \label{thm: main in high-dim}
Suppose that  $u_m \to \infty$ and $m-l_m\to\infty$ as $m\to\infty$. Let $\chi^*$ be a family of tests defined in \eqref{family}.
\begin{itemize} 
\item [(i)] For all $\alpha,\beta\in(0,1)$,
\begin{align} \label{asy approx to cLs/m(alpham,betam)}
\begin{split}
    \cL_{s/m}(\alpham,\betam) & \sim \left(1-\frac{s}{m}\right)\frac{\log u_m}{I_0} + \frac{s}{m}\frac{\log(m-l_m)}{I_1} \\
    &\text{ uniformly in } s\in\{l_m,\ldots,u_m\}. 
    \end{split}
\end{align}
\item [(ii)] If, for all $\alpha,\beta\in(0,1)$,
\begin{align}  
   \Exp_0[T^*(\alpham,\betam)] & \sim \cL_0(\alpham,\betam),  \label{E0 sim L0} \\
   \Exp_1[T^*(\alpham,\betam)] & \sim \cL_1(\alpham,\betam),  \label{E1 sim L1} 
   \end{align}
then $\chi^*$ is asymptotically optimal in the high-dimensional sense.
\item [(iii)] If, for all $\alpha,\beta\in(0,1)$, \eqref{E0 sim L0} holds and, also,  
\begin{align}
    \Exp_1[T^*(\alpham,\betam)] & \ll \frac{(m-u_m)\, \log u_m}{u_m}, \label{high-dim, ST or mod-ST, alternative}
\end{align}
then $\chi^*$ is asymptotically optimal in the high-dimensional sense.
\end{itemize}
\end{theorem}
\begin{IEEEproof}
See Appendix \ref{proofs about hign-dim}. \\
\end{IEEEproof}

\begin{corollary} \label{corollary of SPRTs, GMTs  in high-dim}
If  $u_m \to \infty$ and $m-l_m\to \infty$, then 
\begin{itemize}
    \item the family of SPRTs, $\widetilde{\chi}$, defined in \eqref{family SPRTs}, is asymptotically optimal in the high-dimensional sense,
    \item the family of GMTs, $\widehat{\chi}$, defined in Theorem \ref{Theorem, optimality of GMT}, is asymptotically optimal in the high-dimensional sense.
    \end{itemize}
\end{corollary}
\begin{IEEEproof}
    See Appendix \ref{proofs about hign-dim}. \\
\end{IEEEproof}
    
\begin{corollary} \label{corollary of  3STs in high-dim}
If  $u_m \to \infty$ and $m-l_m\to \infty$ so that 
   $ \log(m-l_m)=\Theta(\log u_m)$, 
    then the family of 3-Stage Tests, $\check{\chi}$, defined in Theorem \ref{thm: asy opt of 3ST}, is asymptotically optimal in the high-dimensional sense.
\end{corollary}
\begin{IEEEproof}
    See Appendix \ref{proofs about hign-dim}. \\
\end{IEEEproof}

\begin{corollary} \label{corollary, ST in high-dim}
If  $u_m\to\infty$  so that    $u_m\ll m$ as $m\to\infty$, then
\begin{itemize}
    \item [(i)] for all $\alpha,\beta\in(0,1)$,
    \begin{equation} \label{AA to EASS with sparsity}
        \cL_{s/m}(\alpham,\betam) \sim \frac{\log u_m}{I_0} \text{ uniformly in } s\in\{l_m,\ldots,u_m\},
    \end{equation}
    \item [(ii)] the family of STs, $\chi'$, and  the family of mod-STs, $\chi''$, defined in Theorem \ref{Theorem, ST}, are both  asymptotically optimal in the high-dimensional sense.
\end{itemize}
\end{corollary}
\begin{IEEEproof}
    See Appendix \ref{proofs about hign-dim}. \\
\end{IEEEproof}

\myremark Corollary \ref{corollary, ST in high-dim}  generalizes  \cite[Theorem 3, Corollary 2]{Malloy_Nowak_2014}  in that it establishes the asymptotic optimality of ST  in the high-dimensional sense 
\begin{itemize}
\item  without any assumption on $l_m$,  which, e.g., can be equal to 0 for every $m \in \bN$,
\item whenever $u_m \to \infty$ so that $u_m \ll m$,
\item assuming only \eqref{KL information numbers}, i.e.,  a finite first moment for the log-likelihood ratio  in \eqref{def: LLR}  under each hypothesis,
\item allowing for distinct familywise error control for two kinds of errors.
\end{itemize}
On the other hand, \cite[Theorem 3, Corollary 2]{Malloy_Nowak_2014}  establishes the asymptotic optimality  of ST in the high-dimensional sense 
\begin{itemize}
\item when the number of signals is a priori known, i.e., $l_m=u_m$ for every $m\in\bN$,
\item when  $u_m\to\infty$ so that 
\begin{equation} \label{Mally and Nowak's cond on lm and um}
    u_m\lesssim \frac{m}{(\log m)^2}, 
\end{equation}
\item assuming a finite second moment for the log-likelihood ratio in \eqref{def: LLR},
\item controlling the probability of at least one error, of any kind. \\
\end{itemize}

\myremark Corollary \ref{corollary, ST in high-dim}  shows that under a sparse setup, i.e., when $u_m\to\infty$ so that $u_m\ll m$ as $m\to \infty$, the optimal expected average sample size for any true number of signals is of the order of logarithm \textit{of the maximum possible number of signals}. On the other hand, under a setup that is neither sparse nor dense, i.e., when $u_m=\Theta(m)$ and $m-l_m=\Theta(m)$ as $m\to\infty$, which implies 
$$\log u_m\sim \log (m-l_m)\sim \log m,$$
then from \eqref{asy approx to cLs/m(alpham,betam)} it follows that 
$$ \frac{\log m}{I_0\vee I_1} \lesssim \cL_{s/m}(\alpham,\betam)   \lesssim   \frac{\log m}{I_0\wedge I_1} $$
uniformly for every $s\in\{l_m,\ldots,u_m\}$, i.e., the optimal expected average sample size under any true number of signals is of the order of logarithm of the \textit{total number of streams}. This difference between these regimes  is illustrated by the shape of the curves in Figures \ref{Figure, high-dim, known exact number s}(a) and \ref{Figure, high-dim, known upper bound u}(a).

\subsection{Controlling generalized familywise error probabilities}  \label{subsec: generalized_familywise}
We next generalize the results of the previous subsection by considering multiple testing procedures that control  generalized familywise error probabilities \cite{lehmann2012generalizations}. Thus, in what follows, in addition to $l_m$ and $u_m$ that satisfy \eqref{conditions for l_m, u_m},  we also introduce $\kappa_m \in \bN$ and $\iota_m\in\bN$ such that 
$$1 \leq \iota_m< u_m \quad \text{ and } \quad 1 \leq \kappa_m< m-l_m,$$
and, for any test $(T,D)\in\cE$, we introduce its $\kappa_m$-\textit{generalized familywise type-I error probability}, i.e., the probability of at least $\kappa_m$ type-I errors,  when the subset of signals is $\cA\subseteq[m]$, as
\begin{equation*}
    \text{$\kappa_m$-GFWE-I}_\cA(T,D)  \equiv \Pro_\cA\left( \exists \, B\subseteq\cA^c : |B|=\kappa_m \; \text{ and } \; D^j=1 \;\; \forall \, j\in B \right),
\end{equation*}
and its $\iota_m$-\textit{generalized familywise type-II error probability},  i.e., the probability of at least $\iota_m$ type-II errors, when the subset of signals is $\cA\subseteq[m]$, as
\begin{equation*}
    \text{$\iota_m$-GFWE-II}_\cA(T,D)  \equiv \Pro_\cA\left( \exists \, B\subseteq\cA : |B|=\iota_m \; \text{ and }  \;  D^j=0 \;\; \forall \, j\in B \right).
\end{equation*}
For any $\alpha,\beta\in(0,1)$ and $m\in\bN$, we further introduce  the class of tests for which these two error metrics are bounded above by $\alpha$ and $\beta$ respectively as 
\begin{equation*}
    \begin{aligned}
    \cE^{G}_m(\alpha,\beta)\equiv \big\{(T,D)\in\cE: \text{$\kappa_m$-GFWE-I}_\cA(T,D)\leq\alpha & \text{ and } \text{$\iota_m$-GFWE-II}_\cA(T,D)\leq\beta, \\
    & \forall\,\cA\subseteq[m], \; l_m\leq |\cA|\leq u_m \big\}.
    \end{aligned}
\end{equation*}
This can be expressed  in terms of the family in \eqref{family} in the following way (see Lemma \ref{lemma, exact value of alphamG and betamG} in Appendix \ref{proofs about hign-dim}):
\begin{align}  \label{inverse binomial, alpha, beta} 
\begin{split}
    \cE_m^G(\alpha,\beta) &=\cE \left(\alpham^G,\betam^G \right), 
\end{split}
\end{align}
where $\alpham^G$ is  the  largest  $p\in(0,1)$ such that 
\begin{align} 
    \mathsf{B}(m-l_m,p;\kappa_m) & \leq \alpha, \label{inverse binomial, alpha} 
    \end{align}
and  $\betam^G$  the  largest  $p\in(0,1)$ such that
\begin{align}   
    \mathsf{B}(u_m,p;\iota_m) & \leq \beta, \label{inverse binomial, beta}
\end{align}
where $\mathsf{B}(n,p;k)$ is  the probability that a Binomial random variable with parameters $n \in \bN$ and $p\in (0,1)$
is \textit{greater than or equal to} $k \in \{0,1, \ldots, n\}$. 

For any $\alpha,\beta\in(0,1)$ and $m\in\bN$, by \eqref{EASS} and \eqref{inverse binomial, alpha, beta} it follows that the optimal expected average sample size in $\cE_m^G(\alpha,\beta)$ when the number of signals is $s$, where $s\in\{l_m,\ldots,u_m\}$, is 
$$ \inf_{(T,D)\in\cE_m^G(\alpha,\beta)} \Exp_{s/m}[T]=   \cL_{s/m}(\alpham^G,\betam^G).$$
This leads us to the following definition of asymptotic optimality. 

\begin{definition} \label{def: AO in high-dim sense, generalized}
A family of tests, $\chi^*$,  defined as  in \eqref{family},  is  asymptotically optimal in the high-dimensional sense under generalized error control if, for any $\alpha,\beta\in (0, 1/2)$,  as $m \to \infty$,
\begin{equation} \label{def of AO in high-dim, uniform, generalized}
    \Exp_{s/m}[T^*(\alpha^G_m,\beta^G_m)]\sim \cL_{s/m}(\alpham,\betam) \; \text{ uniformly in } s\in\{l_m,\ldots,u_m\},
\end{equation}
i.e., 
\begin{equation} \label{def of AO in high-dim, max, generalized}
    \max_{s\in \{l_m, \ldots, u_m\}} 
    \frac{\Exp_{s/m} \left[T^*(\alpha^G_m,\beta^G_m) \right]}{\cL_{s/m} \left(\alpha^G_m,\beta^G_m \right) } \longrightarrow 1.
\end{equation}
\end{definition}

\begin{theorem} \label{theorem for GFWE, SPRTs, GMTs and 3STs}
Suppose that  $u_m \to \infty$ and $m-l_m\to\infty$ so that  $\iota_m\ll u_m$ and $\kappa_m\ll m-l_m$ as $m\to\infty$.   Let $\chi^*$ be a family of tests defined in \eqref{family}.
\begin{itemize}
\item [(i)] For all $\alpha,\beta\in(0,1/2)$, 
\begin{equation} \label{kappa, iota comparable}
\begin{aligned}
    \cL_{s/m}(\alpham^G,\betam^G) 
    & \sim \left(1-\frac{s}{m}\right) \frac{\log (u_m/\iota_m)}{I_0} + \frac{s}{m} \frac{\log\big((m-l_m)/\kappa_m\big)}{I_1}
\end{aligned}
\end{equation}
uniformly in $s\in\{l_m,\ldots,u_m\}$. 

\item [(ii)] If,  for all $\alpha,\beta\in(0,1/2)$,
\begin{align} 
    \Exp_0[T^*(\alpham^G,\betam^G)] & \sim \cL_0(\alpham^G,\betam^G), \label{GFWE, for ST and mod-ST, null} \\
     \Exp_1[T^*(\alpham^G,\betam^G)] & \sim \cL_1(\alpham^G,\betam^G), \label{GFWE, for ST and mod-ST, alt}
\end{align}
then $\chi^*$ is asymptotically optimal in the high-dimensional sense under generalized error control.
\item [(iii)] If, for all $\alpha,\beta\in(0,1/2)$, \eqref{GFWE, for ST and mod-ST, null} holds and, also, 
\begin{align} 
    \Exp_1[T^*(\alpham^G,\betam^G)] & \ll \frac{(m-u_m)\log(u_m/\iota_m)}{u_m}, \label{GFWE, for ST and mod-ST, alternative}
\end{align}
then  $\chi^*$ is asymptotically optimal in the high-dimensional sense under generalized error control.
\end{itemize}
\end{theorem}
\begin{IEEEproof} 
Appendix  \ref{proofs about hign-dim}. \\
\end{IEEEproof}

\myremark (i) The formula for the optimal asymptotic performance in  \eqref{kappa, iota comparable}  agrees with the one in the  case of classical familywise error control, i.e.,   \eqref{asy approx to cLs/m(alpham,betam)}, once the maximum number of signals $u_m$ is replaced by $u_m/ \iota_m$   and the the maximum number  of noises $m-l_m$ is replaced by $(m-l_m)/ \kappa_m$. This suggests that the effect of generalized familywise error control is essentially to reduce the ``effective" maximum numbers of signals and noises. 

(ii) The optimal asymptotic performance in \eqref{kappa, iota comparable} reduces to the one in  the case of classical familywise error control, i.e.,   \eqref{asy approx to cLs/m(alpham,betam)}, when $\iota_m$ and $\kappa_m$ are bounded as $m \to \infty$ or, more generally, when $\iota_m \to \infty$ and $\kappa_m \to \infty$ so that
$$\log\iota_m\ll \log u_m \quad \text{ and } \quad \log\kappa_m\ll \log (m-l_m).
$$ \\

We end this section with the analogues of Corollaries \ref{corollary of SPRTs, GMTs in high-dim}  \ref{corollary of 3STs in high-dim} and \ref{corollary, ST in high-dim}.

\begin{corollary} \label{corollary, GFWE, for SPRT, GMT}
If  $u_m \to \infty$, $m-l_m\to\infty$, $\iota_m\ll u_m$, $\kappa_m\ll m-l_m$ as $m\to\infty$, then
\begin{itemize}
    \item the family of SPRTs, $\widetilde{\chi}$, defined in \eqref{family SPRTs}, is asymptotically optimal in the high-dimensional sense under generalized error control, 
    \item the family of GMTs, $\widehat{\chi}$, defined in Theorem \ref{Theorem, optimality of GMT}, is asymptotically optimal in the high-dimensional sense under generalized error control.
    \end{itemize}
\end{corollary}
\begin{IEEEproof}
See Appendix \ref{proofs about hign-dim}. \\
\end{IEEEproof}

\begin{corollary} \label{corollary, GFWE, for  3ST}
If $u_m \to \infty$, $m-l_m\to\infty$, $\iota_m\ll u_m$, $\kappa_m\ll m-l_m$ so that 
$$ \log \left( \frac{u_m}{\iota_m} \right)=\Theta\left(\log \frac{m-l_m}{\kappa_m} \right) \text{ as } m\to\infty, $$  
then the family of  3-Stage Tests, $\check\chi$, defined in Theorem \ref{thm: asy opt of 3ST} is asymptotically optimal in the high-dimensional sense under generalized error control.
\end{corollary}
\begin{IEEEproof}
See Appendix \ref{proofs about hign-dim}. \\
\end{IEEEproof}

\begin{corollary} \label{corollary, GFWE, for ST and mod-ST}
If $u_m\to\infty$, $u_m\ll m$,
$\iota_m\ll u_m$, $\kappa_m\ll m-l_m$ as $m\to\infty$, then
\begin{itemize}
\item [(i)] for all $\alpha,\beta\in(0,1/2)$, 
\begin{equation*}
\cL_{s/m}(\alpham^G,\betam^G)\sim \frac{\log(u_m/\iota_m)}{I_0} \text{ uniformly in $s\in\{l_m,\ldots,u_m\}$, }
\end{equation*}
\item [(ii)] the family of mod-STs,  $\chi''$, defined in Theorem \ref{Theorem, ST},   is asymptotically optimal in the high-dimensional sense under generalized error control if also
\begin{equation} \label{GFWE, condition for mod-ST}
\frac{\log(m/\kappa_m)}{\log(u_m/\iota_m)}\ll \frac{m}{u_m},
\end{equation}
\item [(iii)] the family of STs, $\chi'$, defined in Theorem \ref{Theorem, ST},      is asymptotically optimal in the high-dimensional sense under generalized error control if also
\begin{equation} \label{GFWE, condition for ST}
\frac{\log(m/\kappa_m)}{\log(u_m/\iota_m)}\ll \sqrt{\frac{m}{u_m}}.
\end{equation}
\end{itemize}
\end{corollary}
\begin{IEEEproof}
See Appendix \ref{proofs about hign-dim}. \\
\end{IEEEproof}

\myremark (i) Unlike in the case of classical familywise error control in Corollary \ref{corollary, ST in high-dim}, the condition $u_m\ll m$ does not suffice for the proof of asymptotic optimality of ST and  mod-ST in the high-dimensional sense under  generalized error control, unless $\log\iota_m\ll\log u_m$. 

(ii) Sufficient conditions for   \eqref{GFWE, condition for mod-ST} and \eqref{GFWE, condition for ST}, that do not depend on $\kappa_m$ and $\iota_m$, are 
$$ u_m\lesssim \frac{m}{\log m}  \qquad \text{and}
\qquad   u_m\lesssim \frac{m}{(\log m)^2} $$
respectively.  To see this, it suffices to observe  that, since  $\kappa_m\geq 1$ and $u_m/\iota_m\to\infty$, 
$$ \frac{\log(m/\kappa_m)}{\log(u_m/\iota_m)}\ll \log m. $$

\section{Numerical studies} \label{sec: numerical study}
In this section we present the results of  two numerical studies in which we compare the  General Multistage Test (GMT) with the Sequential  Probability Ratio Test (SPRT), the Fixed-Sample-Size Test (FSST), the Sequential Thresholding (ST), and the Modified Sequential Thresholding (mod-ST).  

\subsection{Binary testing}
In the first study we consider the Gaussian mean testing problem  of Subsection \ref{subsec: a gaussian example}, with $\eta=0.5$, in two  setups regarding  the  error probabilities: a symmetric one,  where $\alpha=\beta=10^{-6}$, and an  asymmetric one, where  $\alpha=10^{-12},\,\beta=10^{-2}$. In both  setups, 
\begin{itemize}
\item the FSST is implemented using the formulas in \eqref{exact formulas in testing Gaussian mean}; specifically, in the symmetric case we have $\mathsf{FSST}(\alpha,\beta)=(91,0)$, and  in the asymmetric case, $\mathsf{FSST}(\alpha,\beta)=(88,0.2509)$,
\item the  GMT  is implemented  according to the design of Theorem \ref{Theorem, optimality of GMT} with  $\gamma_{0,0}$ and $\gamma_{1,0}$ selected to minimize \eqref{GMT, ESS under P0} and \eqref{GMT, ESS under P1} respectively;  thus,  in the symmetric case it coincides with the 3-Stage Test 
($\widehat{K_0}=\widehat{K_1}=0$), whereas in the asymmetric case it  offers two additional opportunities to accept the null hypothesis ($\widehat{K_0}=0$, $\widehat{K_1}=2$),
\item ST and mod-ST are designed according to \eqref{ST and MST, type-II error in latter stages}-\eqref{ST and MST, joint}, with its maximum possible number of stages,  $K$, selected to match those of the GMT; thus, in the symmetric case we have $K=3$, and in the asymmetric case $K=5$,
\item the SPRT,  defined in  \eqref{def: SPRT}, is implemented  with $A=|\log\alpha|$ and $B=|\log\beta|$.
\end{itemize}

For each of the two setups, we compute  the expected sample size (ESS) of each of the five tests not only when the mean of the Gaussian sequence, $\mu$, is in $\{-0.5,\,0.5\}$, but also for 100 equally-spaced values in $[-0.6,0.6]$. For  each value of $\mu$, the ESS  of each of the  multistage tests is computed using exact formulas. Since there is not an exact formula for the ESS of  the SPRT, it is computed for each value of $\mu$ using  $10^5$  Monte-Carlo simulation runs, which leads to a \textit{relative} standard error well below $1\%$  in all cases. 

\begin{figure} [ht]
  \centering
  \subfloat[$\alpha=\beta=10^{-6}$]{\includegraphics[width=0.5\textwidth]{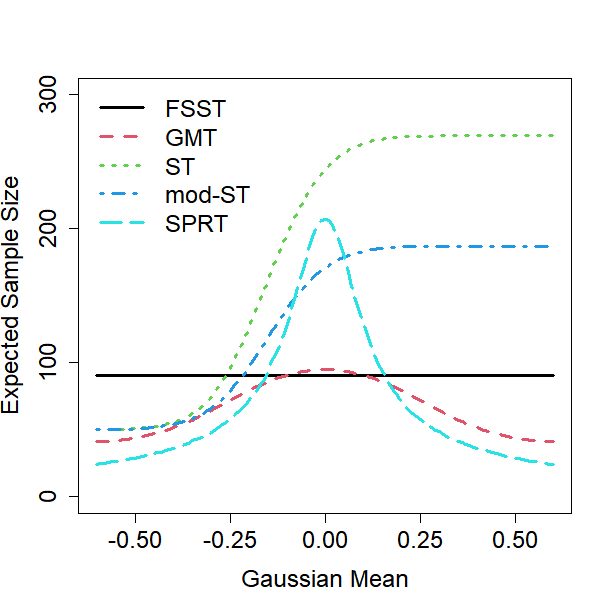}}
  \subfloat[$\alpha=10^{-12},\;\beta=10^{-2}$]{\includegraphics[width=0.5\textwidth]{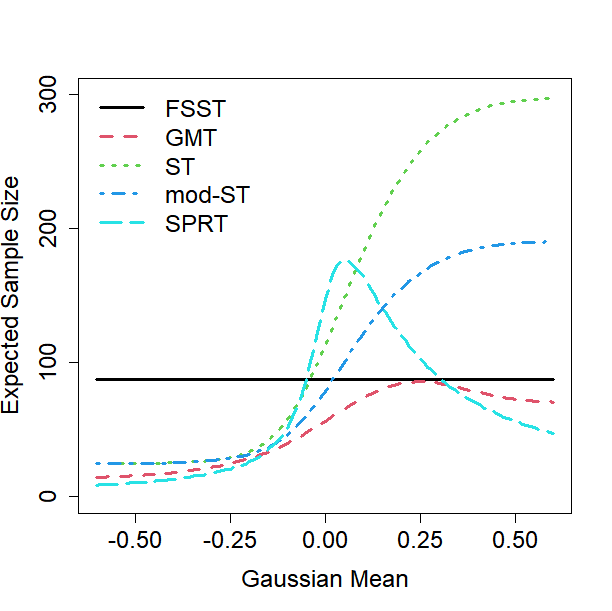}}
  \caption{Expected sample size against the true  mean for the Gaussian mean testing problem \eqref{Normal testing problem} with $\eta=0.5$.}
  \label{Figure, one-dim, four tests}
\end{figure}

\begin{table}[ht]
    \centering
    \begin{tabular}{|c|c|c|c|c|c|}
    \hline
    \multicolumn{2}{|l|}{} & GMT & ST & mod-ST & SPRT \\
    \hline
    \multirow{3}{*}{$\alpha=\beta=10^{-6}$} & $\mu=-0.5$ & 0.49 & 0.56 & 0.56 & 0.32  \\
    \cline{2-6}
    & worst-case & 1.05 & 2.98 & 2.07 & 2.29 \\
    \cline{2-6}
    & $\mu=0.5$ & 0.49 &  2.98 & 2.07 & 0.32 \\
    \hline
    \multirow{3}{*}{$\alpha=10^{-12}, \beta=10^{-2}$} 
    & $\mu=-0.5$ & 0.18 & 0.29 & 0.29 & 0.12 \\
    \cline{2-6}
    & worst-case & 0.98 & 3.39 & 2.17 & 2.02 \\
    \cline{2-6}
    & $\mu=0.5$ & 0.83 & 3.37 & 2.16 & 0.64 \\
    \hline
    \end{tabular}
    \caption{Ratio of the expected sample size over
    $\n(\alpha, \beta)$  under the two hypotheses, i.e., when $\mu=\pm 0.5$, and in the worst-case  with respect to $\mu$. }
    \label{Table of ratios}
\end{table}

The ESS for each  test is  plotted  against  $\mu$  in Figure \ref{Figure, one-dim, four tests}. The ratio  of the ESS  for each  test  over $\n(\alpha, \beta)$   is presented in Table \ref{Table of ratios}  when $\mu=\pm 0.5$ and in the worst-case with respect to $\mu$. Based on   Figure \ref{Figure, one-dim, four tests} and Table \ref{Table of ratios} we can make the following observations:
\begin{itemize}
\item In the symmetric setup, the ESS of GMT (resp. SPRT) under both hypotheses is about  half  (resp. a third)  of $\n(\alpha, \beta)$.  In the asymmetric setup, the ESS of GMT (resp. SPRT)  is 18\%  (resp.  12\%)  of $\n(\alpha, \beta)$  under the null hypothesis and  83\%  (resp.  64\%)  of  $\n(\alpha, \beta)$ under the alternative hypothesis. 
\item  In both setups,  the worst-case ESS of  the SPRT is more than double that of  GMT, whereas the worst-case ESS of  the latter is about the same as $\n(\alpha, \beta)$.  
\item The ESS of mod-ST  coincides with that of ST  when $\mu$ is around the null, while being substantially smaller for larger values of $\mu$. Nevertheless, even for mod-ST, the ESS for large values of $\mu$ 
is much larger even compared to  $\n(\alpha, \beta)$.  
\item The ESS  of GMT is smaller than that of  ST and mod-ST even when $\mu$ is around the null. 
\end{itemize}

Based on these observations, we can conclude that, at least   when one of $\alpha$ and $\beta$ is small enough, 
\begin{itemize}
\item the GMT provides a robust alternative to the SPRT from a statistical point of view, in addition to not requiring continuous monitoring of the sampling process. 
\item ST and even  mod-ST do not seem to offer any advantage compared to GMT even under the null hypothesis.
\end{itemize}
One may ask whether  ST and mod-ST could benefit in this comparison by using more stages than GMT.  To answer this question, in  Figure \ref{Figure, ST only}   we plot the ESS of ST and mod-ST against $\mu$ for every value of  $K$ from 1 to 4 in the symmetric setup and  from 1 to 6 in the asymmetric setup. From  these figures we can see that increasing the number of stages   leads, for both tests, to  a   large increase of the ESS   for  larger values of $\mu$, especially in the case of ST, and a relatively small reduction of the ESS  when $\mu$ is around the null.

\begin{figure} [ht]
  \centering
  \subfloat[ST, $\alpha=\beta=10^{-6}$]
  {\includegraphics[width=0.5\textwidth]{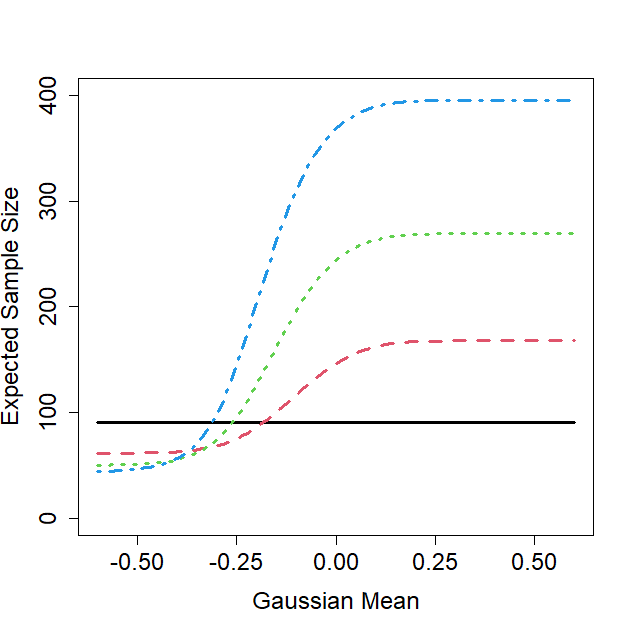}}
  \subfloat[mod-ST, $\alpha=\beta=10^{-6}$] 
  {\includegraphics[width=0.5\textwidth]{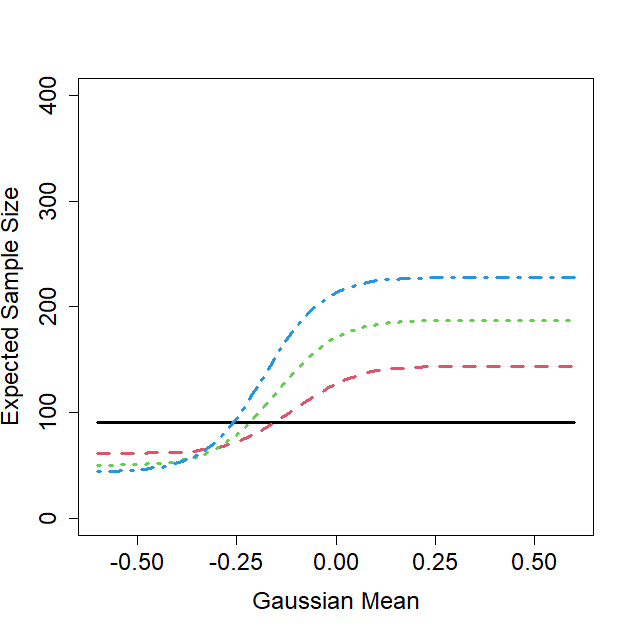}} \\
  \subfloat[ST, $\alpha=10^{-12}, \, \beta=10^{-2}$]{\includegraphics[width=0.5\textwidth]{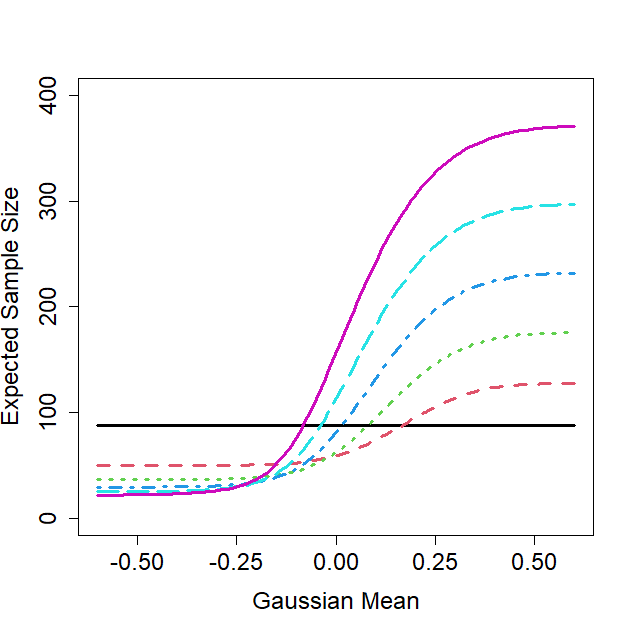}}
  \subfloat[mod-ST, $\alpha=10^{-12}, \, \beta=10^{-2}$] {\includegraphics[width=0.5\textwidth]{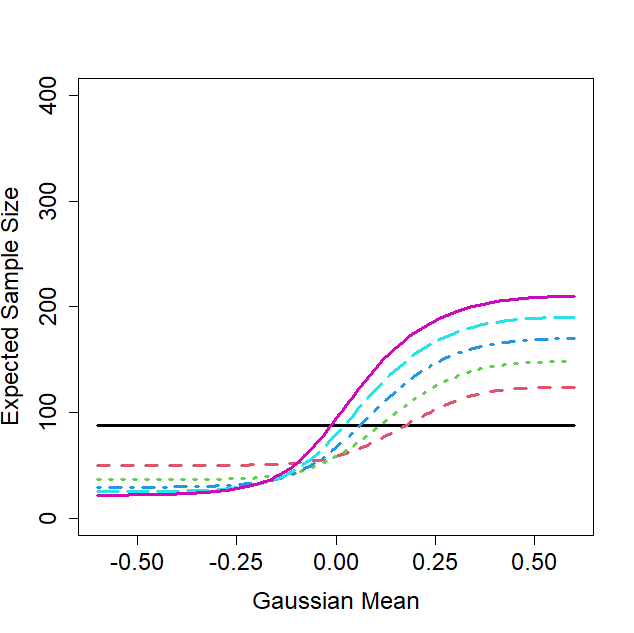}} \\
  \caption{ESS of ST (left column) and mod-ST (right column) with different $K$'s, against the true Gaussian mean $\mu$, when $\alpha=\beta=10^{-6}$ (first row) and when $\alpha=10^{-12},\,\beta=10^{-2}$ (second row). Looking at the RHS of each figure, from bottom to top, the curves correspond to $K=2,\ldots,4$ in the first row, and $K=2,\ldots,6$ in the second row. In each figure, the flat line corresponds to the case  $K=1$, where both ST and mod-ST  reduce to the FSST. 
  }
  \label{Figure, ST only}
\end{figure}

\subsection{High-dimensional testing}
\label{subsec: second study}
In the second study we consider  again the Gaussian mean testing problem  of Subsection \ref{subsec: a gaussian example}, with $\eta=0.5$, but now in the context of Section \ref{sec: problem formulation about high-dim}.  We consider control of familywise type-I and type-II error probabilities below  $\alpha=0.05$ and $\beta=0.05$, respectively. We fix the number of  data streams to be  $m=10^6$. Since we consider only one value for $m$, in what follows we simply write  $l$ and $u$, instead of $l_m$ and $u_m$,  to denote the lower and the upper bound on  the number of signals, and thus, the values of   $\alpha_m$ and $\beta_m$ in \eqref{alpham, betam} take the following form:
\begin{equation} \label{new_alpha_beta}
    1-(1-\alpha)^{1/(m-l)} \quad  \text{and} \quad 1-(1-\beta)^{1/u}.
\end{equation}

We consider two setups regarding the prior information on the number of signals.   In the first one,  this  number  is assumed to be known, i.e., $l=u$, and  for each of its possible values, i.e., for each of $1\leq u\leq m-1$, we compute  the  ESS of each of the five tests under consideration with respect to the mixture distribution $\Pro_\pi$, defined in  \eqref{mixture}, with  $\pi=u/m$.    In the second setup,  only an upper bound on the number of signals is assumed to be known, i.e.,  $l=0$, and for each possible  value of the upper bound, i.e., for each of $1\leq u\leq m$, we compute  the ESS  of each  test under consideration with respect to the mixture distribution $\Pro_{\pi}$, defined in  \eqref{mixture}, with $\pi=u/2m$. This  corresponds to  the average ESS  over the $m$ data streams \textit{and over the  $u+1$ possible values for the  number of signals, $\{0,1,\ldots,u\}$}. Indeed, for any stopping time $T$ we have 
\begin{align} \label{target function based on a uniform weight}
\begin{split}
    \frac{1}{u+1}\, \sum_{j=0}^{u}  \Exp_{j/m}[T] &=
    \frac{1}{u+1} \sum_{j=0}^u \left( \left(1-\frac{j}{m}\right)\Exp_0[T] + \frac{j}{u}\Exp_1[T]\right) \\    
    &=  \left(1-\frac{u}{2m}\right)\Exp_0[T] + \frac{u}{2m}\Exp_1[T]    =\Exp_{u/2m}[T]. 
\end{split}
\end{align}

In both setups, FSST, GMT, SPRT  are designed in exactly the same  way as in the first study, with $\alpha$ and $\beta$ replaced by the corresponding quantities in \eqref{new_alpha_beta}. ST and mod-ST are designed according to \eqref{ST and MST, type-II error in latter stages}-\eqref{ST and MST, joint} with $\alpha$ and $\beta$ also replaced by  the corresponding quantities in \eqref{new_alpha_beta}. However,  $K$  is not selected, as in the first study,  to match the maximum number of stages of GMT.  Instead, for each value of $u$  we select the value of  $K$ that does not exceed 10 and   minimizes the criterion of each setup, i.e., the expected sample size under the mixture distribution $\Pro_\pi$ with $\pi=u/m$ in the first setup   and with $\pi=u/2m$ in the second.   Although it is possible, we do not perform a similar tuning to select the parameters of the GMT, so the comparison can be regarded as favorable for ST and mod-ST.

In Figure \ref{Figure, high-dim, number of stages} we plot the maximum number of stages for each of the three multistage tests against $u/m$. For the GMT, this number  is equal to 
\begin{itemize} 
\item $3$ when  $u$ and   $m-l$ are close, 
which is the case when   $u$ is around  $m/2$ (resp.  close to $m$)  in the first (resp. second) setup,
\item 5 when $u$ and $m-l$ are very different, which is the case when $u$ is close to  $0$ or $m$ (resp. close to $0$) in the first (resp. second) setup, 
\item 4 in all other cases.
\end{itemize}
Regarding  ST and  mod-ST, we observe first of all that they both reduce to the FSST when $u$ is large. Specifically,  for ST (resp. mod-ST), this is the case when   $u/m$  is larger than about 0.3 (resp. 0.4) in the first setup  and  0.55 (resp. 0.7)  in the second.  On the other hand, as $u$ decreases, their maximum  number  of stages increases up to 9, in comparison to at most 5 for GMT.  
\begin{figure}[ht]
  \centering
  \subfloat[]{\includegraphics[width=0.5\textwidth]{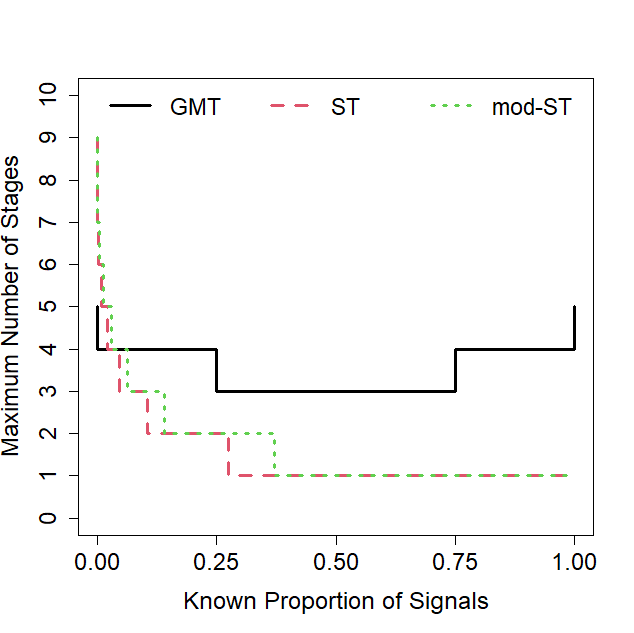}}
  \subfloat[]{\includegraphics[width=0.5\textwidth]{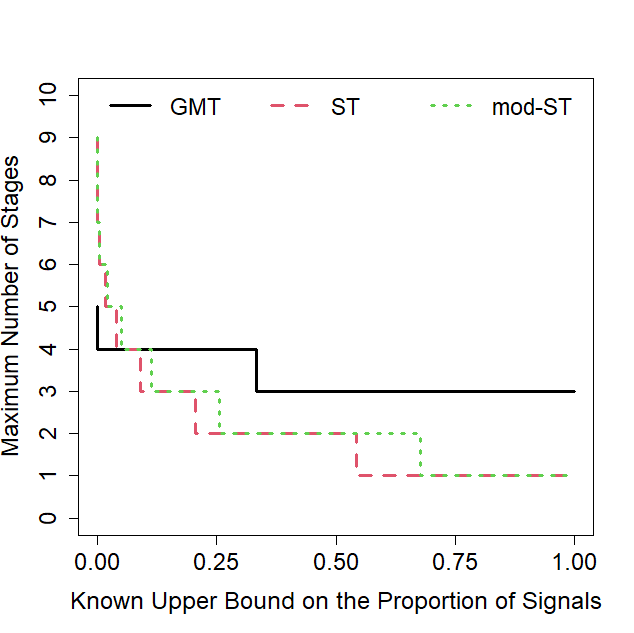}}
  \caption{The maximum number of stages in the multistage tests, against $u/m$.}
  \label{Figure, high-dim, number of stages}
\end{figure}

For each setup and each possible value of $u$,  the ESS of FSST, GMT, ST and mod-ST is computed using exact formulas,  while that of the SPRT using Monte-Carlo simulation with $10^5$ repetitions whose relative standard error is well below $1\%$ in all cases.  The results for the first setup are  presented  in Figure \ref{Figure, high-dim, known exact number s} and for the second in  Figure \ref{Figure, high-dim, known upper bound u}.   Based on these figures  we can make the following observations:

\begin{itemize}
\item All  curves in Figures \ref{Figure, high-dim, known exact number s}(a) and \ref{Figure, high-dim, known upper bound u}(a) decrease sharply for very small values $u$.  This  is consistent with the second remark after Corollary \ref{corollary, ST in high-dim}. The same remark also explains the fact that the curves that correspond to  GMT and SPRT in Figure \ref{Figure, high-dim, known exact number s}(a) (resp. \ref{Figure, high-dim, known upper bound u}(a)) are relatively flat when  $u/m$  is not very close to $0$ or $1$  (resp. not very close to 0). 
\item The ESS of  the GMT (resp. SPRT)  is smaller roughly by a factor of 2  (resp.  4)  relative to that of the FSST.  This does not contradict the results of the previous study, as all  these curves correspond to  weighted averages of the ESS when $\mu=\eta$ and when $\mu=-\eta$, i.e.,  we do not consider values of $\mu$ between $-\eta$ and $\eta$. 
\item  The ESS of  mod-ST  is similar to that of ST when $u$ is very small, but becomes smaller as  $u$ increases,  until  $u/m$ reaches a value of about  0.4 in the first setup and 0.75 in the second, at which both tests reduce to the  FSST.
\item For very small values of $u$, the ESS of mod-ST, but  not that of ST,  is slightly smaller than that of GMT, although the difference is too small to be visible. Note, however, that  for such values of $u$,  mod-ST and ST use   9 stages, whereas  GMT uses 5. 
\end{itemize}

\begin{figure}[ht]
  \centering
  \subfloat[]{\includegraphics[width=0.5\textwidth]{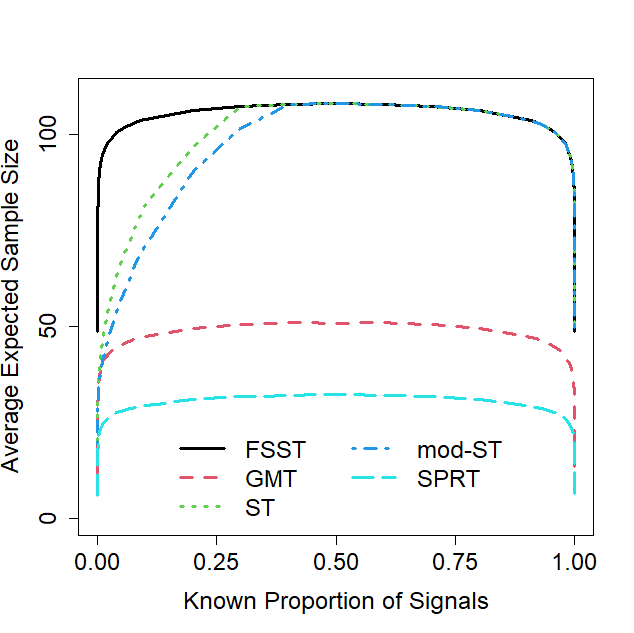}}
  \subfloat[]{\includegraphics[width=0.5\textwidth]{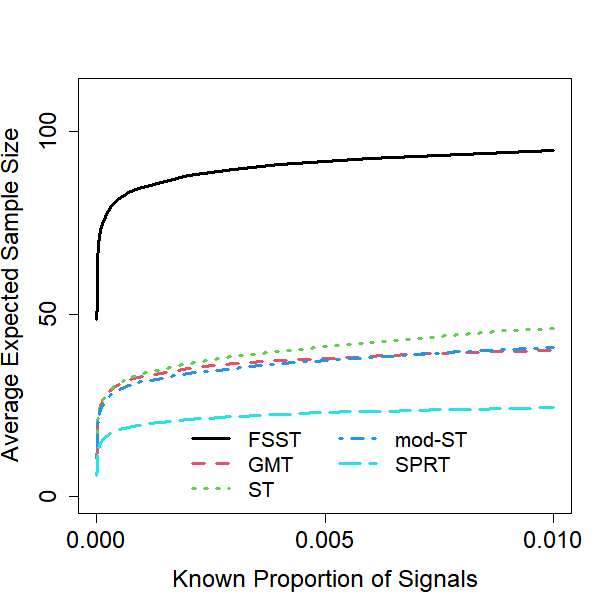}}
  \caption{$\Exp_{u/m}[T(\alpham,\betam)]$ against $u/m$ when $\alpha=\beta=0.05$, $m=10^6$, and $l=u\in\{1,\ldots,m-1\}$. (b) is the left $1\%$ of (a).
  }
  \label{Figure, high-dim, known exact number s}
\end{figure}

\begin{figure}[ht]
  \centering
  \subfloat[]{\includegraphics[width=0.5\textwidth]{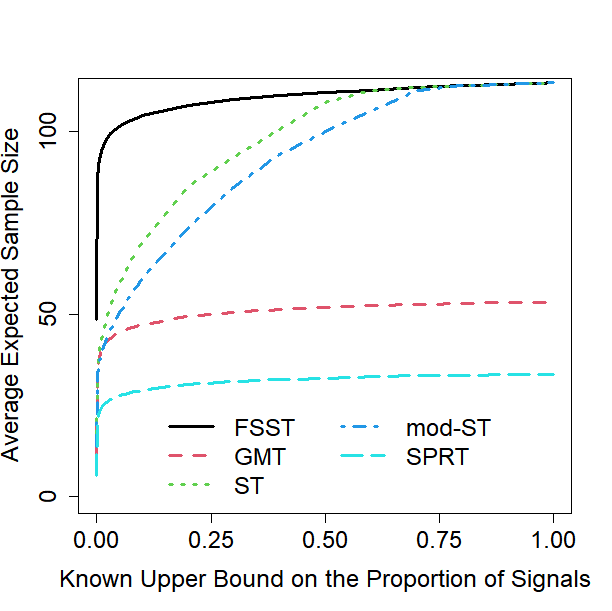}}
  \subfloat[]{\includegraphics[width=0.5\textwidth]{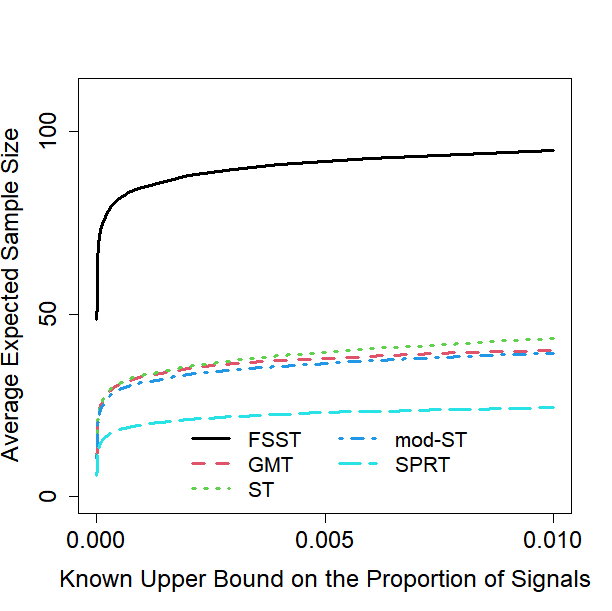}}
  \caption{$\Exp_{u/2m}[T(\alpham,\betam)]$ against $u/m$ when $\alpha=\beta=0.05$, $m=10^6$, $l=0$ and $u\in\{1,\ldots,m\}$. (b) is the left $1\%$ of (a).}
  \label{Figure, high-dim, known upper bound u}
\end{figure}

\section{Generalizations} \label{sec: generalizations}
For  the sake of simplicity and  clarity, we have focused on the  fundamental problem of testing two simple hypotheses about the distribution of a sequence of iid random elements. However,  the methods and results of this paper remain valid with minor modifications  for certain testing problems with non-iid data or  composite hypotheses.

\subsection{Non-iid data} \label{subsec: non-iid data}
Suppose that $X$ is not necessarily an iid sequence and 
consider the problem of testing two simple hypotheses about   its distribution, $\Pro$:
$$H_0 : \Pro=\Pro_0 \quad  \text{versus} \quad 
H_1: \Pro=\Pro_1, $$
where $\Pro_0$  and $\Pro_1$ are mutually absolutely continuous when restricted to $\cF_n$ for every $n \in \bN$. Then,  the  log-likelihood ratio statistic  
$$ \Lambda_n\equiv \log \frac{d\Pro_1(X_1,\ldots,X_n)}{d\Pro_0(X_1,\ldots, X_n)}
$$
is not  necessarily of the form \eqref{def: LLR}. However,  all results in this work  still hold (apart from those that refer to ST, which additionally require independent observations) as long as there exist positive and finite numbers, $I_0$ and $I_1$, such that 
\begin{equation} \label{asy SLLN}
    \Pro_0(\bar\Lambda_n\to -I_0)=\Pro_1(\bar\Lambda_n\to I_1)=1,    
\end{equation}
where, as before, $ \bar\Lambda_n\equiv \Lambda_n/n$, and  real-valued functions,  $\psi_0$ and $\psi_1$,  that satisfy  the following asymptotic versions  of  the inequalities in  \eqref{LD upper bounds},
\begin{equation} \label{asy LD}
\begin{aligned}
    & \limsup_n\, \frac{1}{n} \log \Pro_0(\bar\Lambda_n>c)\leq -\psi_0(c), \quad \forall\, c \geq -I_0, \\
    & \limsup_n\, \frac{1}{n} \log \Pro_1(\bar\Lambda_n\leq c)\leq -\psi_1(c), \quad \forall\, c  \leq I_1,
\end{aligned}
\end{equation}
and the four  properties stated after \eqref{def: psi}. Indeed, these  conditions  imply that  the asymptotic approximation   to the optimal expected sample size under each  hypothesis in \eqref{optimal performance} remains valid (see, e.g.,  \cite[Section 3.4]{Tartakovsky_Book}), and  also that the non-asymptotic upper bounds \eqref{non-asy 1} and \eqref{non-asy 2} remain valid to a first-order asymptotic approximation  as $\alpha,\beta\to 0$.
As a result, all  proofs in this work remain valid,  without essentially any modification. \\

\myremark
The above conditions  are satisfied, for example, when testing the transition matrix of a finite-state Markov chain, or the correlation coefficient  of a first-order autoregression. For more details,  we refer to  \cite{PaperI}.

\subsection{Composite hypotheses}
Let $\xi_0(x), x\in\bR$ be a density with respect to some $\sigma$-finite measure, $\nu$, on $\bR$, and for each $\theta \in \bR$, set 
$$\ b(\theta) \equiv \log\int_\bR \xi_0(x)\exp\{\theta x\} \nu (dx).$$
Suppose that the effective domain of $b$, 
$$\Theta\equiv \left\{  \theta\in  \bR:  b(\theta) <\infty\right\},$$ is an open interval  and that  $X\equiv \{X_n: n \in \bN\}$ is a sequence of iid  random variables with  density
\begin{equation} \label{def: exponential family}
\xi_\theta(x) = \xi_0(x) \exp\{\theta x - b(\theta) \}, \quad x\in \bR
\end{equation} 
with respect to $\nu$,  for some  $\theta\in \Theta$.  For each $\theta \in \Theta$, let $\Pro_\theta$ and $\Exp_\theta$  denote the probability measure and the expectation under $\xi_\theta$, and,  for any $u,v\in\Theta$, set
\begin{equation} \label{I(u,v)}
    I(u,v)\equiv \Exp_u\left[ \log \frac{\xi_u(X_1)}{\xi_v(X_1)} \right] =(u-v)b'(u)-\big( b(u)-b(v) \big).
\end{equation}
Since $b(\cdot)$ is strictly convex in $\Theta$ (see, e.g., \cite[Chapter 2.2.1]{Dembo_Zeitouni_LDPBook}), $ I(u,v)$ is  positive and finite   for every $u,v\in\Theta$.
 
Given  arbitrary  $\theta_0, \theta_1 \in\Theta$ such that $\theta_0  < \theta_1$,  consider the one-sided testing problem
\begin{equation} \label{one-sided testing}
    H_0:\theta\leq \theta_0 \quad \text{ versus } \quad H_1: \theta \geq  \theta_1.
\end{equation}
For this problem,  the  SPRT in \eqref{def: SPRT} can be defined using the sum of the observations, $S_n\equiv X_1+\ldots+X_n$ in the place of  $\Lambda_n$,  and  all multistage tests in this work can be defined using the average of the observations,  $\bar{X}_n\equiv S_n/n$ in the place of  $\bar{\Lambda}_n$.  

Moreover, if we require that 
\begin{itemize}
\item[(i)]  the maximum type-I error probability does not exceed some  $\alpha\in(0,1)$, and
\item[(ii)] the maximum type-II error probability does not exceed some $\beta\in (0,1)$,
\end{itemize} 
then the class of tests of interest  becomes 
\begin{equation}
    \bcE(\alpha,\beta) \equiv \left\{ (T,D)\in\cE: \sup_{\theta\leq \theta_0}\,\Pro_\theta(D=1) \leq \alpha \,\text{ and }\, \sup_{\theta \geq \theta_1}\, \Pro_\theta(D=0) \leq \beta \right\},
\end{equation}
and all multistage tests in this work belong to  $\bcE(\alpha,\beta)$ if they are implemented according to the proposed designs as long as   $\n(\alpha,\beta)$ in  \eqref{def: n*(alpha,beta)} is defined as follows:
\begin{align} \label{def: n*(alpha,beta)_composite}  
    \min \, \left\{n \in \bN: \;  \exists \; c \in \bR \; \text{ so that } \; \Pro_{\theta_0}( \bar X_n > c)\leq \alpha \; \text{ and } \; \Pro_{\theta_1}( \bar X_n \leq c) \leq \beta \right\}.
\end{align}
Besides, all asymptotic  results in this work    remain valid as long as  we replace $\cL_i$ with $\bcL_{\theta_i}$ for both $i\in\{0,1\}$, where,  for each $\theta\in \Theta$,  $\bcL_\theta(\alpha,\beta)$ denotes the smallest expected sample size under $\Pro_\theta$ in $\bcE(\alpha,\beta)$, i.e.,
\begin{equation}
    \bcL_\theta(\alpha,\beta) \equiv \inf_{(T,D)\in\bcE(\alpha,\beta)} \Exp_\theta[T]. 
\end{equation}
In view of the fact (see, e.g., \cite[Chapter 5.4]{Tartakovsky_Book}) that, as $\alpha, \beta \to 0$, 
\begin{align} \label{composite hyp, AO}
\begin{split}
   \sup_{\theta \leq \theta_0}  \bcL_{\theta}(\alpha,\beta)  &\sim \bcL_{\theta_0}(\alpha,\beta)\sim \frac{|\log\beta|}{I(\theta_0,\theta_1)}, \\
    \sup_{\theta \geq \theta_1}  \bcL_{\theta}(\alpha,\beta)  &\sim \bcL_{\theta_1}(\alpha,\beta)\sim \frac{|\log\alpha|}{I(\theta_1,\theta_0)},
\end{split}
\end{align}
this means that the  GMT in this context achieves the \textit{worst-case} expected sample size under each hypothesis asymptotically  as the error probabilities go to zero at arbitrary rates. \\
    
\myremark If, in addition to (i) \textit{and instead of (ii)}, we require that
\begin{itemize}
\item[(ii')]   the maximum expected sample size under  $H_0$ not exceed some $M\in\bN$,
\end{itemize}
then the parameters of each of the above tests will depend only on  $\theta_0, \alpha, M$. In this case, all these tests  can be implemented  not only with `` limited knowledge of the alternative distribution" \cite[Section V.B]{Malloy_Nowak_2014}, but  without  \textit{any}  knowledge of the alternative distribution. Indeed, in this formulation,   the alternative hypothesis can take the form $H_1: \theta >\theta_0$.  We stress, however, that  this is the case for all  tests in this work, i.e., this is not a special property of ST or of  any other multistage test.

\section{Conclusions and open problems} \label{sec: conclusion}
In this work we propose and solve a high-dimensional signal recovery problem that  generalizes the problem considered  in  \cite{Malloy_Nowak_2014}. Specifically,  as in the latter work, we consider multiple, independent data streams, each generating iid data,  pose the same  binary testing problem for each of them,  and require that the decision for each of these testing problems be based only on  observations from the corresponding data stream. However, in the present work we do not assume  that  the number of signals  and noises is  a priori known. Instead, we only require  upper bounds on them and  consider an asymptotic regime in which the maximum numbers of signals and noises go to infinity, while the two  familywise error probabilities, classical or generalized, are fixed. 

For this problem,   we introduce a novel multistage test, which we call the General Multistage Test (GMT), that achieves asymptotically  the optimal, average over all data streams,  expected sample size  uniformly in the unknown number of signals.  Moreover, we show that the  multistage test proposed in \cite{Malloy_Nowak_2014}, Sequential Thresholding (ST), as well as a  modification of it that we introduce in this work, achieve  the same  asymptotic optimality property only subject to a sparsity condition on the maximal number of signals. These theoretical results are supported by simulation studies, in which the GMT has similar performance as ST and its modification when the number of signals is very small, and performs dramatically better otherwise.

The above theoretical results in the high-dimensional setup are based on an asymptotic analysis for the corresponding binary testing problem as the type-I and type-II error probabilities go to zero. For this problem we show that the GMT achieves the optimal expected sample size under both hypotheses, among all sequential tests with the same error control,  as the two error probabilities decay \textit{at arbitrary rates}. To the best of  our knowledge, this is the first multistage in the literature with this property.  On the other hand, ST and its  modification  are shown to achieve the optimal expected sample size only under the null hypothesis, and at the price of an inflated expected sample size under the alternative hypothesis, which can be  much worse even than that  of the corresponding  fixed-sample-size test. 

The proposed multistage test in this work, GMT,  utilizes \textit{deterministic} stage sizes.  In the case of composite hypotheses, in order to achieve an  asymptotic optimality property simultaneously  for every possible parameter value,  at least one stage size will have to be  non-deterministic and to  depend on the already collected observations.  Such an asymptotic optimality property was established for a  test with  3 stages in    \cite{Lorden_1983}, under   constraint \eqref{constraint on alpha and beta for A.O. of 3ST} on  the decay rates of $\alpha$ and $\beta$. A modification of the GMT that achieves such a property without constraints on the decay rates of $\alpha$ and $\beta$ remains an open problem that we will  consider elsewhere.
 
The multistage tests in this work can be applied with minor modifications if the average log-likelihood ratio,  $\bar{\Lambda}$,  is replaced by a different test statistic. 
Moreover,  all asymptotic upper bounds can be extended as long as that test statistic satisfies exponential bounds similar to the ones in  \eqref{LD upper bounds} or \eqref{asy LD}.     Of course, the resulting multistage tests will  not, in general, be  asymptotically optimal, but they  may offer  robustness against misspecification and/or they may be more computationally tractable, especially in the case of dependent data.  

Finally, one can combine the multiple testing problem we consider here with other error metrics, such as  false discovery/non-discovery rates \cite{benjamini1995controlling}, whereas another direction of interest is a  ``centralized'' formulation of the   multiple testing problem, which  allows using observations from all data streams to  decide when to stop sampling and which hypothesis to select for each  testing problem \cite{cohen2015active,huang2018active, lambez2021anomaly,Song_PriorInfo, Song_General,Aris2022}.  Even under the assumption of independence among the various data streams, such information can be  useful in the presence of non-trivial upper bounds on the numbers of signals and noises  or in the case of generalized error control.

\appendix
\subsection{Proofs for the Fixed-Sample-Size Test} 
\label{proofs about FSST}
\begin{IEEEproof} [Proof of Proposition \ref{prop: sprt_mixture_optimality}] 
By the definition of asymptotic optimlity and \eqref{optimal performance}, for any $\epsilon>0$ there exist $\alpha_\epsilon$ and $\beta_\epsilon$ in $(0,1)$ so that,
 for any $\alpha \in (0,\alpha_\epsilon)$ and $\beta \in (0,\beta_\epsilon)$,
$$ 1 \leq \frac{\Exp_i[T^*(\alpha,\beta)]}{\cL_i(\alpha,\beta)}\leq 1+\epsilon, \quad i \in \{0,1\}, $$
$$ 1-\epsilon\leq \frac{\Exp_0[T^*(\alpha,\beta)]}{|\log\beta|/I_0}, \; \frac{\Exp_1[T^*(\alpha,\beta)]}{|\log\alpha|/I_1} \leq 1+\epsilon,$$
and, consequently, for any $\pi\in[0,1]$, 
$$ 1\leq \frac{\Exp_\pi[T^*(\alpha,\beta)]}{\cL_\pi(\alpha,\beta)}\leq \frac{(1-\pi)\, \Exp_0[T^*(\alpha,\beta)]+\pi \, \Exp_1[T^*(\alpha,\beta)]}{(1-\pi) \, \cL_0(\alpha,\beta)+\pi \, \cL_1(\alpha,\beta)}\leq 1+\epsilon,$$
$$ 1-\epsilon\leq \frac{(1-\pi)\, \Exp_0[T^*(\alpha,\beta)]+\pi \, \Exp_1[T^*(\alpha,\beta)]}{(1-\pi)|\log\beta|/I_0 + \pi|\log\alpha|/I_1}\leq 1+\epsilon. $$
\end{IEEEproof}

The following proposition shows the well-definedness of \eqref{def: n*(alpha,beta)} and \eqref{def: c*(alpha,beta)}.
\begin{proposition} \label{well-definedness of n* and c*}
For any $\alpha,\beta\in(0,1)$, the set 
\begin{equation} \label{the set in n*}
    \left\{n \in \bN: \;  \exists \; c \in \bR \; \text{ so that } \; \Pro_0( \bar\Lambda_n > c)\leq \alpha \; \text{ and } \; \Pro_1( \bar\Lambda_n \leq c) \leq \beta \right\} 
\end{equation}
is non-empty, and the set
\begin{equation} \label{the set in c*}
    \left\{ c\in\bR: \; \Pro_0(\bar\Lambda_{\n(\alpha,\beta)}>c)\leq \alpha \; \text{ and } \; \Pro_1(\bar\Lambda_{\n(\alpha,\beta)}\leq c)\leq \beta \right\}
\end{equation}
is either a singleton or a left-closed interval.
Thus, $\n(\alpha,\beta)$ in \eqref{def: n*(alpha,beta)} and $\thre(\alpha,\beta)$ in \eqref{def: c*(alpha,beta)} are well-defined.
\end{proposition}
\begin{IEEEproof}
Fix $\alpha,\beta\in(0,1)$. By \eqref{KL information numbers} and the Weak Law of Large Numbers, for any $c\in(-I_0, I_1)$, we have
$$\Pro_0(\bar\Lambda_n>c)\to 0 \quad \text{ and } \quad \Pro_1(\bar\Lambda_n\leq c)\to 0, \quad \text{ as } n\to\infty, $$
and thus the set in \eqref{the set in n*} is non-empty. Besides, since the set in \eqref{the set in n*} is a subset of $\bN$ and is lower bounded, its minimum, $\n(\alpha,\beta)$, is well-defined and finite. Moreover, by the definition of $\n(\alpha,\beta)$, the set in \eqref{the set in c*} is non-empty. We next show that the set in \eqref{the set in c*} is either a singleton or a left-closed interval, and thus its minimum, $\thre(\alpha,\beta)$, is well-defined. We do this by showing that if the set in \eqref{the set in c*} is not a singleton, then it must be a left-closed interval. Indeed, if $c_1<c_2$ belong to this set, then for any $c_1<c<c_2$, we have
\begin{equation*}
    \begin{aligned}
    \Pro_0(\bar\Lambda_{\n(\alpha,\beta)}>c) & \leq \Pro_0(\bar\Lambda_{\n(\alpha,\beta)}>c_1)\leq \alpha, \\
    \Pro_1(\bar\Lambda_{\n(\alpha,\beta)}\leq c) & \leq \Pro_1(\bar\Lambda_{\n(\alpha,\beta)}\leq c_2)\leq \beta,
    \end{aligned}
\end{equation*}
which, by definition, imply that $c$ also belongs to this set. This proves that this set must be an interval. To show that it contains its left endpoint, we apply the right-continuity of the cumulative distribution function, which implies that (i) its left endpoint cannot be $-\infty$ because otherwise,
$$ 1 = \Pro_0(\bar\Lambda_{\n(\alpha,\beta)}>-\infty) = \lim_{c\to-\infty}\Pro_0(\bar\Lambda_{\n(\alpha,\beta)}>c) \leq \alpha,  $$
a contradiction; and (ii) for any $c\in\bR$, if $c+\epsilon$ belongs to it for any $\epsilon>0$, then $c$ also belongs to it because 
\begin{equation*}
    \begin{aligned}
    \Pro_0(\bar\Lambda_{\n(\alpha,\beta)}>c) & = \lim_{\epsilon\downarrow 0} \Pro_0(\bar\Lambda_{\n(\alpha,\beta)}>c+\epsilon)\geq \alpha, \\
    \Pro_1(\bar\Lambda_{\n(\alpha,\beta)}\leq c) & \leq \Pro_1(\bar\Lambda_{\n(\alpha,\beta)}\leq c+\epsilon)\leq \beta.
    \end{aligned}
\end{equation*}
\end{IEEEproof}

Two properties of the fixed-sample-size test that follow straightforwardly from its definition and we use extensively in the proofs about the multistage tests are that,
for any $\alpha_1,\alpha_2,\beta_1,\beta_2\in(0,1)$,
\begin{align}
    & \text{if $\alpha_1\geq\alpha_2$ and $\beta_1\geq\beta_2$, then $\n(\alpha_1,\beta_1)\leq \n(\alpha_2,\beta_2)$,} \label{obs 1} \\
    & \text{if $\n(\alpha_1,\beta_1)=\n(\alpha_2,\beta_2)$ and $\alpha_1\geq\alpha_2$, $\beta_1\leq \beta_2$, then $\thre(\alpha_1,\beta_1)\leq \thre(\alpha_2,\beta_2)$.} \label{obs 2} 
\end{align} 

\begin{IEEEproof} [Proof of Theorem \ref{Theorem, two non-asy upper bounds on n*(alpha,beta)}]
For any $\alpha,\beta\in(0,1)$ and $c\in(-I_0,I_1)$, we define
$$ \n(\alpha,\beta,c) \equiv \min \, \left\{n \in \bN: \; \Pro_0( \bar\Lambda_n > c)\leq \alpha \; \text{ and } \; \Pro_1( \bar\Lambda_n \leq c) \leq \beta \right\},$$
whose well-definedness is also proved in the proof of Proposition \ref{well-definedness of n* and c*}.
By definition,
\begin{equation*}
    \n(\alpha,\beta) \leq \min_{c\in(-I_0,I_1)}\n(\alpha,\beta,c).
\end{equation*}
It then suffices to show that, for any $\alpha,\beta\in(0,1)$ and $c\in (-I_0,I_1)$,
\begin{equation} \label{non-asy upper bound on n*(alpha,beta,t)}
    \n(\alpha,\beta,c) \leq 
    \max\left\{ \frac{|\log\beta|}{\psi_1(c)},\frac{|\log\alpha|}{\psi_0(c)} \right\} + 1.
\end{equation}
Indeed, this implies that
\begin{align}
    \n(\alpha,\beta) &\leq \min_{c\in (-I_0,I_1)} \max\left\{ \frac{|\log\alpha|}{\psi_0(c)}, \frac{|\log\beta|}{\psi_1(c)} \right\}+1. \label{non-asy upper bound on n*(alpha,beta)}
\end{align}
Since $\psi_0(c)$ (resp. $\psi_1(c)$) is strictly increasing (resp. strictly decreasing) and continuous in $c\in(-I_0,I_1)$, and its range is $(0,\infty)$ for $c\in(-I_0,I_1)$, the minimum in \eqref{non-asy upper bound on n*(alpha,beta)} is attained when the two terms in the maximum are equal, i.e., at $c=g^{-1}\left(|\log\alpha|/|\log\beta|\right)$, which proves \eqref{non-asy 1}.
The inequality \eqref{non-asy 2} then follows by noticing that 
$$ h_0(\alpha,\beta)\vee h_1(\alpha,\beta)\geq \min_{c\in(-I_0,I_1)} \{ \psi_0(c) \vee \psi_1(c) \}= \cC. $$
        
To prove \eqref{non-asy upper bound on n*(alpha,beta,t)}, we fix $\alpha,\beta\in(0,1)$ and $c\in (-I_0,I_1)$ and, for ease of notation, we write $\n(\alpha,\beta,c)$ in short as $\n$. Then, by the definition of minimum,
\begin{equation*}
    \text{either } \quad \Pro_1(\bar\Lambda_{\n-1}\leq c) > \beta \quad \text{ or } \quad \Pro_0(\bar\Lambda_{\n-1} > c) >\alpha,
\end{equation*}
which implies
$$ 1 < \max\left\{ \frac{\log\beta}{\log \Pro_1(\bar\Lambda_{\n-1}\leq c)}, \; \frac{\log \alpha}{\log \Pro_0(\bar\Lambda_{\n-1}>c)} \right\}, $$
or equivalently
\begin{equation*}
    \n-1  < \max\left\{ \frac{\log\beta}{\log \Pro_1(\bar\Lambda_{\n-1}\leq c)/(\n-1)}, \; \frac{\log \alpha}{\log \Pro_0(\bar\Lambda_{\n-1}>c)/(\n-1)} \right\}.
\end{equation*}      
Applying the inequalities in  \eqref{LD upper bounds} completes the proof. \\
\end{IEEEproof} 

\begin{IEEEproof} [Proof of Corollary \ref{Corollary, asy upper bounds on n*(alpha,beta)}] 
We only prove (i), as the proof of (ii) is similar.  By  \eqref{optimal performance} it follows that, as $\alpha,\beta\to 0$ so that $|\log\alpha|\ll |\log\beta|$, 
$$ \frac{|\log \alpha|}{I_1}  \sim  \cL_1(\alpha,\beta) \ll   \cL_0(\alpha,\beta) \sim  \frac{|\log \beta|}{I_0} .$$ By the properties of  $\psi_1$ and $g$ it follows that, as $\alpha,\beta\to 0$ so that $|\log\alpha|\ll |\log\beta|$,
$$h_1(\alpha,\beta) \to \psi_1(-I_0+) = I_0$$
and, by \eqref{non-asy 1},
$$ \n(\alpha, \beta)\lesssim \frac{|\log \beta|}{I_0}. $$
Combining these two completes the proof. \\
\end{IEEEproof}

\subsection{Proofs for the 3-Stage Test} \label{proofs about 3ST}
\begin{IEEEproof} [Proof of Proposition \ref{thm: asy opt of 3ST}] We prove the  asymptotic optimality of the 3-Stage Test only under the null, as the corresponding result under the alternative  can be proved similarly. 
For any $\alpha,\beta\in(0,1)$ and $\epsilon_0,\epsilon_1\in(0,1)$, by the design of the test we have
\begin{equation*}
   \check{T}(\alpha,\beta)  \leq 
\check N_{0,0} +     N \cdot  1\left\{ \check N_{0,0} < N, \;
\bar\Lambda_{\check N_{0,0}}> C_{0,0}\right\}.
\end{equation*}
When  $\check N_{0,0} < N$, we have 
$$N_{0,0}=\check N_{0,0} \geq \frac{|\log\beta|}{(1-\epsilon_0) I_0},$$
thus,  by the selection of   $C_{0,0}$ and $N_{0,0}$ in \eqref{Lornde's way of controlling error probs at n0, n1} we have 
$$C_{0,0}=-\frac{|\log\beta|}{N_{0,0}}\geq -(1-\epsilon_0)\, I_0, $$
and, consequently,
\begin{equation*}
\begin{aligned}
    \Exp_0[\check{T}(\alpha,\beta)] 
    &\leq  \check N_{0,0} + N \, \Pro_0\left(\bar\Lambda_{\check N_{0,0}}>-(1-\epsilon_0)\,I_0\right). 
\end{aligned}
\end{equation*}

Applying \eqref{non-asy 2}, we  obtain
\begin{equation*}
\begin{aligned}
    \Exp_0[\check T(\alpha, \beta)] 
    & \leq \left( \frac{|\log(\beta/2)|}{(1-\epsilon_0)\,I_0} + 1\right) +\left(\frac{|\log((\alpha \wedge \beta)/2)|}{\cC}+1\right)\, \Pro_0\left( \bar\Lambda_{\check N_{0,0}}>-(1-\epsilon_0)\,I_0 \right) \\
    & = \frac{|\log(\beta/2)|}{I_0}\left( \frac{1}{1-\epsilon_0} + \frac{|\log((\alpha \wedge \beta)/2)|}{|\log(\beta/2)|}\, \frac{I_0}{\cC}\, \Pro_0\left( \bar\Lambda_{\check N_{0,0}}>-(1-\epsilon_0)\,I_0 \right) \right)+2.
\end{aligned}
\end{equation*} 
If $\epsilon_0$ is selected so that \eqref{Lorden's constraint on epsilon0, epsilon1} holds as $\alpha,\beta\to 0$, then we can conclude that, as $\alpha,\beta\to 0$ so that $|\log \alpha|/|\log\beta|$ does not go to infinity, 
\begin{equation*}
    \Exp_0[\check{T}(\alpha, \beta)]
    \lesssim \frac{|\log\beta|}{I_0}.
\end{equation*} 

It remains to show that we can always find   an  $\epsilon_0$ 
that satisfies \eqref{Lorden's constraint on epsilon0, epsilon1} under our standing  assumption of \eqref{KL information numbers}. Indeed, this implies the  Weak Law of Large Numbers, according to which  $\bar\Lambda_n \to -I_0$ in probability under $\Pro_0$. This means that  for any $\epsilon\in(0,1)$ there exists $N_\epsilon\in\bN$ such that $\Pro_0\left( \bar\Lambda_n> -(1-\epsilon)\,I_0 \right)\leq \epsilon$ for every $n\geq N_\epsilon$. Thus, we get a mapping:
$$ \epsilon\in(0,1) \quad \longmapsto \quad N_\epsilon\in\bN. $$
Without loss of generality, we assume that $N_\epsilon$ is decreasing in $\epsilon$ and $N_\epsilon\to \infty$ as $\epsilon\to 0$. Now, if for any $\alpha,\beta\in(0,1)$ we set $\epsilon_0=\epsilon_0(\alpha,\beta)$ equal to 
$$\sup\{\epsilon\in(0,1): N_{\epsilon}\geq |\log(\beta/2)|/I_0\},$$ 
then as $\alpha,\beta\to 0$ we have $\epsilon_0\to 0$ and, since $\check N_{0,0}\geq |\log(\beta/2)|/I_0$,
$$ \Pro_0\left( \bar\Lambda_{\check N_{0,0}}> -(1-\epsilon_0)\,I_0 \right)\leq \epsilon_0\to 0. $$
\end{IEEEproof}

\subsection{Proofs for the General Multistage Test} \label{proofs about GMT}
\begin{IEEEproof} [Proof of Proposition \ref{GMT, make sense and error control}]
Fix $\alpha,\beta\in(0,1)$. We first show that if the GMT is designed according to \eqref{N}-\eqref{GMT, general design, 10} and conditions \eqref{cond}-\eqref{cond on c's} are satisfied, then it belongs to $\cE(\alpha,\beta)$. Indeed, by the union bound its type-I error probability is   upper bounded by 
\begin{equation} \label{upper bound on P0(GMT=1)}
    \Pro_0\left( \bar\Lambda_{N_{1,0}}> C_{1,0} \right) + \sum_{j=1}^{K_1} \Pro_0\left( \bar\Lambda_{N_{1,j}}> C_{1,j} \right) + \Pro_0\left( \bar\Lambda_N > C \right),
\end{equation}
which, by \eqref{N}, \eqref{GMT, general design, 1}, \eqref{GMT, general design, 10} and the definition of ${\sf{FSST}}$ in \eqref{sf FSS}, is further upper bounded by 
\begin{equation*}  
    \left(\frac{3\alpha}{4}-\sum_{j=1}^{K_1}\left(\frac{\alpha}{4}\right)^j\right) + \sum_{j=1}^{K_1} \left(\frac{\alpha}{4}\right)^j + \frac{\alpha}{4} = \alpha.
\end{equation*}
In a similar way we can establish the upper bound on the   type-II error probability. 

We next show that when the free parameters of GMT in  \eqref{GMT_free_parameters}  satisfy  \eqref{gamma0 >= 3alpha/4 and gamma1 >= 3beta/4}-\eqref{M0 <= M0hat and M1 <= M1hat}, then  conditions  \eqref{cond}-\eqref{cond on c's} hold.

First of all,    for any $x \in (0,1)$ we have
\begin{align*}
    3x/4-  \sum_{j=1}^\infty \left(x/4\right)^j = 3x/4-   (x/4)/(1-x/4)>x/4.
\end{align*}
Thus, by the selection of $N_{0,0}$ and $N_{1,0}$ according to \eqref{GMT, general design, 00}-\eqref{GMT, general design, 10} and \eqref{obs 1}, we have 
\begin{align} \label{summation}
\begin{split}
    N_{0,0}  &\leq \n\left( \gamma_{0,0}, \, 3\beta/4 -\sum_{j=1}^{\infty} (\beta/4)^j \right) \leq \n(\gamma_{0,0},\beta/4), \\
    N_{1,0}  &\leq \n\left(  3\alpha/4-\sum_{j=1}^{\infty} (\alpha/4)^j, \gamma_{1,0} \right) \leq \n(\alpha/4, \gamma_{1,0}).
\end{split}
\end{align}
Consequently, for  
$$N_{0,0} \vee N_{1,0}\leq N= \n(\alpha/4,\beta/4)$$ to hold, by \eqref{obs 1} it suffices that 
\begin{equation*}  
    \gamma_{0,0}\geq\alpha/4, \quad \text{ and } \quad \gamma_{1,0}\geq\beta/4,
\end{equation*}
which is implied by \eqref{gamma0 >= 3alpha/4 and gamma1 >= 3beta/4}. 

Since,  by \eqref{obs 1}, the \textit{active} type-II (resp. type-I) error probability at the $j^{th}$ opportunity to  accept (resp. reject) the null hypothesis is strictly decreasing in $j$, condition \eqref{all gammai's strictly decrease} suffices for
$$N_{i,0}\leq N_{i,1}\leq\cdots\leq N_{i,K_i}, \quad \;  i\in\{0,1\}.$$ 

For $N_{i, K_i}\leq N$ to hold for both  $ i  \in \{0,1\}$, by \eqref{obs 1} again it suffices that 
\begin{align}  \label{first condition on M0 and M1}
\begin{split}
    K_0 &\leq \max\left\{ j\in\bN: \n\left(\gamma_{0,j},\left(\beta/4\right)^j\right)\leq \n\left(\alpha/4, \beta/4\right) \right\}, \\
    K_1 &\leq \max\left\{ j\in\bN: \n\left(\left(\alpha/4\right)^j, \gamma_{1,j} \right)\leq \n\left(\alpha/4, \beta/4\right) \right\}. 
     \end{split}
\end{align}

If $N_{0,0}=N_{1,0}$, by \eqref{obs 2} it follows that  condition \eqref{gamma0 >= 3alpha/4 and gamma1 >= 3beta/4} suffices for $C_{0,0}\leq C_{1,0}$.

If $K_1\geq 1$ and $N_{0,0}=N_{1,j}$ for some $j\in [K_1]$,  by \eqref{obs 2}  it follows that for 
$C_{0,0}\leq C_{1,j}$ to hold it suffices that
\begin{equation*} 
    \gamma_{0,0}\geq (\alpha/4)^j \quad \text{ and } \quad \gamma_{1,j} \geq 3\beta/4,
\end{equation*}
and by \eqref{gamma0 >= 3alpha/4 and gamma1 >= 3beta/4} and \eqref{all gammai's strictly decrease} it is clear that for the latter to hold  it suffices that
$\gamma_{1,K_1} \geq 3\beta/4$, or equivalently
\begin{equation} \label{condition for c0<=c1j}
    K_1\leq \max\left\{ j\in\bN: \gamma_{1,j} \geq 3\beta/4 \right\}.
\end{equation}

Similarly, if $K_0\geq 1$ and $N_{1,0}=N_{0,j}$ for some $j\in \{0\}\cup[K_0]$, then  for $C_{1,0} \leq C_{0,j}$ to hold it suffices that 
\begin{equation} \label{condition for c1<=c0j}
    K_0\leq \max\left\{ j\in\bN: \gamma_{0,j} \geq 3\alpha/4 \right\}.
\end{equation}

Finally, if $K_0,K_1\geq 1$ and  $N_{0,j} =N_{1,k}$ for some 
$j\in \{0, \ldots, K_0\}$  and  $k\in \{0,\ldots,  K_1\}$, then 
by \eqref{obs 2} again  it follows that  for $C_{0,j} \leq C_{1,k}$  to hold    it suffices that
\begin{equation*} 
    \gamma_{0,j}\geq (\alpha/4)^k \quad \text{ and } \quad \gamma_{1,k}\geq (\beta/4)^j,  
\end{equation*}
which is implied by  \eqref{condition for c0<=c1j} and \eqref{condition for c1<=c0j}. Since condition \eqref{M0 <= M0hat and M1 <= M1hat} is a combination of \eqref{first condition on M0 and M1}, \eqref{condition for c0<=c1j}, and \eqref{condition for c1<=c0j}, the proof is complete. \\
\end{IEEEproof}

\begin{IEEEproof} [Proof of Theorem \ref{Theorem, optimality of GMT}]
We only prove (i) as the proof of (ii) is analogous.
By  \eqref{summation} and  the upper bound in  \eqref{non-asy 1} we obtain 
\begin{align*}
    N_{0,0}\leq \n(\gamma_{0,0},\beta/4)\leq \frac{|\log(\beta/4)|}{h_1(\gamma_{0,0},\beta/4)} + 1.  
 \end{align*}
For each $j \in [K_0]$, by the selection of  $\gamma_{0,j}$ according to  \eqref{GMT, opt condition} and the upper bound in  \eqref{non-asy 2} we have
\begin{align*}
    N_{0,j}=\n\left(  (\beta/ 4 )^j, ( \beta/4 )^j \right) &\leq j\,  \frac{|\log(\beta/4)|}{\cC} + 1.
\end{align*}
Finally, by the selection of $K_0$ as $\widehat{K_0}$  in \eqref{M} it follows that either 
$$  \n\left(\alpha/4, \beta/4 \right) < \n\left((\beta/4)^{K_0+1}, (\beta/4)^{K_0+1}\right) $$
or 
$$ (\beta/4)^{K_0+1}<3\alpha/4,$$
and the latter inequality implies that
$$ (\beta/4)^{K_0+2}  <\alpha/4.
$$
As a result, by \eqref{obs 1} it follows that,  in either case,  
$$ N=\n\left(\alpha/4, \beta/4 \right) \leq \n\left(\left( \beta/4 \right)^{K_0+2},\left(\beta/4 \right)^{K_0+2}\right) \leq (K_0+2) \, \frac{|\log(\beta/4)|}{\cC}+1,$$
where for  the second inequality we use the upper bound in  \eqref{non-asy 2}. Applying the above inequalities to \eqref{GMT, ESS under P0}, we obtain
\begin{align} \label{GMT, ESS0}
    & \; \Exp_0[\GMT(\alpha, \beta)] \nonumber\\
    \leq & \; \left(\frac{|\log(\beta/4)|}{h_1(\gamma_{0,0},\beta/4)}+1\right) + \left(\frac{|\log(\beta/4)|}{\cC}+1\right) \cdot \gamma_{0,0} + 
    \sum_{\substack{2\leq j\leq K_0 \\ \text{or } j = K_0+2}}
    \left(j\frac{|\log(\beta/4)|}{\cC} +1\right) \cdot \left(\frac{\beta}{4}\right)^{j-1}  \nonumber\\
    \leq & \; \frac{|\log(\beta/4)|}{h_1(\gamma_{0,0},\beta/4)} + \frac{|\log(\beta/4)|}{\cC}\left( \gamma_{0,0} + \sum_{j=2}^{\infty} j\, \left(\frac{\beta}{4}\right)^{j-1} \right)  + \left( 1+\gamma_{0,0}+\sum_{j=2}^{\infty} \left(\frac{\beta}{4}\right)^{j-1} \right)  \nonumber\\
    \leq & \; \frac{|\log(\beta/4)|}{h_1(\gamma_{0,0},\beta/4)}+ \frac{|\log(\beta/4)|}{\cC} \left(\gamma_{0,0}+\beta\right)  + \left(1+\gamma_{0,0}+\beta\right).
\end{align}
If $\gamma_{0,0}$ is selected so that \eqref{condition on gamma} holds, then from Corollary \ref{Corollary, asy upper bounds on n*(alpha,beta)}.(ii) we conclude that, as $\alpha, \beta \to 0$, 
\begin{equation*}
    \Exp_0[\check{T}(\alpha, \beta)]\lesssim \frac{|\log\beta|}{I_0}.
\end{equation*} 
\end{IEEEproof}

\begin{IEEEproof} [Proof of Proposition \ref{proposition, GMT}] By  \eqref{hi in Gaussian case}  and \eqref{GMT, ESS0} we have
$$ \Exp_0[\GMT] \leq \frac{|\log(\beta/4)|}{I} \left( \left( 1+\sqrt{\frac{|\log\gamma_{0,0}|}{|\log(\beta/4)|}} \right)^2 + 4 (\gamma_{0,0}+\beta) \right) + (1+\gamma_{0,0}+\beta). $$
If $\gamma_{0,0}$ is selected to satisfy \eqref{condition on gamma}, then, as $\alpha,\beta\to 0$, 
\begin{equation} \label{GMT, higher-order term}
\begin{aligned}
    \Exp_0[\GMT] & \lesssim \frac{|\log\beta|}{I} \left( 1+ 2\sqrt{\frac{|\log\gamma_{0,0}|}{|\log\beta|}} + 4 \gamma_{0,0} \right).
\end{aligned}
\end{equation}
Further selecting $\gamma_{0,0}$ according to  \eqref{Higher-order, selection of gamma} completes the proof.  \\
\end{IEEEproof}

\subsection{Proofs for the Sequential Thresholding and its modification} \label{proofs about ST}
We start with a lemma that provides  non-asymptotic upper bounds for the stage sizes of ST and mod-ST, as well as for the expected sample sizes of the two tests under the null hypothesis. 

\begin{lemma} \label{lemma:new for upper bounds of ST and mod-ST}
Suppose that the parameters  of ST and mod-ST are selected according to \eqref{ST and MST, type-II error in latter stages}-\eqref{ST and MST, joint}. 
\begin{itemize}
\item [(i)] For the stage sizes of ST we have, for every $j \in [K]$, 
\begin{align} \label{ST_stage_sizes}
    m_j  &\leq \n\left( \alpha^{1/K}, (\beta/2)^j \right) \leq  j \, \frac{|\log(\beta/2)|}{h_1 \left(\alpha^{1/K},  \beta/2 \right) } + 1.
\end{align} 
\item [(ii)] For the stage sizes of mod-ST we have, for every $j \in [K]$, 
\begin{align}\label{mod-ST_stage_sizes}
    m_1+ \ldots + m_j  &\leq \n\left( \alpha^{j/K}, (\beta/2)^j \right) \leq j \, \frac{|\log(\beta/2)|}{h_1 \left(\alpha^{1/K},  \beta/2 \right) } + 1.
\end{align} 
\item [(iii)]  Both $\Exp_0[\ST(\alpha, \beta)]$ and  $\Exp_0[T''(\alpha, \beta)]$ are bounded above by
\begin{equation} \label{ST, ESS0}
\begin{aligned}
    \frac{|\log(\beta/2)|}{h_1 \left(\alpha^{1/K},  \beta/2\right)} \; \left( 1-\alpha^{1/K} \right)^{-2} + \left( 1-\alpha^{1/K} \right)^{-1}.
\end{aligned}
\end{equation}
\end{itemize}
\end{lemma}
 
 \begin{IEEEproof}
(i)  The second inequality  in \eqref{ST_stage_sizes} follows  by  \eqref{non-asy 1},  according to which,  for every  $j \in [K]$, 
\begin{align} \label{upper_bound}
    \n\left( \alpha^{1/K}, (\beta/2)^j \right) &\leq\frac{|\log(\beta/2)^j|}{h_1 \left(\alpha^{1/K},  (\beta/2)^j \right) } + 1 \leq j \; \frac{|\log(\beta/2)|}{h_1 \left(\alpha^{1/K},  \beta/2 \right) } + 1,
\end{align}
where for the second inequality in \eqref{upper_bound} we use the fact  that  the function $h_1$ in \eqref{definition of hi} is decreasing in its second argument.  For $j\in\{2,\ldots,K\}$, the first inequality in \eqref{ST_stage_sizes} holds with equality by the selection of $(m_j,b_j)$ according to \eqref{mj, bj}.  Finally, by the selection of $(m_1,  b_1)$ according to \eqref{m1, b1}  and the fact that 
$$\sum_{j=2}^\infty (\beta/2)^j=\frac{(\beta/2)^2}{1-\beta/2}\leq \beta/2,$$ we obtain 
\begin{align} \label{upper bound on m1}
    m_1 &= n\left(\alpha^{1/K},\,\beta -  \sum_{j=2}^K (\beta/2)^j\right) \leq \n\left(\alpha^{1/K},\,\beta/2\right),
\end{align}
which proves the first inequality  in \eqref{ST_stage_sizes} for $j=1$. 

(ii)  The second inequality in  \eqref{mod-ST_stage_sizes}  follows by applying  \eqref{non-asy 1}, according to which,  for every  $j \in [K]$, 
\begin{align} \label{where for the equality}
\n\left( \alpha^{j/K}, (\beta/2)^j \right) &\leq
    \frac{|\log(\beta/2)^j|}{h_1 \left(\alpha^{j/K},  (\beta/2)^j \right) } + 1
    = j \; \frac{|\log(\beta/2)|}{h_1 \left(\alpha^{1/K},  \beta/2 \right) } + 1,
\end{align}
where for the equality in \eqref{where for the equality} we use the  following property of the function $h_1$ in \eqref{definition of hi}:
$$h_1(x^r, y^r)=h_1(x,y), \quad \forall \; x,y\in(0,1),\, r>0.$$ 
It remains to prove the first  inequality in  \eqref{mod-ST_stage_sizes}.  When $j=1$, this follows  from \eqref{upper bound on m1}. Thus, it suffices to show that if it  holds for some $j-1 \in [K-1]$, then it will also hold for $j$.  Indeed, by the induction hypothesis and \eqref{obs 1}, 
$$ M_{j-1} \equiv m_1+\cdots+m_{j-1}\leq \n\left(\alpha^{(j-1)/K}, (\beta/2)^{j-1}\right)\leq \n\left(\alpha^{j/K}, (\beta/2)^j\right),$$
and by \eqref{mj defined by joint probs} we obtain
\begin{align*} 
    & \; M_{j-1}+m_j   \\
    \leq & \; M_{j-1} + \min\bigg\{   n\in\bN:  \exists\, b\in\bR \text{ such that } \Pro_0\bigg( \bar\Lambda_{M_{j-1}+n}> b \bigg)\leq \alpha^{j/K} \text{ and }   \\
    &  \qquad \qquad \qquad \qquad \qquad \qquad \qquad \quad \; \qquad  \Pro_1\left( \bar\Lambda_{M_{j-1}+n}\leq b \right)\leq (\beta/2)^j  \bigg\} \\
    = & \; \n\left(\alpha^{j/K}, (\beta/2)^j\right),
\end{align*}
where the equality follows from the definition of $\n$ in \eqref{def: n*(alpha,beta)}. 

(iii)  For both ST and mod-ST,     the expected sample size under the null hypothesis is of the form
\begin{equation} \label{ST, ESS}
    m_1 +\sum_{j=2}^K m_j \; \Pro_0\left( \bigcap_{i=1}^{j-1} \Lambda'_i>b_i \right).
\end{equation} 
From \eqref{ST and MST, joint}, this is 
upper bounded by  
\begin{equation*} 
    m_1+\sum_{j=2}^K m_j  \, \alpha^{(j-1)/K},
\end{equation*} 
and, from  (i) and (ii) of the  lemma,  the latter is further upper bounded by 
\begin{equation} 
\begin{aligned}
    \frac{|\log(\beta/2)|}{h_1 \left(\alpha^{1/K},  \beta/2\right)} \; \sum_{j=1}^K j\,\alpha^{(j-1)/K} + \sum_{j=1}^K \alpha^{(j-1)/K}.
\end{aligned}
\end{equation}
Replacing the upper limit $K$ in each sum by $\infty$, we obtain \eqref{ST, ESS0}. \\
\end{IEEEproof}
\begin{IEEEproof} [Proof of Theorem \ref{Theorem, ST}]
When $K$ is selected so that \eqref{ST, condition on K, transformation} holds, by Corollary \ref{Corollary, asy upper bounds on n*(alpha,beta)}  and  Lemma \ref{lemma:new for upper bounds of ST and mod-ST}(iii) we conclude that 
$$ \Exp_0[\ST(\alpha, \beta)], \;\Exp_0[T''(\alpha, \beta)]\lesssim \frac{|\log\beta|}{I_0},$$
which proves the asymptotic optimality of both ST and mod-ST under the null hypothesis. Moreover,  since for both tests  it is possible to reject the null hypothesis only at the last stage, the expected sample size under the   alternative is bounded below by $ (1-\beta) \cdot (m_1+\ldots+ m_K)$ and, as a result, it is equal, to a first-order asymptotic approximation  as $ \beta \to 0$,   to the maximum possible sample size,  $m_1+\ldots+ m_K$. By Lemma \ref{lemma:new for upper bounds of ST and mod-ST}(ii) with  $j=K$ we obtain  the asymptotic upper bound on the expected sample size  of mod-ST under the alternative, i.e., 
\begin{align} \label{more}
    \Exp_1[T''] &\sim   \sum_{j=1}^K m_j\lesssim  K\,\frac{|\log\beta|}{I_0}.
\end{align}

It remains to show the following asymptotic approximation for the stage sizes of ST:
\begin{align} \label{show}
    m_j &\sim j \frac{|\log\beta|}{I_0}, \quad \forall \; j \in [K].
\end{align}   
Indeed, this implies the  asymptotic approximation to the expected sample size of ST under $H_1$, i.e., 
\begin{align*} 
    \Exp_1[\ST]  &\sim  \sum_{j=1}^K m_j \sim  \frac{|\log\beta|}{I_0} \sum_{j=1}^K j = \frac{|\log \beta|}{I_0} \frac{K(K+1)}{2}
\end{align*}
which, in view of  \eqref{ST, condition on K} and \eqref{optimal performance}, implies 
\begin{align*}
    \Exp_1[T'] \gg  \frac{K+1}{2} \, \cL_1(\alpha, \beta),
\end{align*}
and, in view of \eqref{more}, 
\begin{align*}
    \Exp_1[T''] &\lesssim  \frac{2}{K+1} \,     \Exp_1[T''(\alpha, \beta)].
\end{align*}
The asymptotic upper bounds in \eqref{show} follow from Lemma \ref{lemma:new for upper bounds of ST and mod-ST}(i).  Therefore, it suffices to establish the corresponding  asymptotic lower bounds.  From  \eqref{m1, b1}, \eqref{mj, bj}, and \eqref{obs 1}  it follows that, for any $\alpha, \beta \in (0,1)$, 
\begin{align*}
    m_1 &= n\left(\alpha^{1/K},\,\beta -  \sum_{j=2}^K (\beta/2)^j\right)  \\
    &\geq  \n(\alpha^{1/K},\beta)  \geq \cL_0(\alpha^{1/K},\beta),
\end{align*} 
and, for every $j \in \{2, \ldots, K\}$, 
\begin{align*}
    m_j &= \n\left( \alpha^{1/K}, (\beta/2)^j \right) \geq  \cL_0\left(\alpha^{1/K}, (\beta/2)^j \right).  
\end{align*} 
Then, by  the asymptotic approximation to the optimal expected sample size in \eqref{optimal performance} it follows that,  as $\alpha, \beta \to 0$ so that $\alpha^{1/K} \to 0$,  
\begin{align*}
    m_j &\gtrsim j \frac{|\log\beta|}{I_0}, \quad \forall \; j \in [K],
\end{align*} 
which completes the proof. \\
\end{IEEEproof}

\begin{IEEEproof} [Proof of Proposition \ref{Proposition: higher-order, ST and mod-ST}]
By \eqref{hi in Gaussian case} and  \eqref{ST, ESS0} we have 
$$ \Exp_0[\ST],\; \Exp_0[T''] \leq \frac{|\log(\beta/2)|}{I} \left( 1+\sqrt{\frac{|\log\alpha^{1/K}|}{|\log(\beta/2)|}} \right)^2 \left( 1-\alpha^{1/K} \right)^{-2} + \left( 1-\alpha^{1/K} \right)^{-1}. $$
 If $\alpha^{1/K}$ satisfies \eqref{ST, condition on K, transformation} as $\alpha,\beta\to 0$, then 
\begin{equation} \label{ST, higher-order term}
    \Exp_0[\ST],\; \Exp_0[T'']
    \lesssim \frac{|\log\beta|}{I} \left( 1+ 2\sqrt{\frac{|\log\alpha^{1/K}|}{|\log\beta|}} + 2\alpha^{1/K} \right).
\end{equation}
If, further, $\alpha^{1/K}$ is selected as in \eqref{ST, selection of K}, we obtain  \eqref{higher-order upper bound on ST and MST}. \\
\end{IEEEproof}

\subsection{Proofs for the High-dimensional Signal Recovery Problem} \label{proofs about hign-dim}
To prove the results in Section \ref{sec: problem formulation about high-dim}, we state and prove two supporting lemmas. 
\begin{lemma} \label{lemma for equivalence of classes}
For any $\alpha, \beta \in (0,1)$ and $m\in\bN$,
$$\cE_m(\alpha,\beta) =\cE\big(\alpham, \betam\big),$$  where $\alpha_m$ and $\beta_m$ are given by \eqref{alpham, betam}.     
\end{lemma}

\begin{IEEEproof}
Fix $\alpha,\beta\in(0,1)$ and $m\in\bN$.
For any $\cA\subseteq [m]$ with $l_m\leq |\cA|\leq u_m$,   $j\in [m]$, $i \in \{0,1\}$ and
$(T,D) \in \cE$, we have
\begin{align*}
    \Pro_\cA(D^j=i) &=\begin{cases}
    \begin{aligned}
    & \Pro_1(D=i), \quad && \text{if } j \in \cA \\
    & \Pro_0(D=i), \quad && \text{if } j\in \cA^c,
\end{aligned}
\end{cases}
\end{align*}
and, by the assumption of  independence among the data streams, we obtain %
\begin{equation*} \label{high-dim, simplification of error probs}
\begin{aligned}
    \text{FWE-I}_\cA(T,D) &= \Pro_\cA\left( \bigcup_{j\in \cA^c} \{D^j=1\} \right)
    =1-\Pro_\cA\left( \bigcap_{j\in \cA^c} \{D^j=0 \}\right)\\ 
    &=1-\big( 1-\Pro_0(D=1) \big)^{m-|\cA|} \leq 1-\big( 1-\Pro_0(D=1) \big)^{m-l_m},
\end{aligned}
\end{equation*}
and, similarly,
\begin{equation*}
    \text{FWE-I}_\cA(T,D) \leq 1-\big( 1-\Pro_1(D=0) \big)^{u_m}.
\end{equation*}
Therefore,  $(T,D) \in \cE_m(\alpha, \beta)$ 
if and only if 
\begin{align*}
1-\big( 1-\Pro_0(D=1) \big)^{m-l_m} & \leq\alpha,\\
1-\big( 1-\Pro_1(D=0) \big)^{u_m} & \leq\beta,
\end{align*}
or, equivalently, 
\begin{align*}
\Pro_0(D=1) &\leq  1- (1-\alpha)^{1/(m-l_m)}=\alpham ,\\
\Pro_1(D=0) &\leq  1-(1-\beta)^{1/u_m}= \betam,
\end{align*}
i.e.  $(T,D) \in \cE(\alpham, \betam)$ . \\
\end{IEEEproof} 

\begin{lemma} \label{lemma: log alpham ~ log(m-lm)}
Fix $\alpha, \beta \in (0,1)$ and for any $m\in\bN$, let $\alpha_m, \beta_m$ be defined as in \eqref{alpham, betam}. 
\begin{itemize}
\item[(i)]  If as $m\to\infty$, $u_m\to \infty$, then 
$ |\log \beta_m| \sim  \log u_m.$
\item[(ii)]  If as $m\to\infty$,  $m-l_m \to \infty$, then
$|\log \alpha_m| \sim  \log (m- l_m)$.
\item [(iii)] If as $m\to\infty$, $u_m, m-l_m\to\infty$, then
\begin{align*}
 \cL_0(\alpham,\betam) &\sim \frac{\log u_m}{I_0}, \\
\cL_1(\alpham,\betam) &\sim \frac{\log(m-l_m)}{I_1}. 
\end{align*}
\end{itemize}
\end{lemma}

\begin{IEEEproof} [Proof of Lemma \ref{lemma: log alpham ~ log(m-lm)}]
(i) For any fixed $\beta \in (0,1)$ we can  find some constant $C\in\bR$ based on $\beta$ such that
$$ \exp\{-C \beta\} \leq 1-\beta \leq \exp\{-\beta\}.$$
Thus, for any $m \in \bN$, by the definition of 
$\beta_m$ in  \eqref{alpham, betam} we obtain 
$$ 1-\exp\{-\beta/u_m\}\leq  \beta_m \leq 1-\exp\{-C \beta / u_m\}. $$
For $u_m$ large enough, $C\beta/u_m<1$. Moreover,  for every $x\in(0,1),$ 
 $$ x/2\leq x-x^{2}/2 \leq 1-e^{-x} \leq x. $$
 Therefore, for $u_m$ large enough, 
$$ \beta/2u_m\leq  \beta_m \leq C\beta/u_m. $$
Taking logarithms  and letting $u_m\to \infty$ completes the proof.

(ii) The proof is similar to that of (i) and is omitted. 

(iii) This follows from (i), (ii) and Proposition \ref{prop: sprt_mixture_optimality}. \\
\end{IEEEproof}  

\begin{IEEEproof} [Proof of Theorem \ref{thm: sharpness}]
We only prove the first statement, as the second can be proved similarly. Thus, we  assume that    $u_m\not\to\infty$ as $m\to \infty$, which means that there exists  an $M\in\bN $ and a strictly  increasing sequence of positive integers $(m_k)$   such that $m_k  \to \infty$ as $k\to\infty$ and 
$$u_{m_k} = M, \quad \forall \; k \in \bN.$$
Fix arbitrary $\alpha, \beta \in (0,1)$. Then, by the definition of  $\alpham$ and $\beta_m$ in \eqref{alpham, betam} we have 
\begin{align} \label{subseq}
    \beta_{m_k}  &= 1-(1-\beta)^{1/M}, \quad \forall \; k \in \bN,
\end{align} 
which is a constant,
whereas  $\alpha_{m_k} \to 0$ as $k \to \infty$. In particular, since 
$$ m_k-l_{m_k}\geq m_k-u_{m_k} = m_k-M \to \infty,$$
by Lemma \ref{lemma: log alpham ~ log(m-lm)}.(ii) it follows that
$$|\log  \alpha_{m_k}| \sim \log m_k.$$ 
Thus, even though $(\beta_{m_k})$ is bounded, we have (see, e.g., \cite[Lemma F.2]{Song_General})
$$ \Exp_1[\widetilde T(\alpha_{m_k},\beta_{m_k})] \lesssim   \frac{|\log\alpha_{m_k}|}{I_1}  \sim  \frac{\log m_k }{I_1},$$
which further  implies 
\begin{equation} \label{core step}
\begin{aligned}
    \cL_{M/m_k}(\alpha_{m_k},\beta_{m_k}) 
    &\leq   \frac{M}{m_k}\, \Exp_1[\widetilde T(\alpha_{m_k},\beta_{m_k})] + \left(1- \frac{M}{m_k} \right) \,  \Exp_0[\widetilde T(\alpha_{m_k},\beta_{m_k})] \\
    &\lesssim \frac{M}{m_k} \frac{\log m_k  }{I_1}  + \left(1- \frac{M}{m_k} \right) \,  \Exp_0[\widetilde T(\alpha_{m_k},\beta_{m_k})]\\
    & \sim  \Exp_0[\widetilde T(\alpha_{m_k},\beta_{m_k})] .
\end{aligned}
\end{equation}
Now, if  $\chi^*$  is  a family of tests that is asymptotically optimal in the high-dimensional sense, then by the definition of this notion of asymptotic optimality in Definition \ref{def of AO in high-dim, uniform}, setting  $s=u_{m_k}=M$ for every $k\in\bN$  we have,  as $k \to \infty$,  
\begin{align*}
    \cL_{M/m_k}(\alpha_{m_k},\beta_{m_k})
    &\sim\Exp_{M/m_k}[T^*(\alpha_{m_k},\beta_{m_k})] \\
    &=   \frac{M}{m_k} \,  \Exp_0[T^*(\alpha_{m_k},\beta_{m_k})]  + \left(1-\frac{M}{m_k}\right) \Exp_0[T^*(\alpha_{m_k},\beta_{m_k})]  \\
    &\geq   \left(1-\frac{M}{m_k}\right) \Exp_0[T^*(\alpha_{m_k},\beta_{m_k})]  \\
    &\sim  \Exp_0[T^*(\alpha_{m_k},\beta_{m_k})].
\end{align*}
Combining these two asymptotic bounds and  \eqref{subseq},
which all hold for arbitrary $\alpha, \beta \in (0,1)$, we conclude that
$$ \Exp_0\left[T^*\left(\alpha_{m_k},1-(1-\beta)^{1/M}\right)\right]\lesssim \Exp_0\left[\widetilde T\left(\alpha_{m_k},1-(1-\beta)^{1/M} \right) \right], \quad \forall \; \alpha, \beta \in (0,1),$$
or equivalently,
$$ \Exp_0\left[T^*\left(\alpha_{m_k},\beta\right)\right]\lesssim \Exp_0[\widetilde T\left(\alpha_{m_k},\beta \right)], \quad \forall \; \alpha, \beta \in (0,1).$$ 
This completes the proof. \\
\end{IEEEproof}

\begin{IEEEproof} [Proof of Theorem \ref{thm: main in high-dim}]
(i) By Lemma \ref{lemma: log alpham ~ log(m-lm)}, $u_m ,m-l_m\to\infty$ implies $\alpham,\betam\to 0$ as $m\to\infty$.  Therefore,  by Proposition \ref{prop:  sprt_mixture_optimality}  we have:
\begin{equation*}
\begin{aligned}
    \cL_{s/m}(\alpham,\betam)     & \sim \left(1-\frac{s}{m}\right)\frac{\log u_m}{I_0} + \frac{s}{m} \frac{\log(m-l_m)}{I_1} \text{ uniformly in } s\in\{l_m,\ldots,u_m\}.
\end{aligned}
\end{equation*}

To show that the family $\chi^*(\alpha, \beta)$ is asymptotically optimal in a high-dimensional sense, by \eqref{def of AO in high-dim, max}  it follows that  it suffices to show that, as $m\to\infty$,
\begin{equation*}
    \sup_{\pi\in \left[ l_m /m, \, u_m/m\right]} 
    \frac{\Exp_\pi[T^*(\alpham,\betam)]}{\cL_\pi(\alpham,\betam)} \to 1, \quad \forall\,\alpha,\beta\in(0,1).
\end{equation*}

(ii) When \eqref{E0 sim L0}-\eqref{E1 sim L1}  hold, this can be shown as in the proof of Proposition \ref{prop:  sprt_mixture_optimality}.  

(iii) Alternatively, when \eqref{E0 sim L0} and \eqref{high-dim, ST or mod-ST, alternative} hold, for any $\pi$ in  $[l_m/m, u_m/m]$, we have
\begin{align}
\frac{\Exp_\pi[T^*(\alpham,\betam)]}{\cL_\pi(\alpham,\betam)} & \leq 
    \frac{(1-\pi) \, \Exp_0[T^*(\alpham,\betam)]+\pi \,  \Exp_1[T^*(\alpham,\betam)]}{(1-\pi) \, \cL_0(\alpham,\betam)}  \nonumber\\
    & \leq \frac{\Exp_0[T^*(\alpham,\betam)]}{\cL_0(\alpham,\betam)}+\frac{u_m \, \Exp_1[T^*(\alpham,\betam)]}{(m-u_m) \, \cL_0(\alpham,\betam)}. \label{opt only under H0}
    \end{align}
As $m\to\infty$ so that $u_m,m-l_m\to\infty$, by Lemma \ref{lemma: log alpham ~ log(m-lm)} it follows that $\alpham,\betam\to0$ and $\cL_0(\alpham,\betam)\sim \log u_m/I_0$. Therefore, the  first term in    \eqref{opt only under H0} goes to 1 as $m\to \infty$  by \eqref{E0 sim L0}  and  the second term  goes to 0    as $m\to \infty$ by \eqref{high-dim, ST or mod-ST, alternative}. \\
\end{IEEEproof}

\begin{IEEEproof} [Proof of Corollary \ref{corollary of SPRTs, GMTs  in high-dim}]
According to \eqref{optimal performance} (resp. Theorem \ref{Theorem, optimality of GMT}), the family of SPRTs (resp. GMTs) satisfies \eqref{E0 sim L0} and \eqref{E1 sim L1}. \\
\end{IEEEproof}

\begin{IEEEproof} [Proof of Corollary \ref{corollary of  3STs in high-dim}] According to Theorem \ref{thm: asy opt of 3ST}, the family of 3-Stage Tests satisfies \eqref{E0 sim L0} and \eqref{E1 sim L1}   if also $|\log\alpham|=\Theta(|\log\betam|)$ as $m \to \infty$, which, from Lemma \ref{lemma: log alpham ~ log(m-lm)}, is equivalent to $\log(m-l_m)=\Theta(\log u_m)$. \\
\end{IEEEproof}

To prove Corollary \ref{corollary, ST in high-dim}, we need the following lemma.
\begin{lemma} \label{lemma for um<<m implies ...}
If $u_m \to \infty $ so that  $u_m\ll m$ as $m\to\infty$,  then 
$$ \frac{u_m}{\log u_m} \ll \frac{m}{\log m} \text{ as } m\to\infty. $$
\end{lemma}
\begin{IEEEproof}
For  $u_m >e$ we have 
\begin{equation*}
\begin{aligned}
    \frac{u_m}{\log u_m} & = \frac{u_m}{\log u_m} \cdot 1\{u_m\geq \sqrt{m} \} + \frac{u_m}{\log u_m} \cdot 1\{u_m< \sqrt{m}\} \\
    & \leq \frac{2u_m}{\log m} \cdot 1\{u_m \geq  \sqrt{m} \} + \sqrt{m} \cdot 1\{u_m<\sqrt{m} \} \\
    & \leq \max\left\{ \frac{2u_m}{\log m}, \sqrt{m} \right\},
\end{aligned}
\end{equation*}
where $1\{\cdot\}$ is the  indicator function. As $m \to \infty$, $\sqrt{m} \ll m/ \log m$. If also $ u_m \to \infty $ so that  $u_m\ll m$, then  it is clear that 
$$\max\left\{ \frac{2u_m}{\log m}, \sqrt{m} \right\} \ll 
\frac{m}{\log m},$$
and this completes the proof. \\
\end{IEEEproof}

\begin{IEEEproof} [Proof of Corollary \ref{corollary, ST in high-dim}] 
(i)  When $u_m\to\infty$ and $u_m\ll m$ as $m\to\infty$, by \eqref{asy approx to cLs/m(alpham,betam)} and Lemma \ref{lemma for um<<m implies ...} we have, as $m\to\infty$,
\begin{equation*} 
\begin{split}
    \frac{\log u_m}{I_0} & \sim \left(1-\frac{u_m}{m}\right)\frac{\log u_m}{I_0} \lesssim \cL_{s/m}(\alpham,\betam)\\
    & \lesssim \frac{\log u_m}{I_0} + \frac{u_m}{m} \frac{\log m}{I_1} \sim \frac{\log u_m}{I_0} \text{ uniformly in } s\in\{l_m,\ldots,u_m\}.
\end{split}
\end{equation*} 

(ii) By Theorem \ref{thm: main in high-dim}, it suffices to show that, for any $\alpha,\beta\in(0,1)$, \eqref{E0 sim L0} and \eqref{high-dim, ST or mod-ST, alternative} hold for $\chi'$ and $\chi''$. To this end, we first observe that $u_m\to\infty$ and $u_m\ll m$ as $m\to\infty$ imply that $m-l_m \sim m\to\infty$, since 
$$m\geq m-l_m\geq m-u_m\sim m. $$ 
By Theorem \ref{Theorem, ST}, \eqref{E0 sim L0} is satisfied by $\chi'$ and $\chi''$ as long as, for every $m\in\bN$, the maximum number of stages, $K_m$, can be selected so that, as $m\to\infty$,
\begin{equation*}
    \frac{|\log\alpham|}{|\log\betam|} \ll \;  K_m \ll  |\log\alpham|.
\end{equation*}
In view of Lemma \ref{lemma: log alpham ~ log(m-lm)}, this is equivalent to
\begin{equation} \label{condition on Km, 1}
\begin{aligned}
    \frac{\log m}{\log u_m} \ll \; & K_m \ll  \log m,
\end{aligned}
\end{equation} 
which is always feasible given $u_m\to\infty$. 

By \eqref{ST, asy upper bound on E1[T']}  and Lemma \ref{lemma: log alpham ~ log(m-lm)} we have
\begin{align*}
    \Exp_1[\ST(\alpham,\betam)] 
    &\sim \frac{K_m(K_m+1)}{2} \frac{|\log \beta_m|}{I_0}    \\
    &\sim \frac{K_m(K_m+1)}{2} \frac{\log u_m}{I_0}\leq K_m^2 \, \frac{\log u_m}{I_0},
\end{align*}
while by \eqref{ST, asy upper bound on E1[T'']} and Lemma \ref{lemma: log alpham ~ log(m-lm)} we have
$$ \Exp_1[T''(\alpham,\betam)]\lesssim K_m \frac{|\log\betam|}{I_0}\sim K_m\frac{\log u_m}{I_0}. $$
Thus, \eqref{high-dim, ST or mod-ST, alternative} is satisfied by $\chi'$ if 
$$ K_m^2 \log u_m \ll \frac{(m-u_m)\log u_m}{u_m}\sim \frac{m\log u_m}{u_m}$$
or, equivalently, 
\begin{equation} \label{condition on Km, 2, for ST}
    K_m\ll \sqrt{\frac{m}{u_m}},
\end{equation}
while \eqref{high-dim, ST or mod-ST, alternative} is satisfied by $\chi''$ if 
\begin{equation} \label{condition for Km, 2, for mod-ST}
    K_m \ll \frac{m}{u_m}.
\end{equation}
A condition that guarantees the existence of $(K_m)$ that satisfies \eqref{condition on Km, 1} and \eqref{condition on Km, 2, for ST} simultaneously is 
\begin{equation} \label{for ST}
    \frac{\log m}{\log u_m} \ll \sqrt{\frac{m}{u_m}},
\end{equation}
while a condition that guarantees the existence of $(K_m)$ that satisfies \eqref{condition on Km, 1} and \eqref{condition for Km, 2, for mod-ST} simultaneously is 
\begin{equation} \label{for mod-ST}
    \frac{\log m}{\log u_m} \ll \frac{m}{u_m}.
\end{equation}
By Lemma \ref{lemma for um<<m implies ...}, both \eqref{for ST} and \eqref{for mod-ST} are implied by the condition $u_m\ll m$. The proof is complete. \\
\end{IEEEproof}

\begin{lemma} \label{lemma, exact value of alphamG and betamG}
For any $\alpha,\beta\in(0,1)$ and $m\in\bN$, 
$$\cE_m^G(\alpha,\beta)=\cE(\alpham^G,\betam^G),$$
 where $\alpham^G$ and $\betam^G$ are defined in \eqref{inverse binomial, alpha, beta}. 
\end{lemma}
\begin{IEEEproof} [Proof of Lemma \ref{lemma, exact value of alphamG and betamG}]
Fix $\alpha,\beta\in(0,1)$ and $m\in\bN$, and recall the definition of the function $\mathsf{B}(\cdot,\cdot;\cdot)$, that is used in the definition of $\alpham^G$ and $\betam^G$  in \eqref{inverse binomial, alpha, beta}, which is increasing in its first argument when  the other two arguments are fixed. 

For any $\cA\subseteq [m]$ with $l_m\leq  |\cA| \leq u_m$  and $(T,D)\in\cE$ we have
\begin{align*}
    & \;  \kappa_m-\text{GFWE-I}_\cA(T,D)  \\
    = & \;   \Pro_\cA\big( (T,D)  \; \text{makes at least}  \; \kappa_m \; \text{type-I errors out of the}  \; |\cA^c| \; \text{noises} \big) \\
    = & \; \mathsf{B}(|\cA^c|, \Pro_0(D=1); \kappa_m)   \leq  \mathsf{B}(m-l_m, \Pro_0(D=1); \kappa_m),
\end{align*}
and similarly we obtain 
\begin{align*}
    \iota_m-\text{GFWE-II}_\cA(T,D) 
    & \leq  \mathsf{B}(u_m, \Pro_1(D=0); \iota_m).
\end{align*}
Therefore, $(T,D)\in \cE_m^G(\alpha,\beta)$ if and only if 
\begin{equation*}
    \begin{aligned}
    \mathsf{B}(m-l_m, \Pro_0(D=1); \kappa_m) & \leq \alpha \\
    \mathsf{B}(u_m, \Pro_1(D=0); \iota_m) & \leq \beta,
    \end{aligned}
\end{equation*}
or, equivalently, $(T,D)\in\cE\left(\alpham^G,\betam^G\right)$. \\
\end{IEEEproof}

\begin{lemma} \label{prop: bounds on alphamG and betamG}
Let  $\alpha,\beta\in(0,1/2)$,  $\iota_m\leq u_m/2$, and $\kappa_m\leq (m-l_m)/2$. Then: 
\begin{align}
    \frac{1}{e} \;   \frac{\kappa_m }{m-l_m}  \, \alpha^{1/\kappa_m}& \leq \alpham^G \leq  e^2 \;   \frac{\kappa_m \, }{m-l_m} \alpha^{1/\kappa_m},  \label{bounds on alphamG} \\
    \frac{1}{e} \;   \frac{\iota_m}{u_m} \, \beta^{1/\iota_m}& \leq \betam^G \leq e^2 \;   \frac{ \iota_m }{u_m}\,  \beta^{1/\iota_m}, \label{bounds on betamG}
\end{align}
and, as $(m-l_m)/\kappa_m \to \infty$ and $u_m/ \iota_m \to \infty$, 
\begin{equation} \label{|log alphamG| sim}
    |\log \alpha_m^G| \sim \log \left( \frac{m-l_m}{ \kappa_m} \right) \quad \text{and} \quad |\log \beta_m^G| \sim \log \left( \frac{u_m}{\iota_m} \right).
\end{equation}
\end{lemma}

\begin{IEEEproof}
We only prove \eqref{bounds on alphamG} as \eqref{bounds on betamG} can be proved similarly. First of all, for any $p \in (0,1)$ we have
\begin{equation*} 
    {{m-l_m}\choose{\kappa_m}} \, p^{\kappa_m} \, (1-p)^{m-l_m} \leq \mathsf{B}(m-l_m,p;\kappa_m) \leq {{m-l_m}\choose{\kappa_m}} \, p^{\kappa_m},
\end{equation*}
which, using the bounds
$$ \left(\frac{n}{k}\right)^k \leq{n\choose k}\leq \left(\frac{en}{k}\right)^k, \quad \forall\;1\leq k\leq n, $$
can be further lower and upper bounded as follows:
\begin{equation} \label{bounds on binomial prob}
    \left(\frac{m-l_m}{\kappa_m}\right)^{\kappa_m} p^{\kappa_m} (1-p)^{m-l_m} \leq  \mathsf{B}(m-l_m,p;\kappa_m)   \leq \left(\frac{e \, (m-l_m)}{\kappa_m}\right)^{\kappa_m} p^{\kappa_m}.
\end{equation}

The lower bound on $\alpham^G$ in \eqref{bounds on alphamG} follows from the fact that $\alpham^G$ is defined as the \textit{largest} $p\in(0,1)$ such that $\mathsf{B}(m-l_m, p; \kappa_m)\leq \alpha$, where the latter is satisfied by any $p\leq \kappa_m\alpha^{1/\kappa_m}/e(m-l_m)$ according to \eqref{bounds on binomial prob}.

We next show the upper bound on $\alpham^G$ in \eqref{bounds on alphamG}.  Let  $p\in(0,1)$  such that $\mathsf{B}(m-l_m, p; \kappa_m)\leq \alpha$. From \eqref{bounds on binomial prob}, we have 
\begin{equation} \label{upper bound (m-lm)p}
    \begin{aligned}
    p \leq \frac{\kappa_m}{m-l_m}\alpha^{1/\kappa_m} (1-p)^{-\frac{m-l_m}{\kappa_m}}.
    \end{aligned}
\end{equation}
Moreover, the condition that $\alpha<1/2$ implies that $\kappa_m$ is strictly greater than the median of $\text{Binomial}(m-l_m,p)$, where the latter is either $\lfloor(m-l_m)p\rfloor$ or $\lceil(m-l_m)p\rceil$, and thus we have $\kappa_m\geq (m-l_m)p$, or equivalently, $p\leq\kappa_m/(m-l_m)$. This combined with the other condition that $\kappa_m\leq (m-l_m)/2$ implies $p\leq 1/2$, and thus $1-p\geq e^{-2p}$. Applying these two inequalities to \eqref{upper bound (m-lm)p}, we have 
$$ p\leq \frac{\kappa_m}{m-l_m} \alpha^{1/\kappa_m} e^{2\frac{(m-l_m)p}{\kappa_m}} \leq e^2 \frac{\kappa_m}{m-l_m} \alpha^{1/\kappa_m}. $$ 
Since this upper bound applies to any $p\in(0,1)$ such that $\mathsf{B}(m-l_m,p;\kappa_m)\leq \alpha$ and $\alpham^G$ is the largest such $p$, the upper bound is proved.

Finally, we show \eqref{|log alphamG| sim}.
Note that the lower and upper bounds in \eqref{bounds on alphamG} or \eqref{bounds on betamG} differ by a constant factor, namely $e^3$, and $\alpha^{1/\kappa_m}\in(\alpha,1)$, $\beta^{1/\iota_m}\in(\beta,1)$, where $\alpha,\beta$ are fixed. Therefore, as long as $\kappa_m/(m-l_m) \to 0$ and $\iota_m/u_m\to0$, we have $\alpham^G,\betam^G\to 0$ and
$$ \alpham^G=\Theta\left(\frac{\kappa_m}{m-l_m}\right), \quad \beta_m^G=\Theta\left(\frac{\iota_m}{u_m}\right). $$
\eqref{|log alphamG| sim} follows. \\
\end{IEEEproof}

\begin{IEEEproof} [Proof of Theorem \ref{theorem for GFWE, SPRTs, GMTs and 3STs}]
This is similar to the proof of Theorem \ref{thm: main in high-dim} and is omitted. \\
\end{IEEEproof}

\begin{IEEEproof} [Proof of Corollary \ref{corollary, GFWE, for SPRT, GMT}]
This is similar to the proof of Corollary \ref{corollary of SPRTs, GMTs in high-dim} and is omitted. \\
\end{IEEEproof}

\begin{IEEEproof} [Proof of Corollary \ref{corollary, GFWE, for  3ST}]
This is similar to the proof of Corollary \ref{corollary of  3STs in high-dim} and is omitted. \\
\end{IEEEproof}

\begin{IEEEproof} [Proof of Corollary \ref{corollary, GFWE, for ST and mod-ST}]
(i) This can be proved similarly as Corollary \ref{corollary, ST in high-dim}.(i).

(ii) $\&$ (iii) Now we prove the high-dimensional asymptotic optimality under generalized error control for $\chi'$ and $\chi''$.
By Theorem \ref{theorem for GFWE, SPRTs, GMTs and 3STs}, it suffices to show that \eqref{GFWE, for ST and mod-ST, null} and \eqref{GFWE, for ST and mod-ST, alternative} hold for $\chi'$ or $\chi''$ under condition \eqref{GFWE, condition for ST} or \eqref{GFWE, condition for mod-ST} respectively. Similar as the proof of Corollary \ref{corollary, ST in high-dim}.(ii),
\eqref{GFWE, for ST and mod-ST, null} is satisfied for both $\chi'$ and $\chi''$ if the sequence of maximum number of stages, $(K_m)$, is selected so that, as $m\to\infty$, 
\begin{equation} \label{GFWE, Km, 0}
    \frac{\log(m/\kappa_m)}{\log(u_m/\iota_m)}\sim \frac{|\log\alpham^G|}{|\log\betam^G|}\ll K_m\ll |\log\alpham^G| \sim \log(m/\kappa_m),
\end{equation}
which is always feasible given $u_m/\iota_m\to\infty$.

As in the proof of Corollary \ref{corollary, ST in high-dim}(ii), \eqref{GFWE, for ST and mod-ST, alternative} is satisfied by $\chi'$ if 
\begin{equation*}
    \Exp_1[T'(\alpham^G,\betam^G)] \lesssim K_m^2 \;  \frac{\log(u_m/\iota_m)}{I_0}\ll \frac{m\log(u_m/\iota_m)}{u_m},
\end{equation*}
or, equivalently,
\begin{equation} \label{GFWE, Km, 1, ST}
    K_m\ll \sqrt\frac{m}{u_m},
\end{equation}
and \eqref{GFWE, for ST and mod-ST, alternative} is satisfied by $\chi''$ if 
$$ \Exp_1[T''(\alpham^G,\betam^G)]\lesssim K_m\frac{\log(u_m/\iota_m)}{I_0} \ll \frac{m\log(u_m/\iota_m)}{u_m} $$
or, equivalently,
\begin{equation} \label{GFWE, Km, 1, mod-ST}
    K_m\ll \frac{m}{u_m}.
\end{equation}
A condition that guarantees the existence of $(K_m)$ that satisfies both \eqref{GFWE, Km, 0} and \eqref{GFWE, Km, 1, ST} is \eqref{GFWE, condition for ST}, while a condition
for the existence of $(K_m)$ that satisfies both
\eqref{GFWE, Km, 0} and \eqref{GFWE, Km, 1, mod-ST} is \eqref{GFWE, condition for mod-ST}. \\
\end{IEEEproof}

\bibliographystyle{unsrt}
\bibliography{main}

\end{document}